\documentclass[aps,showpacs,nofootinbib,superscriptaddress]{revtex4-1}
\usepackage{epsf}
\usepackage{bm}
\usepackage{graphicx}
\usepackage{amsfonts}
\usepackage[mathscr]{euscript}
\usepackage{amsmath}
\usepackage{color}
\usepackage[normalem]{ulem}

\newcommand{\stkout}[1]{\ifmmode\text{\sout{\ensuremath{#1}}}\else\sout{#1}\fi}

\newcommand{\beas}{\begin{eqnarray*}}
\newcommand{\eeas}{\end{eqnarray*}}
\newcommand{\bea}{\begin{eqnarray}}
\newcommand{\eea}{\end{eqnarray}}
\def\gtap{\ \raise.3ex\hbox{$>$\kern-.75em\lower1ex\hbox{$\sim$}}\ }
\def\ltap{\ \raise.3ex\hbox{$<$\kern-.75em\lower1ex\hbox{$\sim$}}\ }

\begin{document}
\title{Angular distributions in  electroweak pion production off nucleons: 
 Odd parity hadron terms, strong relative phases, and model dependence} 
\author{J.E. Sobczyk}\affiliation{Instituto de F\'\i sica Corpuscular (IFIC), Centro Mixto
CSIC-Universidad de Valencia, Institutos de Investigaci\'on de
Paterna, Apartado 22085, E-46071 Valencia, Spain}\author{E. Hern\'andez} \affiliation{Departamento de F\'\i sica Fundamental 
e IUFFyM,\\ Universidad de Salamanca, E-37008 Salamanca, Spain} 
\author{S.X. Nakamura}\affiliation{
Laborat\'orio de F\'{\i}sica Te\'orica e Computacional-LFTC, Universidade 
Cruzeiro do Sul,
S\~{a}o Paulo, SP, 01506-000, Brazil}
\author{J.~Nieves}
\affiliation{Instituto de F\'\i sica Corpuscular (IFIC), Centro Mixto
CSIC-Universidad de Valencia, Institutos de Investigaci\'on de
Paterna, Apartado 22085, E-46071 Valencia, Spain} 
\author{T.~Sato}\affiliation{RCNP, Osaka University, Ibaraki, Osaka 
567-0047, Japan}


\begin{abstract}

 The study of pion production in nuclei is important 
for signal and background determinations in current
and  future  neutrino oscillation experiments. The first step, however,  is
to understand the   pion production reactions at the free nucleon level.
We present an exhaustive study of the charged-current   and 
neutral-current   
neutrino and antineutrino  pion production off nucleons, paying  special 
attention to the angular distributions of the outgoing pion. We show, using
 general arguments, that parity violation and time-reversal odd correlations 
  in the weak differential cross sections are generated from the interference 
  between different contributions to the hadronic current that are not 
  relatively real. Next, we present a detailed comparison of three, state 
  of the art, microscopic models for electroweak pion production off nucleons, 
  and we also confront their predictions with polarized electron  data, as a 
  test of the vector content of these models. We also illustrate the 
  importance of carrying out a comprehensive  test 
   at the level of outgoing pion angular distributions, going beyond 
   comparisons done for partially integrated cross sections, where 
   model differences cancel to a certain extent. Finally, we observe 
   that all charged  and neutral current   
distributions show  sizable anisotropies, and identify channels for which 
 parity-violating effects are clearly visible. Based on the above results, we 
 conclude  that the use of  
  isotropic distributions for the pions in the center of mass  of the 
 final pion-nucleon system, as assumed by some of the Monte Carlo event 
 generators, needs to be improved by incorporating the findings of
microscopic calculations.  

\end{abstract}

\maketitle

\section{Introduction}

Knowledge of neutrino interaction cross sections is an important and
necessary ingredient in any neutrino measurement, and it is crucial for
reducing systematic errors affecting present and future neutrino
oscillation experiments. This is because 
neutrinos do not ionize the materials they are passing through, and hence neutrino detectors are
based on neutrino-nucleus interactions~\cite{Morfin:2012kn, Formaggio:2013kya,
  Alvarez-Ruso:2014bla, Katori:2016yel,Mosel:2016cwa, Alvarez-Ruso:2017oui}.

The precise determination of neutrino
oscillation parameters requires an accurate understanding of the detector
responses and this can only be achieved if nuclear effects are
under control. Before addressing the nuclear
effects, one  first  needs to fully understand  the  reaction 
mechanisms  at the hadron level. All this represents a challenge for both
 hadron and nuclear physics. From a hadron physics perspective, neutrino
 reactions
 allow us to investigate the axial structure of the
nucleon and baryon resonances, enlarging our knowledge of hadron structure
beyond what is presently inferred from experiments with hadronic and
electromagnetic probes.

Pion production is one of
the main reaction mechanisms for neutrinos with energies of a few GeV~\cite{Formaggio:2013kya}. 
The MiniBooNE~\cite{AguilarArevalo:2010bm} 
and MINER$\nu$A~\cite{Eberly:2014mra,McGivern:2016bwh} collaborations have reported high 
quality data for weak pion production in the $\Delta(1232)$ region from 
$CH_2$ and $CH$ targets, respectively. Although the best theoretical 
calculations have been unable
to reproduce  MiniBooNE data, the models implemented in event generators  have 
been more successful~\cite{Alvarez-Ruso:2017oui}. All approaches combine pion 
production off nucleons and pion final state interaction (FSI) models 
based on the analysis of previous data. The most recent MINER$\nu$A data have features
similar to the MiniBooNE data, however, event generators are unable to reproduce
 simultaneously the magnitude of both data
sets. 

Some of the differences  for pion production cross sections in nuclei
found   in different approaches
have their origin in the differences already existing in the production models
used at the free nucleon level. Thus, the first step towards putting  neutrino induced 
pion production on nuclear targets on a firm ground is to have a realistic 
model at the nucleon level.  
From this perspective, in this work we make an exhaustive study of  charged 
current (CC) and neutral current (NC) 
neutrino and antineutrino  pion production reactions off nucleons, paying  
special attention to the angular distributions of the outgoing pion. We show, 
using general arguments, that the possible dependencies on the azimuthal 
angle ($\phi^*_\pi$) measured in the final pion-nucleon center of mass (CM) system 
 are
 $1,\cos\phi^*_\pi,\cos 2\phi^*_\pi, \sin\phi^*_\pi$ and 
$\sin 2\phi^*_\pi$, and that the two latter ones  give rise to parity
 violation
 and time-reversal odd correlations  in the weak differential cross sections, 
 as already found in Refs.~\cite{Hernandez:2007qq, Hernandez:2006yg}. Here,  
 we make a detailed discussion of the origin of the parity-violating  
 contributions, and explicitly show that they  are generated from the 
 interference between different contributions to the hadronic current 
 that are not relatively real.
Next, we present a detailed comparison of three, state of the art,
 microscopic models for electroweak pion production off nucleons. One is the 
dynamical coupled-channel model (DCC) developed at Argonne
National Laboratory (ANL) and  Osaka University~\cite{Matsuyama:2006rp, 
Kamano:2013iva,Nakamura:2015rta}. This approach provides a unified
treatment of all resonance production processes. It satisfies unitarity and 
its predictions have
been extensively and successfully compared to data on $\pi N$ and
$\gamma N$ reactions up to invariant masses slightly above 2 GeV.
The second model included in this comparison is the one initiated by
T. Sato and T.S.H. Lee (SL) to describe pion production 
by photons and electrons~\cite{Sato:1996gk,Sato:2000jf}
and also by neutrinos~\cite{Sato:2003rq,Matsui:2005ns,Sato:2009de}, in the  
$\Delta(1232)$ region. In fact, one can consider the   DCC model  as an extension 
 of the SL model to higher $\pi N$ invariant masses. The last model we consider
 was initially developed 
by E. Hern\'andez, J. Nieves and M. Valverde (HNV) in Ref.~\cite{Hernandez:2007qq}, and it is based on the approximate chiral symmetry of QCD. 
The model was later improved in Refs.~\cite{Hernandez:2013jka,
Alvarez-Ruso:2015eva,Hernandez:2016yfb}, incorporating among other effects
 a partial restoration of unitarity, through the implementation of Watson 
 theorem in the $P_{33}$ pion-nucleon channel. A brief
description of these models will be given below, while further details can be
consulted in the above given references. 

Though in this work we are mainly
interested in  neutrino induced reactions, we shall dedicate
a full section  to pion electroproduction. In this way, 
we can make a direct comparison of the vector part of the different models and data. 
Since  the quality of the data is very good in this case, we can use this 
comparison to extract relevant information on the vector part of the 
models\footnote{To make this information more meaningful for the case 
of pion production by neutrinos, we will select  kinematical regions
 as close as possible to the ones examined in the case of pion
electroproduction. Thus,  most of the results that we are going to show
correspond to pion production by electron neutrinos. However, in order to
compare with actual experimental data, we will also show results for pion
production by muon neutrinos. In fact, cross sections are equal for
 NC processes, while there is not much difference for CC reactions for neutrino energies above 1\,GeV.}.  
We will show that the bulk of the DCC model predictions for electroproduction
 of pions in the $\Delta$ region could be reproduced,  with a reasonable
  accuracy, by the simpler HNV model. Given the high degree of complexity
   and sophistication of the DCC approach, we find that this validation 
   is remarkable. The HNV model 
   might be more easily implemented in the Monte Carlo event generators 
   used for neutrino oscillation analyses, and this would contribute 
   to a better theoretical control of such analyses.

Furthermore, we show that the DCC and HNV models agree reasonably well for
CC and NC neutrino and antineutrino total
cross sections, as well as for the corresponding differential cross sections
with respect to the outgoing lepton variables. With respect to the pion 
 angular dependence of the weak cross sections, we will observe, 
 first of all, that  CC and  NC  distributions show  clear anisotropies. 
 This means that using  an isotropic distribution for the pions in the 
 CM of the final pion-nucleon system, as assumed by some 
 of the Monte Carlo event generators, is not supported by the results 
 of the DCC and HNV models.  We will also illustrate the importance of 
 carrying out a comprehensive test of the different models at the level 
 of outgoing pion angular distributions, going beyond comparisons 
 done for partially integrated cross sections, where model differences 
 tend to cancel. 
Finally, we will discuss the pion azimuthal angular distributions, where  
parity violation shows up mainly through the  
 $\sin\phi^*_\pi$ term mentioned above and discussed in detail in what follows. We will show that parity violation is 
 quite significant for  NC neutrino reactions producing charged pions, and
  especially for the $\nu_e n\to e^- n\pi^+$ and  
  $\bar \nu_e p\to e^+ p\pi^-$  CC processes, where background non-resonant
   contributions are sizable. The azimuthal distributions for these weak 
   processes could provide  information on the relative phases 
of  different hadronic current contributions that would be complementary to that inferred from 
polarized electron scattering.

The work is organized as follows: In Sec.~\ref{sec:models}, we give a brief
description of the DCC and HNV models. In Sec.~\ref{sec:diffxs} we discuss  different
expressions for neutrino induced pion production differential cross sections. One
 of them  makes explicit the dependence on the pion azimuthal angle, which is 
 easily related to the violation of parity.  Next, we discuss how parity violation  originates from the interference of different
contributions to the hadronic current that are not relatively real. In
Sec.~\ref{sec:neutrino_comparison}, we present an extensive collection of results for total and differential
cross sections for pion production by neutrinos and antineutrinos. Sec.~\ref{sec:electropi} is
dedicated to pion electroproduction. Finally in Sec.~\ref{sec:conclusions} we
present an exhaustive summary of this study. In addition, we include four 
appendixes.  In Appendix~\ref{app:lt}, we give the Lorentz transformation 
from the laboratory system to the CM of the 
final pion-nucleon, paying special attention to the form of the different four-vectors in the
latter system. In Appendix~\ref{sec:app-varios}, we compile 
some auxiliary equations that help  determine the dependence on the pion azimuthal angle of the 
electro-weak pion production  off the nucleon. In Appendix~\ref{app:helicity}, we give the CC differential cross section for pion
production by neutrinos as a sum  over cross sections for virtual $W$ of different polarization. For that
purpose, we introduce and evaluate the  helicity components of the lepton and
hadron tensors. The final expression, evaluated for massless leptons, is 
analogous to the corresponding one commonly used for  pion electroproduction,
 which is rederived in
Appendix~\ref{app:helicity2}.

\section{Brief description of the DCC and HNV models}
\label{sec:models}
\subsection{DCC model}
The DCC model~\cite{Matsuyama:2006rp, Kamano:2013iva} was designed to
 describe meson-baryon scattering 
and electroweak meson production in the nucleon resonance region in a unified manner. 
To describe the hadron states up to invariant masses $W\ltap 2$~GeV,
the model includes stable two-particle channels
$\pi N, \eta N, K\Lambda, K \Sigma$ and unstable particle channels
$\rho N, \sigma N, \pi \Delta$,
the latter being the doorway states to the three-body $\pi\pi N$ state.
The $T$-matrix for the meson-baryon scattering 
is obtained by solving the coupled-channel Lippmann-Schwinger equation,
\begin{eqnarray}
  \langle\alpha, \vec{p}\, ' |T(W)|\beta,\vec{p}\,\rangle  & = & \langle 
  \alpha,\vec{p}\,'|V(W)|\beta,\vec{p}\,\rangle 
    + \sum_\gamma \int d^3{k}\,\langle \alpha,\vec{p}\,'|V(W)|\gamma,\vec{k}\,\rangle 
    G_\gamma^0(\vec{k},W)\langle \gamma,\vec{k}\,|T(W)|\beta,\vec{p}\,\rangle ,
\end{eqnarray}
where  $\alpha,\beta$ and $\gamma$ denote meson-baryon 
two-body states and $\vec p$ etc., the three-momenta in their CM.
The energy ($W$) dependent effective potential is split into three contributions ,
\begin{eqnarray}
  V(W) = v_{\rm non-res} + \Gamma\frac{1}{W - m^0_{\rm res}}\Gamma^\dagger + Z(W) \ .
\end{eqnarray}
The $v_{\rm non-res}$ term  consists of non-resonant
meson-baryon interactions that include $t$-channel meson exchange and $u$-,
and $s$-channel baryon exchange mechanisms.
The second term  includes bare $N^*$ and $\Delta$ 
excitation $s$-channel processes, with $m^0_{\rm res}$ and  $\Gamma$ the 
bare mass and  bare decay vertex of an unstable resonance
. The last term $Z(W)$ is a particle-exchange diagram
including $\pi\pi N$ intermediate states. 
The Green function $G_\gamma^0(\vec{k},W)$
is the meson($M_\gamma$)-baryon($B_\gamma$) propagator for a 
channel $\gamma$ and is written as
\begin{eqnarray}
  G_\gamma^0(\vec{k},W) & = &
  \frac{1}{W - [E_{B_\gamma}(\vec{k}\,) + E_{M_\gamma}(\vec{k}\,) +
  \Sigma_\gamma(\vec{k},W)]
    + i\epsilon} \ .
\end{eqnarray}
The decay of an unstable particle channel into $\pi\pi N$ is included
in $\Sigma_\gamma(\vec{k},W)$. 
By considering $Z(W)$ and $\Sigma_\gamma$,
the $T$-matrix satisfies not only  two-body unitarity but also  three-body
unitarity~\cite{Matsuyama:2006rp}.

The electroweak meson production amplitudes from the DCC model are given as
\begin{eqnarray}
  \langle \alpha,\vec{p}\,'|J^\mu(q)|N(\vec{p}\,)\rangle  & = & 
  \langle \alpha,\vec{p}\,'|j^\mu(q)|N(\vec{p}\,)\rangle 
    + \sum_\gamma \int d^3{k}\,\langle \alpha,\vec{p}\,'|T(W)|\gamma,\vec{k}\,\rangle 
    G_\gamma^0(\vec{k},W)\langle \gamma,\vec{k}\,|j^\mu(q)|N(\vec{p}\,)\rangle ,
\end{eqnarray}
where the electroweak meson production current ($j^\mu$) consists of
a non-resonant meson production current $j^\mu_{\rm non-res}$ including
$s$-, $t$- and $u$-channel exchange mechanisms similar to 
$v_{\rm non-res}$, and a nucleon resonance excitation contribution:
\begin{eqnarray}
  j^\mu = j^\mu_{\rm non-res} + \Gamma \frac{1}{W - m^0_{\rm res}}
   \Gamma^\mu \ .
\end{eqnarray}

One of the present authors, T.S., initiated a development of a dynamical 
approach,
 referred in this work as the SL 
model, with the aim of providing a reasonable description of  $\pi N$ 
scattering and
electroweak pion production in the $\Delta(1232)$ region in a unified manner~\cite{Sato:1996gk,Sato:2000jf,Sato:2003rq,Matsui:2005ns,Sato:2009de}.  
The aim of the SL model was to study  the  electroweak pion production of the 
$\Delta(1232)$ resonance. Therefore, the only meson-baryon
channel included is the  $\pi N$ state and the model cannot be applied  beyond
the $\Delta(1232)$ resonance region.
The DCC approach described in the above paragraph can be viewed as
an extension of the SL model to a higher resonance region, and it  has been developed
through the analysis of  the large available  data sample on differential
 cross sections and
polarization observables for pion- and photo-induced meson production 
reactions 
($\sim$23,000 data points).
The resonance masses, widths, and electromagnetic couplings for 
$N\to N^*, \Delta$ transitions have been extracted from the partial wave
amplitudes of the model at the pole positions.
The DCC approach was extended to describe the neutrino-induced meson
production reactions in Refs.~\cite{Nakamura:2015rta,Nakamura:2016cnn}.
The vector current at finite $q^2$ (four-momentum transfer square) and the isovector
couplings of the isospin 1/2 resonances are determined by analyzing data 
for 
pion electroproduction and the photo reaction on the neutron.
The axial  couplings for the $N\to N^*,\Delta$ transitions are determined 
by the pion coupling constants, assuming partial conservation of the axial 
current (PCAC), while 
dipole  $q^2$-dependence is assumed for the axial form factors.
In this work, we use a 10\% weakened bare axial coupling constant , 
$g_{AN\Delta}({\rm new})=0.9\, g_{AN\Delta}({\rm original})$,
for the $N\to\Delta(1232)$ transition, as compared to  the value 
used in \cite{Nakamura:2015rta,Nakamura:2016cnn}. While 
$g_{AN\Delta}({\rm original})$ was obtained using PCAC, $g_{AN\Delta}({\rm
new})$ is chosen so as to give a better reproduction of the neutrino cross
section data of Ref.~\cite{Rodrigues:2016xjj} that have been obtained from a
reanalysis of old ANL and
Brookhaven National Laboratory (BNL)  data.

\subsection{HNV model}
The HNV model  was originally introduced in Ref.~\cite{Hernandez:2007qq} to describe 
pion production by neutrinos in the $\Delta$ resonance region. In its
first version, it included  the dominant  direct  and  
crossed $\Delta-$pole terms plus a set of background terms.
The weak $N\to \Delta$ transition matrix element was 
parametrized in terms of four vector $C^V_{3-6}$ and
four axial $C^A_{3-6}$ form factors. Vector form factors were known  from the study of pion
electroproduction (in fact $C_6^V$ was set exactly to zero from conservation of the
vector current (CVC) ), while axial form factors were mostly
unknown.  The term proportional to $C_5^A$ gives the dominant contribution. Assuming the 
pion pole dominance of the pseudoscalar $C_6^A$ form
factor, the PCAC hypothesis  gives 
$C_6^A$  in terms of $C_5^A$. In the absence of good experimental data that
allowed for an independent determination of all axial form factors, 
Adler's model \cite{Adler:1968tw}, in which $C^A_3=0$ and $C^A_4=-\frac14 C_5^A$, 
was adopted.  Thus,  $C_5^A$ remained as the only unknown form-factor and its
value at $q^2=0 $ and its $q^2$ dependence were fitted to experiment.

\begin{figure}[tbh]
\centerline{\includegraphics[height=6cm]{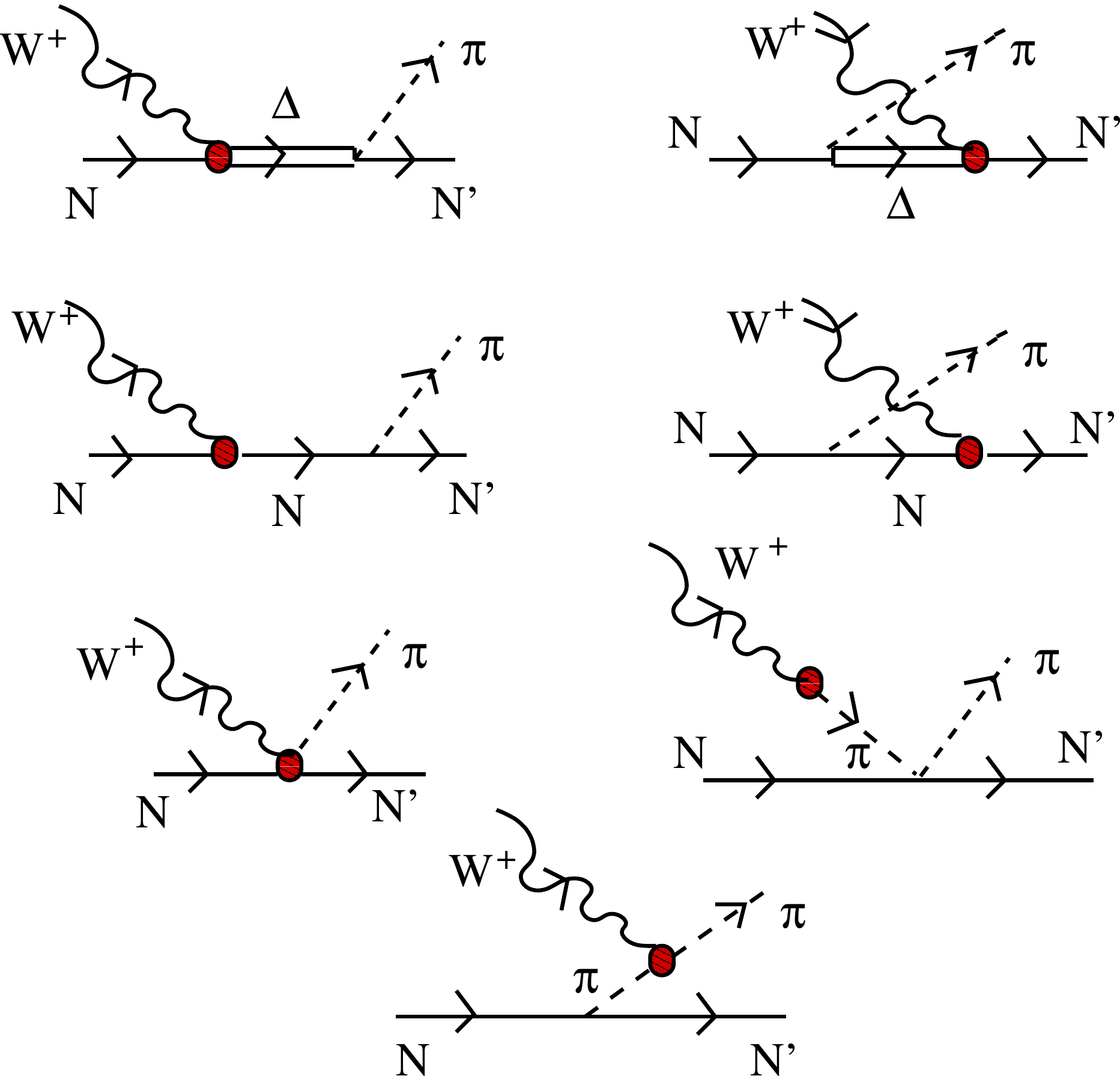}}
\caption{\footnotesize Model for the $W^+N\to N^\prime\pi$
  reaction as introduced in Ref.~\cite{Hernandez:2007qq}. It contains the Delta
   ($\Delta P$) and crossed Delta pole ($C\Delta P$) terms 
  (first row), the nucleon ($N P$) and crossed nucleon pole ($CN P$) terms
  (second row), the contact $CT$ and pion pole ($PP$) terms (third row), and the
  pion in flight ($PF$) term (fourth row).  }
  \label{fig:diagramas}
\end{figure}

The background terms are required and fixed by
chiral symmetry and they were
 obtained from the leading order predictions of a  
SU(2) nonlinear sigma model. The
weak vertexes were supplemented with well established form factors in a way
that  preserved CVC and PCAC. The Feynman diagrams  for the different
contributions to $W^+N\to N^\prime\pi$ (corresponding to a CC
process induced by neutrinos) are depicted in Fig.~\ref{fig:diagramas}. All sort of details can be found in
Ref.~\cite{Hernandez:2007qq}. 
NC pion production by neutrinos as well as antineutrino induced processes were 
also discussed in \cite{Hernandez:2007qq}. NC amplitudes were also
given in terms of the resonant and background contributions introduced 
above, though in this case nucleon strange form-factors needed to be considered. 
Some preliminary results were also shown in Ref.~\cite{Hernandez:2006yg}, 
where  NC neutrino and antineutrino pion production reactions were suggested 
as a  way to distinguish $\nu_\tau-$neutrinos from antineutrinos, below the 
$\tau-$production threshold,
but above the pion production one.

\begin{figure}[tbh]
\includegraphics[height=1.5cm]{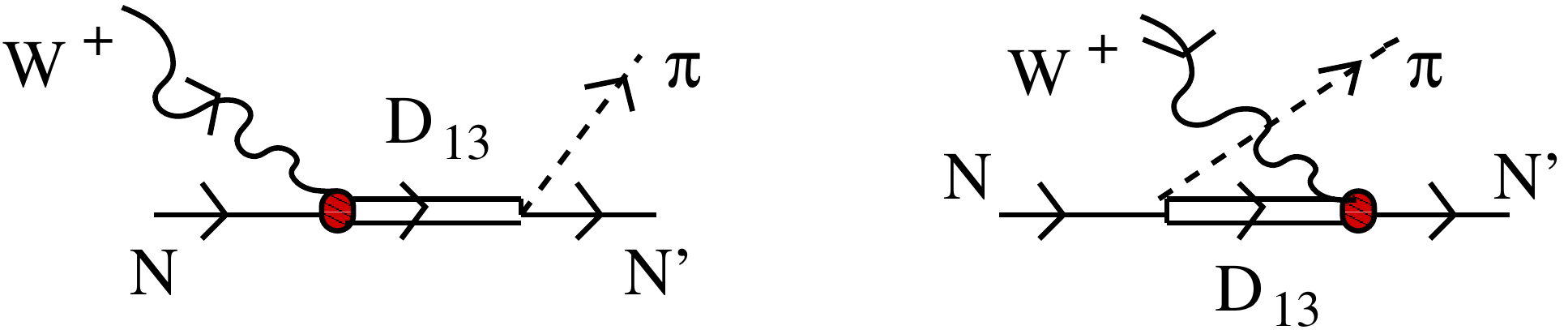}
\caption{$D_{13}(1520)$ contributions to 
 $W^+N\to N^\prime\pi$ introduced in Ref.~\cite{Hernandez:2013jka}. Both $D_{13}$ 
 ($DP$) and crossed $D_{13}$ pole ($CDP$) 
 terms are considered. }
 \label{fig:d13}
\end{figure}

To extend the HNV model to neutrino energies up to 2 GeV, in 
Ref.~\cite{Hernandez:2013jka}, the authors included the two contributions
depicted in 
Fig.~\ref{fig:d13}, which are driven by the exchange of 
the spin-3/2  $D_{13}(1520)$ resonance.
According to Ref.~\cite{Leitner:2008wx}, this is the only extra resonance  giving
 a significant contribution in that neutrino energy
region. All the details concerning the $DP$ and  $CDP$ 
 contributions  can be reviewed in the Appendix
of Ref.~\cite{Hernandez:2013jka}.

In Ref.~\cite{Alvarez-Ruso:2015eva} the HNV model was  partially unitarized 
 by
imposing Watson theorem.  Watson theorem is a consequence of unitarity  and time reversal
invariance. It implies that, below the two-pion production threshold, 
 the phase
of the electro or weak pion production amplitude should be given  
by the  $\pi N\to\pi N$  elastic phase shifts 
$\left[\delta_{L_{2J+1,2T+1}}(W_{\pi N})\right]$, with $W_{\pi N}$ the final 
$\pi N$ invariant mass. 
The
procedure followed in Ref.~\cite{Alvarez-Ruso:2015eva} was inspired  by that implemented by M.G. Olsson in 
Ref.~\cite{Olsson:1974sw}. To correct the interference
between the dominant $\Delta P$ term  and the 
background (including here not only the nonresonant background, but also the 
$C\Delta P$, $DP$ and $CDP$ terms),
the authors introduced two independent vector and axial phases, that are 
functions of $q^2$ and  $W_{\pi N}$. 
The amplitude was changed as
\begin{equation}
\label{eq:watson}
 T_B+T_{\Delta P}\to T_B+e^{i\delta_V}T^V_{\Delta P}+e^{i\delta_A}T^A_{\Delta P} 
\end{equation}
where the vector $\delta_V$ and axial $\delta_A$  Olsson  phases
 were fixed by 
requiring that the dominant vector and axial multipoles with the $\Delta(1232)$ 
quantum
numbers have  the correct phase $\delta_{P_{33}}(W_{\pi N})$. See 
Ref.~\cite{Alvarez-Ruso:2015eva} for details.

Very recently~\cite{Hernandez:2016yfb}, the HNV model has been supplemented 
with additional local terms. The aim was to improve the description of the 
$\nu_\mu n\to\mu^- n\pi^+$
channel, for which most theoretical models give predictions much below
experimental data. As discussed in Ref.~\cite{Hernandez:2016yfb}, this channel gets a
large contribution from the $C\Delta P$ term and then it is   sensitive to
the spin 1/2 component of the Rarita Schwinger (RS) covariant $\Delta$
propagator. Starting from the case of zero width, the $\Delta$ propagator was modified in that reference as 
\begin{eqnarray}
\frac{P_{\mu\nu}(p_\Delta)}{p_\Delta^2-M_\Delta^2} &\to&
\frac{P_{\mu\nu}(p_\Delta)+ c \left(P_{\mu\nu}(p_\Delta) -
\frac{p^2_\Delta}{M_\Delta^2}P_{\mu\nu}^{\frac32}(p_\Delta)\right)}{p_\Delta^2-M_\Delta^2}
\nonumber\\&=&  \frac{P_{\mu\nu}(p_\Delta)+ c (p_\Delta^2-M_\Delta^2)
\,\delta P_{\mu\nu}(p_\Delta)}{p_\Delta^2-M_\Delta^2}
= \frac{P_{\mu\nu}(p_\Delta)}{p_\Delta^2-M_\Delta^2} + 
c\, \delta P_{\mu\nu}(p_\Delta)
\label{eq:modifipropa}
\end{eqnarray} 
where $ P_{\mu\nu}$ and $P_{\mu\nu}^{\frac32}$ are, respectively, the 
RS covariant and pure spin-3/2
projectors~\cite{Hernandez:2016yfb}. This modification was motivated by the discussion in 
Ref.~\cite{Pascalutsa:2000kd}, where the authors advocated the use of the so called
consistent $\Delta$ couplings,  derivative couplings  that preserve the
gauge invariance of the free massless spin 3/2 Lagrangian.  
One can convert an inconsistent coupling into a consistent one (see 
Ref.~\cite{Pascalutsa:2000kd}), the net effect being  a change of the $\Delta$
propagator into
\begin{eqnarray}
\frac{\frac{p^2_\Delta}{M_\Delta^2}P_{\mu\nu}^{\frac32}(p_\Delta)}
{p_\Delta^2-M_\Delta^2}
\end{eqnarray} 
where only its spin-3/2 part  contributes. This prescription would correspond to taking
$c=-1$ in Eq.~(\ref{eq:modifipropa}). What one can see from 
Eq.~(\ref{eq:modifipropa})
is that the difference between the usual approach and the one based on the use of
consistent couplings  amounts to the new local term generated by
 $-\delta P_{\mu\nu}(p_\Delta)$. Thus, as long as both  approaches include all
 relevant local terms consistent with chiral symmetry, the strengths of which have to be fitted to data,
they  will give rise to the same 
 physical predictions. To keep the HNV model simple, the authors of 
 Ref.~\cite{Hernandez:2016yfb} just took  $c$ in Eq.~(\ref{eq:modifipropa})
 as a free parameter that was fitted to data. Before that, the
   $\Delta$ width was reinserted in the first term so that  the final 
     modification was
\begin{eqnarray}
 \frac{P_{\mu\nu}(p_\Delta)}
{p_\Delta^2-M_\Delta^2+iM_\Delta\Gamma_\Delta} \to \frac{P_{\mu\nu}(p_\Delta)}
{p_\Delta^2-M_\Delta^2+iM_\Delta\Gamma_\Delta}+ 
c\, \delta P_{\mu\nu}(p_\Delta)
\label{eq:mod}
\end{eqnarray} 
This  amounted to the introduction of  new contact terms originating
from $\delta P_{\mu\nu}(p_\Delta)$ and with a strength controlled by $c$.
In this way  a much better agreement for the $\nu_\mu n\to \mu^- n
\pi^+$ channel was achieved. In the new fit, the value $c=-1.11\pm0.21$, close to
$-1$, was 
obtained. Note, however, that due to the presence of the $\Delta$
width,  the prescription in Eq.~(\ref{eq:mod})  with $c=-1$ does not correspond exactly to the use
of a consistent coupling (see the discussion in 
Ref.~\cite{Hernandez:2016yfb}). Another good feature of this modification was that
the Olsson phases needed to satisfy Watson theorem were smaller in this case.
This means that after the latter modification, the model without the Olsson
phases was closer to satisfying unitarity than before the modification in Eq.~(\ref{eq:mod})
was implemented.

In this work we refer to the HNV model as the original model introduced in  Ref.~\cite{Hernandez:2007qq}
with the modifications discussed above and that were added in
Refs.~\cite{Hernandez:2013jka,Alvarez-Ruso:2015eva,Hernandez:2016yfb}. It
contains the contributions shown in Figs.~\ref{fig:diagramas} and \ref{fig:d13},
the modified $\Delta$ propagator of Eq.~(\ref{eq:mod}), and it implements Watson
theorem through the procedure just sketched here and explained in detail in
Ref.~\cite{Alvarez-Ruso:2015eva}. 
In the case of pion photo or electroproduction, the corresponding HNV model derives directly from the vector part 
of that constructed for weak pion production by neutrinos. The different contributions to the hadronic 
current  are given in the appendix of 
Ref.~\cite{Hernandez:2016yfb}. Watson theorem  as well as the
$\Delta$ propagator modification of Eq.~(\ref{eq:mod}) are also taken into
account in those cases.
\section{Pion production differential cross section. Parity violating terms}
\label{sec:diffxs}
Let us consider the case of a CC process induced by 
neutrinos
\bea
\nu_l(k)+ N(p)\to l^- (k')+N(p')+\pi(k_\pi)
\eea
\begin{figure}[h]
\resizebox{10cm}{!}{\includegraphics[scale=0.5]{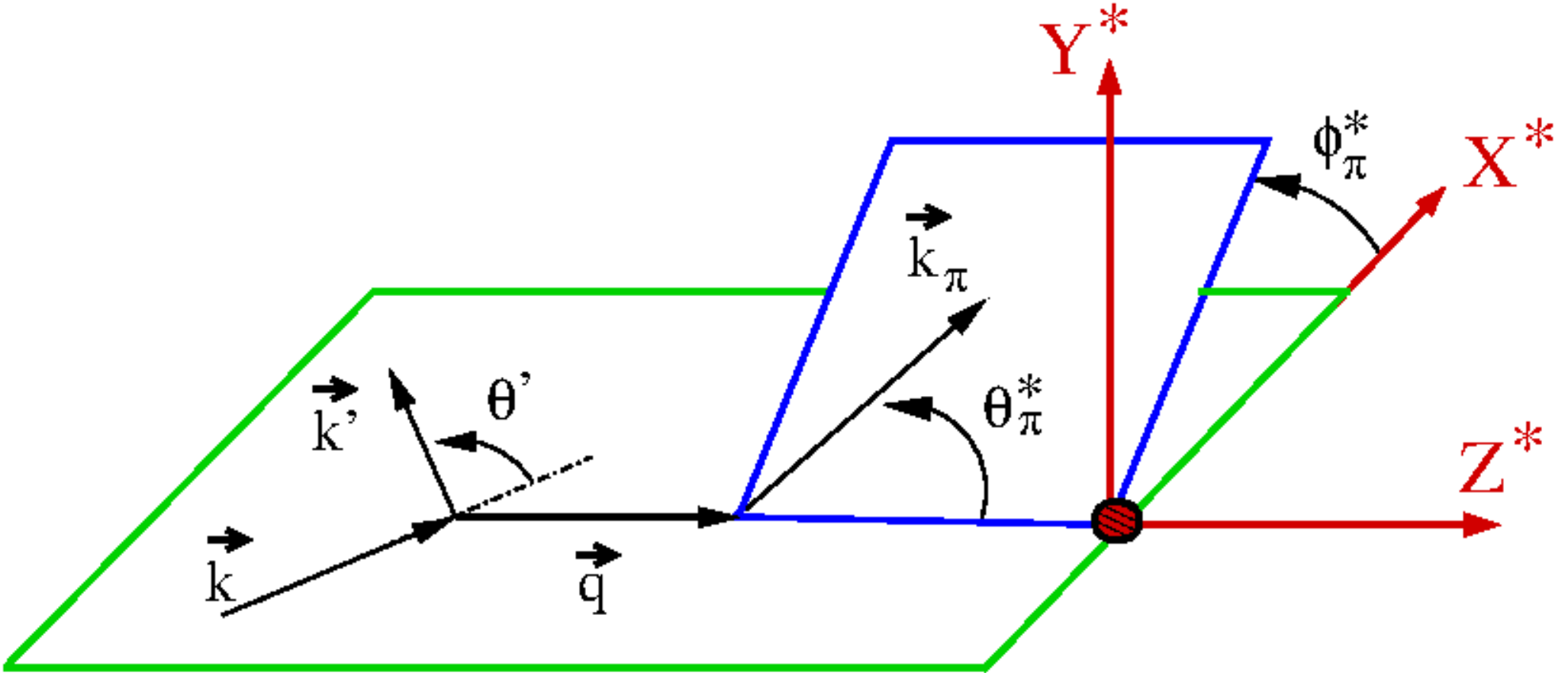}}
\caption{ Definition of the scattering and reaction planes. 
The $X^* Y^* Z^*$ coordinate axes move along with the CM system
of the final pion-nucleon  and their orientation has been  chosen 
in such a way that the lepton momenta lie in the $O^*X^*Z^*$ plane with the positive $Z^*$ axis chosen along
     $\vec q$ and the positive $Y^*$ axis chosen along 
     $\vec k\wedge\vec k^{\,\prime}$. }
  \label{fig:plano}
\end{figure}

The  cross section in the laboratory (LAB) system is given by
\bea
\sigma_{{\rm CC}+}=\frac{G_F^2}{4\pi^2|\vec k|}\int\frac{d^3k'}{E'}\frac{d^3k_\pi}{E_\pi}
\,L^{\mu\nu}(k,k')\,W_{\mu\nu}(q,p,k_\pi),
\eea
where $k^\mu=(|\vec k\,|,0,0,|\vec k\,|)$,
$k^{\prime\,\mu}=(E',\vec k\,')$, $p^\mu=(M,0,0,0)$, 
with $M$ the nucleon mass,  
and $k_\pi^\mu=(E_\pi,\vec k_\pi)$ are respectively the four-momenta of the 
initial lepton, final lepton,
initial nucleon and final pion in the LAB frame. Besides, $q=k-k'$ is the
four-momentum transfer and $G_F=1.1664\times 10^{-11}$ MeV$^{-2}$ is the Fermi
constant. The leptonic tensor is given by
\bea
L^{\mu\nu}(k,k')=k^\mu k^{\prime\,\nu}+k^\nu k^{\prime\,\mu}-g^{\mu\nu}k\cdot
k'+i\epsilon^{\mu\nu\alpha\beta}k'_\alpha k_\beta,
\label{eq:lt}
\eea
where we use $\epsilon_{0123}=+1$ and the metric 
$g^{\mu\nu}={\rm diag}\,(+1,-1,-1,-1)$. The
expression is valid both for CC and NC processes induced by
neutrinos\footnote{ Note that for NC processes there is an
extra factor of $1/4$ in the definition of the cross section when using the
normalization of the NC current used in the HNV model. In the DCC and SL models,
the NC current is defined with an extra factor of 1/2, as compared to the one
 used in the HNV model, and thus there is no need to correct the expression of the
 cross section in that case.}.  For the
case of antineutrinos the antisymmetric part of the leptonic tensor changes 
sign.
The hadronic tensor is given by
\bea
&&W^{\mu\nu}(q,p,k_\pi)=\frac1{4M}
\int\frac{d^3p'}{(2\pi)^3\,2E'_N}\,
\delta^{4}(q+p-p'-k_\pi){\cal H}^{\mu\nu}(p,p',k_\pi)
\label{eq:hadronTensor1}
\eea
with
\bea
{\cal H}^{\mu\nu}(p,p',k_\pi)=\frac12\sum_{s,s'}\langle N'(p',s')\,\pi(k_\pi)
|J^\mu_{{\rm CC}+}(0)|N(p,s)\rangle\langle N'(p',s')\,\pi(k_\pi)
|J^\nu_{{\rm CC}+}(0)|N(p,s)\rangle^*,
\label{eq:hadronTensor2}
\eea
being $s $  the helicity of the initial nucleon, and  $p'=(E'_N,\vec p\,')$ and $s'$, the
four-momentum and helicity of the final nucleon, respectively. $J^\mu_{{\rm CC}+}(0)$ represents the 
hadronic current operator for a CC process induced by neutrinos. For CC reactions induced by antineutrinos, we
need $J^\mu_{{\rm CC}-}(0)=J^{\mu\dagger}_{{\rm CC}+}(0)$, while in the NC case one has to
use the corresponding $J^\mu_{ \rm NC}(0)$ NC operator. 
In every case,
one  trivially finds that  
${\cal H}^{\mu\nu}$ can be written as the sum of a real symmetric and a pure imaginary antisymmetric 
parts
\bea
{\cal H}^{\mu\nu}={\cal H}^{\mu\nu}_s+
i{\cal H}^{\mu\nu}_a,\quad {\cal H}^{\mu\nu}_s =\frac12\Big({\cal H}^{\mu\nu}+
{\cal H}^{\nu\mu}\Big)\in\mathbb R,\qquad
{\cal H}^{\mu\nu}_a = -\frac{i}2\Big({\cal H}^{\mu\nu}-
{\cal H}^{\nu\mu}\Big)\in\mathbb R. \label{eq:hsy-antisy}
\eea

Making use of the invariant nature of
the $L^{\mu\nu}W_{\mu\nu}$ tensor product under a proper Lorentz 
transformation $\Lambda$, we can write
\bea
L^{\mu\nu}(k,k')W_{\mu\nu}(q,p,k_\pi)=L^{\mu\nu}(\Lambda k,\Lambda k^{\prime})
W_{\mu\nu}(\Lambda q, \Lambda p, \Lambda k_{\pi
}),
\eea
For each value of $k'$, the $\Lambda$ Lorentz transformation is chosen such 
that the  transformed momenta correspond
to those measured in the CM of the final pion-nucleon system. The
corresponding axes, that we denote as 
 $ X^* Y^* Z^*$,  are such that $Z^{*+}$ is oriented along $\vec q$, $Y^{*+}$ is oriented
    along $\vec k\wedge\vec k^{\,\prime}$ and $X^{*+}$ is oriented along
    $(\vec k\wedge\vec k^{\,\prime})\wedge\vec q$ (see Fig. 3).
With the above result, and making the 
 change of variables $\stackrel{\to}{\Lambda k_\pi}\to \vec k^*_\pi$, for which
$d^3 k_\pi/E_\pi\to d^3 k^*_\pi/E^*_\pi$, we can rewrite 
 the cross section as
 \bea
\sigma_{{\rm CC}+}=\frac{G_F^2}{4\pi^2|\vec k|}\int\frac{d^3k'}{E'}\frac{d^3k^*_\pi}{E^*_\pi}
\,L^{\mu\nu}(\Lambda k, \Lambda k)\,W_{\mu\nu}( \Lambda q, \Lambda p, k^*_\pi)
\label{eq:sigma}.
\eea 
In Appendix~\ref{app:lt} we give the value for $\Lambda$  and the corresponding transformed
 four-momenta that we shall simply denote as $k^*, k^{\prime *}, q^*,  p^*$ in what follows.  
 One of
the features of the new momenta is that
$k^*, k^{\prime *}, q^*,  p^*$ do not depend on $\phi'$ so that the integral on that variable would
just give rise to a factor of $2\pi$. Another salient feature
  is that the
 second spatial components of 
$k^*$ and $k^{\prime *}$ are  zero. This latter property allows us to immediately identify symmetric 
and antisymmetric non-diagonal components of the lepton tensor 
\bea
&&L^{02}(k^*, k^{\prime *})=-L^{20}(k^*, k^{\prime *})\ ,\ L^{12}(
k^*, k^{\prime *})=-L^{21}( k^*, k^{\prime *})\ ,\ L^{23}(k^*, k^{\prime *})
=-L^{32}( k^*, k^{\prime *}),\nonumber\\
&&L^{01}( k^*, k^{\prime *})=L^{10}( k^*, k^{\prime *})\ ,\ L^{03}( k^*, k^{\prime *})=
L^{30}( k^*, k^{\prime *})\ ,\ L^{13}( k^*, k^{\prime *})=L^{31}( k^*, k^{\prime *}).
\label{eq:tl}
\eea
In the case of $q^*$ and $p^*$, both the first and the second spatial components are
zero, a fact that will be used below. For $k_\pi^*$, which is nothing but 
the four-momentum of the final pion
measured in the CM of the final pion-nucleon system, we shall use 
\bea
 k_\pi^{*\mu}=(E^*_\pi,|\vec k^*_\pi|\sin\theta^*_\pi\cos\phi^*_\pi,|\vec
k^*_\pi|\sin\theta^*_\pi\sin\phi^*_\pi,|\vec k^*_\pi|\cos\theta^*_\pi),
\eea
where the pion angles are  defined with
respect to the $ X^* Y^* Z^*$ axes (see Fig.~\ref{fig:plano}). 

From Eq.~(\ref{eq:sigma}), we can now write the  differential cross 
section
\bea
\frac{d\sigma_{{\rm CC}+}}{d\Omega'dE'd\Omega^*_\pi}=\frac{|\vec k\,'|}{|\vec k|}
\frac{G_F^2}{4\pi^2}\int\frac{|\vec k^*_\pi|^2d|\vec k^*_\pi|}{E^*_\pi}
\,L^{\mu\nu}(k^*,k^{\prime *})\,W_{\mu\nu}(q^*, p^*, k^*_\pi). \label{eq:WL}
\eea

The integral in $|\vec k^*_\pi|$ can be easily done using that
\bea
W_{\mu\nu}( q^*, p^*, k^*_\pi)
=\frac1{4M}\int\frac{d^3 p^{\prime*}}{2E^{\prime\,*}_N}\frac{1}{(2\pi)^3}
\,\delta^4(q^*+p^*-p^{\prime*}-k^*_\pi)\,{\cal H}_{\mu\nu}(p^*, p^{\prime*}=
p^*+ q^*- k^*_\pi,k^*_\pi).
\eea
After the trivial $d^3 p^{\prime*}$ integration, there remains a 
delta of energy conservation that can be used to integrate in $|\vec k^*_\pi|$.
One gets
\bea
\label{eq:phase-space}
\int\frac{|\vec k^*_\pi|^2d|\vec k^*_\pi|}{E^*_\pi}\frac1{E^{\prime\,*}_N}\delta(W_{\pi N}-
E^{\prime\,*}_N-E^*_\pi)=\frac{|\vec k^*_\pi|_0}{W_{\pi
N}},
\eea
with $|\vec k^*_\pi|_0=\lambda^{1/2}(W_{\pi N}^2,M^2,m_\pi^2)/(2W_{\pi
N})$ and $\lambda(a,b,c)=(a+b-c)^2-4ab$.  The differential cross sections can thus be simplified to
\bea
\frac{d\sigma_{{\rm CC}+}}{d\Omega'dE'd\Omega^*_\pi}&=&
\frac{G_F^2|\vec k^*_\pi|_0}{256\pi^5MW_{\pi N}}\frac{|\vec k'|}{|\vec k|}
\,L^{\mu\nu}( k^*,k^{\prime *})\,{\cal H}_{\mu\nu}(p^*, p^{\prime*}=
p^*+ q^*- k^*_\pi,k^*_\pi),
\eea
Changing variables from
$(\theta',E')\to \left(Q^2=-q^2, W_{\pi N}=\sqrt{(q+p)^2}\,\right)$, we  further
obtain
\bea
\frac{d\sigma_{{\rm CC}+}}{dQ^2 dW_{\pi N}d\Omega^*_\pi}&=&
\frac{G_F^2|\vec k^*_\pi|_0}{256\pi^4 M^2 |\vec k|^2}\,
\,L^{\mu\nu}( k^*,k^{\prime *})\,{\cal H}_{\mu\nu}(p^*, p^{\prime*}=
p^*+ q^*- k^*_\pi,k^*_\pi),
\eea
where the trivial dependence on $\phi'$ (final lepton  laboratory azimuthal  angle) has been integrated out giving rise to 
a factor of
$2\pi$.
\subsection{The $\phi^*_\pi$ dependence of the 
${d\sigma_{{\rm CC}+}}/({d\Omega'dE'd\Omega^*_\pi})$ 
and ${d\sigma_{{\rm CC}+}}/({dQ^2 dW_{\pi N}d\Omega^*_\pi})$ differential cross sections
}
The $\phi^*_\pi$ dependence of the differential cross section 
can be isolated using very general
arguments. For that purpose, let us consider the active rotation $\hat R$ defined as
\bea
 \hat R^\mu_{\ \nu}=\left(\begin{array}{cccc}1&0&0&0\\
0&\cos\phi^*_\pi&-\sin\phi^*_\pi&0\\
0&\sin\phi^*_\pi&\cos\phi^*_\pi&0\\
0&0&0&1
\end{array}\right),
\eea
which is such that 
\bea
(\hat R^{-1} k^*_\pi)^\mu=
(E^*_\pi, |\vec k^*_\pi|\sin\theta^*_\pi,0,|\vec k^*_\pi|\cos\theta^*_\pi),
\eea
while $\hat R^{-1} q^*= q^*,\ \hat R^{-1} p^*= p^*$. Thus, making use of the tensor character of 
$W^{\mu\nu}(  q^*,  p^*, k^*_\pi)$, we will have
\bea
W^{\mu\nu}= W^{\mu\nu}( q^*, p^*, k^*_\pi)&=&
W^{\mu\nu}(\hat R\hat R^{-1} q^*,\hat R\hat R^{-1}
 p^*,\hat R\hat R^{-1} k^*_\pi)\nonumber\\
&=&
\hat R^{\mu}_{\ \ \alpha}\hat R^{\nu}_{\ \ \beta}\
W^{\alpha\beta}(\hat R^{-1} q^*,\hat R^{-1} p^*,\hat R^{-1} k^*_\pi)\nonumber\\
&=&
\hat R^{\mu}_{\ \ \alpha}\hat R^{\nu}_{\ \ \beta}\
W^{\alpha\beta}( q^*, p^*,{ \hat R^{-1} k^*_\pi} )=
\hat R^{\mu}_{\ \ \alpha}\hat R^{\nu}_{\ \ \beta}\
\widetilde W^{\alpha\beta} \label{eq:rotation}
\eea
where, for short, we have introduced  the notation
\bea
W^{\mu\nu}= W^{\mu\nu}( q^*, p^*, k^*_\pi)\ \ ,\ \ 
\widetilde W^{\mu\nu}= W^{\mu\nu}( q^*, p^*,{ \hat R^{-1} k^*_\pi} )=
W^{\mu\nu}\big|_{\phi^*_\pi=0}. \label{eq:defWha}
\eea
It is interesting to note that, since the second spatial components of  
  $ q^*, p^*, \hat R^{-1} k^*_\pi $ are zero,  the non-zero
  contributions to the 
$\widetilde W^{a2}$ and $\widetilde W^{2a}$ components of the hadronic 
tensor for $a=0,1,3$  should always involve  terms  constructed using 
the Levi-Civita pseudotensor, $v^a\epsilon^{2\alpha\beta\rho} 
q^*_\alpha p^*_\beta (\hat R^{-1}k^*_{\pi})_\rho$ or 
$\epsilon^{2a\alpha\beta}v_\alpha w_\beta$ with  $v\neq w$ 
being any of the four-vectors $ q^*,  p^*, 
\hat R^{-1} k^*_\pi $.  On the other hand, any component of 
the type $\widetilde W^{ab}$, with $a,b=0,1,3$, cannot contain the  
 Levi-Civita pseudotensor, because the coordinate 2 will appear 
 in the contraction of the pseudotensor with the available vectors,
  and none of them has a spatial component in the $Y^*$ axis.

In the case of photo- or electropion production on unpolarized nucleons, 
and  
since the electromagnetic interaction conserves parity
\footnote{For electromagnetic processes, terms containing the Levi-Civita 
pseudotensor should necessarily involve the polarization (pseudo-vector) 
of the nucleons to prevent parity violation.}, one has 
\bea
\widetilde W^{a2}_{em}=\widetilde W^{2a}_{em}=0,\ a=0,1,3.
\label{eq:egammapi}
\eea

Going back to the  $\phi^*_\pi$ dependence of $W^{\mu\nu}$, we see that 
it is now fully contained in $\hat R$. Thus, performing the rotations in
 Eq.~(\ref{eq:rotation}), the different components of the tensor 
  $W^{\mu\nu}$ can be written in terms of 
  $\widetilde W^{\mu\nu}= W^{\mu\nu}\big|_{\phi^*_\pi=0}$ and the pion 
  azimuthal angle $\phi^*_\pi$. The explicit expressions are given in 
  Eq.~(\ref{eq:th}) of  Appendix~\ref{sec:app-varios}, from where it follows 
  that the possible dependencies are $1,\cos\phi^*_\pi,\cos 2\phi^*_\pi,
\sin\phi^*_\pi$ and $\sin 2\phi^*_\pi$, as discussed  in detail also in 
Refs.~\cite{Hernandez:2007qq, Hernandez:2006yg,Sato:2003rq}. We have then
\bea
\frac{d\sigma_{{\rm CC}+}}{d\Omega'dE'd\Omega^*_\pi}&=&\frac{|\vec k'|}{|\vec k|}
\frac{G_F^2}{4\pi^2}\Big(A^*+B^*\cos\phi^*_\pi+
C^*\cos 2\phi^*_\pi+D^*\sin\phi^*_\pi+E^*\sin 2\phi^*_\pi\Big),\nonumber \\ 
\frac{d\sigma_{{\rm CC}+}}{dQ^2 dW_{\pi N}d\Omega^*_\pi}&=&
\frac{G_F^2W_{\pi N}}{4\pi M |\vec k|^2}\Big(A^*+B^*\cos\phi^*_\pi+
C^*\cos 2\phi^*_\pi+D^*\sin\phi^*_\pi+E^*\sin 2\phi^*_\pi\Big),
\label{eq:dcsabcde}
\eea
with the $A^*,B^*,C^*,D^*$ and $E^*$ structure functions  given by
\bea
A^*&=&\int\frac{|\vec k^*_\pi|^2d|\vec
k^*_\pi|}{E^*_\pi}\,\Big[L^{00}\,\widetilde W_{00}^{(s)}+2L^{03}\,\widetilde W_{03}^{(s)}+L^{33}\,\widetilde W_{33}^{(s)}+\frac12(L^{11}+L^{22})
\,\left(\widetilde W_{11}^{(s)}+\widetilde W_{22}^{(s)}\right)+2i L^{12}\,\widetilde W_{12}^{(a)}\Big],\nonumber\\
B^*&=&\int\frac{|\vec k^*_\pi|^2d|\vec
k^*_\pi|}{E^*_\pi}\,2\Big[L^{01}\,\widetilde W_{01}^{(s)}+L^{13}\,\widetilde W_{13}^{(s)} + iL^{02}\,\widetilde W_{02}^{(a)} + i L^{23}\,\widetilde W_{23}^{(a)}
\Big],\nonumber\\
C^*&=&\int\frac{|\vec k^*_\pi|^2d|\vec
k^*_\pi|}{E^*_\pi\,}\frac12\left[(L^{11}-L^{22})\,\left(\widetilde W_{11}^{(s)}-\widetilde W_{22}^{(s)}\right)\right],\nonumber\\
D^*&=&\int\frac{|\vec k^*_\pi|^2d|\vec
k^*_\pi|}{E^*_\pi}\,2\Big[-L^{01}\,\widetilde W_{02}^{(s)}-
L^{13}\,\widetilde W_{23}^{(s)}+ 
i L^{02}\,\widetilde W_{01}^{(a)} +
i L^{23}\,\widetilde W_{13}^{(a)}\Big],\nonumber\\
E^*&=&\int\frac{|\vec k^*_\pi|^2d|\vec
k^*_\pi|}{E^*_\pi}\,\left[(L^{22}-L^{11})\,\widetilde W_{12}^{(s)}\right],
\label{eq:abcde}
\eea
where we have made use of Eq.~(\ref{eq:tl}), and we have denoted
$L^{\mu\nu}=L^{\mu\nu}(k^*,k^{\prime *})$ for simplicity.
In addition, following Eq.~(\ref{eq:hsy-antisy}),  we have split the hadron tensor into 
symmetric ($\widetilde W_{\mu\nu}^{(s)}$) and antisymmetric  ($\widetilde W_{\mu\nu}^{(a)}$)
parts,
\begin{equation}
\widetilde W_{\mu\nu} = \widetilde W_{\mu\nu}^{(s)} + i\, \widetilde W_{\mu\nu}^{(a)}, \qquad \widetilde W_{\mu\nu}^{(s,a)} \in \mathbb R \label{eq:prop}
\end{equation}
Thus, and thanks to the fact that $L^{a 2}$ ($a=0,1,3$) is purely imaginary 
while the rest of the components of the lepton tensor are real, we trivially 
confirm that all $A^*,B^*,C^*,D^*$ and $E^*$ structure functions are real. 

Besides, since
\begin{equation}
W^{\mu\nu}\pm W^{\nu\mu}= \hat R^{\mu}_{\ \ \alpha}\hat R^{\nu}_{\ \ \beta}\,
\left(\widetilde W^{\alpha\beta}\pm \widetilde W^{\beta\alpha}\right)\, \label{eq:sobreW}
\end{equation}
we have that the symmetric and antisymmetric parts of $W^{\mu\nu}$ are 
determined respectively from $\widetilde W_{\mu\nu}^{(s)}$ and  $\widetilde 
W_{\mu\nu}^{(a)}$ using the same rotation. Therefore, we can conclude that 
the $C^*$ and $E^*$ structure constants are
generated from the contraction of the symmetric parts of the lepton and 
hadronic tensors, while 
$A^*, B^*$ and $D^*$ also get contributions  from the contraction of the 
antisymmetric parts of the lepton and 
hadronic tensors
(see also Eqs.~A8 and A9 of Ref.~\cite{Hernandez:2007qq}). As already
mentioned, the antisymmetric part of the lepton tensor changes sign for the
 case of antineutrino
induced reactions. Note also 
that from Eq.~(\ref{eq:egammapi}), it trivially follows that for electropion
 production off unpolarized nucleons, the $E^*$ structure function vanishes, 
 i.e., there is no $\sin 2\phi^*_\pi$ term in the differential cross section.
  Moreover, the symmetric contribution to $D^*$ will also vanish. Thus, the 
  dependence on $\sin \phi^*_\pi$ will only
survive  for  polarized  electrons, for which  the lepton tensor has 
an antisymmetric part that  leads 
to non-zero $L^{02}$ and $L^{23}$ components (see Eq.~(\ref{eq:tl})).

The above differential
cross sections can be written as a sum over differential cross sections, 
$d\sigma(W^* N\to N'\pi)/d\Omega^*_\pi\big|_{\phi^*_\pi=0}$, for
virtual $W$ of different polarizations. This relation is  given,  in 
the zero lepton mass limit, in Eq.~(\ref{eq:fin1})
of  Appendix~\ref{app:helicity}. Such a limit  is exact for
NC processes and provides an excellent approximation for CC processes induced by electron
neutrinos.

\subsection{Parity violation in the
${d\sigma_{{\rm CC}+}}/({d\Omega'dE'd\Omega^*_\pi})$ 
and ${d\sigma_{{\rm CC}+}}/({dQ^2 dW_{\pi N}d\Omega^*_\pi})$ differential
 cross sections }
\label{sec:pv} 
The terms proportional to $\sin\phi^*_\pi$ and $\sin 2\phi^*_\pi$  in Eq.~(\ref{eq:dcsabcde})  give rise to parity
violation  in the weak ${d\sigma}/({d\Omega'dE'd\Omega^*_\pi})$ 
and ${d\sigma}/({dQ^2 dW_{\pi N}d\Omega^*_\pi})$ differential cross sections~\cite{Hernandez:2007qq, Hernandez:2006yg}. The
reason is the following. After a parity transformation $\vec k,\vec
k\,',\vec q=\vec k-\vec k\,',\vec p$ and 
 $\vec k_\pi$ change direction ($\vec{v}\to \vec{v}_P=-\vec{v}\,$). The new 
 $Z^{*+}_P\equiv \vec {q}_P$ and
 $ X^{*+}_P\equiv(\vec k_P\wedge\vec k'_P)\wedge\vec q_P$ axes also change
 direction accordingly, but $ Y^{*+}_P\equiv\vec k_P\wedge\vec k'_P$ does not. 
 Measured in the new $ X^*_P Y^*_P Z^*_P$ system we have that  the 
 transformed four-vectors 
 $  k^*,   k^{\prime *},  q^*$ and $ p^*$
 have exactly the same components as before the parity transformation, since none 
 of these vectors has components along the $Y^*$ axis. 
   However,  the pion momentum does have  a component along the $Y^*$ axis and therefore the values of $\theta^*_\pi$ and 
   $\phi^*_\pi$ for the reversed pion momentum
 measured with respect to the new $
 X^*_P Y^*_P Z^*_P$ system change now
 as
 \beas
 \theta^*_\pi\to\theta^*_\pi\ \ ,\ \ \phi^*_\pi\to 2\pi-\phi^*_\pi.
 \eeas
 As a result, $L^{\mu\nu}$ and $\widetilde W_{\mu\nu}= 
 W^{\mu\nu}\big|_{\phi^*_\pi=0} $ remain the same and thus
 the $A^*,B^*,C^*,D^*$ and $E^*$ structure functions
 do not change. However, for the $\phi^* _\pi$ dependence
 we have that
 \bea
&& \cos\phi^*_\pi\to\cos(2\pi-\phi^*_\pi)=\cos\phi^*_\pi,\nonumber\\
&& \cos2\phi^*_\pi\to\cos\big(2(2\pi-\phi^*_\pi) \big)=\cos2\phi^*_\pi,\nonumber\\
&& \sin\phi^*_\pi\to\sin(2\pi-\phi^*_\pi)=-\sin\phi^*_\pi,\nonumber\\
&& \sin2\phi^*_\pi\to\sin \big(2(2\pi-\phi^*_\pi) \big)=-\sin2\phi^*_\pi.
 \eea
The sign change in the $\sin\phi^*_\pi$ and 
$\sin2\phi^*_\pi$ terms implies that  the $D^*$ and $E^*$ contributions to the differential cross sections violate
parity. Parity violation in weak production is then reflected by the fact that the pion
angular distributions  above and below the scattering 
plane are different  (see the discussion of Fig.~\ref{fig:ang_inte} in 
Sec.~\ref{sec:diff_cross-sec}). Note, however, that after integrating in $\phi^*_\pi$, the parity
breaking terms cancel, and  one obtains that the ${d\sigma_{{\rm CC}+}}/(
{d\Omega'dE'd\cos\theta^*_\pi})$ 
and ${d\sigma_{{\rm CC}+}}/({dQ^2 dW_{\pi N}d\cos\theta^*_\pi})$ differential 
cross sections are invariant under  parity.

From the discussion below Eq.~(\ref{eq:defWha}), one notices that the structure 
functions $A^*,B^*$ and $C^*$ always involve either symmetric hadron 
tensor terms that do not contain  the Levi-Civita pseudotensor, or  
antisymmetric hadron tensor terms constructed using the  Levi-Civita 
pseudotensor. In turn, $D^*$ and $E^*$ always involve either symmetric
 hadron tensor terms constructed using the  Levi-Civita pseudotensor, 
 or  antisymmetric hadron tensor terms that do not contain  the Levi-Civita
  pseudotensor. Using the terminology of Refs.~\cite{Hernandez:2007qq, 
  Hernandez:2006yg}, the structure functions $A^*,B^*$ and $C^*$ ($D^*$ 
  and $E^*$) are therefore constructed out of the parity conserving 
  (violating) hadron tensors (see for instance Eq.~A1 of
   Ref.~\cite{Hernandez:2007qq} and the related discussion).

A further remark concerns  time-reversal (T). As discussed in 
Refs.~\cite{Hernandez:2007qq, Hernandez:2006yg}, the $\sin\phi^*_\pi$ and $\sin2\phi^*_\pi$ 
terms
encode T-odd correlations. However, 
the existence of these terms does not necessarily mean that there exists
a violation of T-invariance in the process because of the
existence of strong final state interaction effects~\cite{Karpman:1969wx, Cannata:1970br}.

There is a subtlety, worth mentioning, for the case of pion production induced by initial polarized electrons. Following 
the above discussion, one could wrongly conclude that there exists parity violation in these 
processes. This is because, as commented before, though the  $\sin2\phi^*_\pi$ contribution is 
absent, the $L^{02}_{em}$ and $L^{23}_{em}$ terms in $D^*\sin\phi^*_\pi$ survive, since they do 
not involve the vanishing $\widetilde W^{a2}_{em}$ and $\widetilde W^{2a}_{em}$ 
components\footnote{The hadron tensor that describes the  virtual-photon pion production 
off an unpolarized nucleon can never have Levi-Civita pseudotensor contributions, but it can have
 antisymmetric  $\widetilde W_{01}^{(a)}$ and $\widetilde W_{03}^{(a)}$ terms, since they do 
 not involve the Levi-Civita pseudotensor.}.
What happens is that  $L^{02}_{em}$ and $L^{23}_{em}$ change sign under a parity transformation, 
contrary to the weak pion production case. This is because the antisymmetric part of the 
electromagnetic lepton tensor is proportional to the helicity, $h$, of the initial electron 
($\propto h \epsilon^{\mu\nu\rho\sigma} k^*_\rho k^{\prime *}_\sigma$). The helicity is a 
pseudoscalar and it changes sign under parity, which induces also a change of sign  in 
$L^{02}_{em}$ and $L^{23}_{em}$ that  compensates  the 
change  of sign under parity of  $\sin\phi^*_\pi$. As a consequence  $D^*\sin\phi^*_\pi$ 
remains parity invariant. With  respect 
to time reversal, the helicity does not change sign under T, and thus the lepton tensors 
in electro-- and weak--pion production behave in the same way under time reversal transformations, 
and therefore  T--odd correlations exist also in the case of electromagnetic reactions.

%
\subsubsection{ Origin of the parity conserving and parity violating 
contributions to the
hadronic tensor}
\label{sec:originpv}
In this section, we will use  the terminology parity conserving (PC)  and parity violating (PV) terms
to refer to  contributions to the hadronic
tensor that give rise to parity conservation/violation when contracted with the
leptonic tensor. Taking into account the structure of the leptonic tensor, where
the symmetric part is a true tensor while the antisymmetric part is proportional
to the Levi-Civita pseudotensor, it is clear that  (i) any symmetric part in the hadron tensor that
 contains a Levi-Civita pseudotensor or  (ii)
any antisymmetric part  in the hadron tensor that does not contain a Levi-Civita pseudotensor
 are PV ones~\cite{Hernandez:2007qq, Hernandez:2006yg}. We have explicitly seen this in the expressions of $D^*$ and $E^*$ of 
 Eq.~(\ref{eq:abcde}), as we pointed out  above in the main body of
 Subsec.~\ref{sec:pv} (we recall here again  the discussion of  Eq.~(\ref{eq:sobreW}), where we have shown that the symmetric and antisymmetric parts of  the tensors $ W^{\mu\nu}$ and $\widetilde W^{\mu\nu}$ are connected by the rotations of Eq.~(\ref{eq:rotation})). As we are going to show in the following, the PV terms
originate from the interference between different contributions to the hadronic current that are not
relatively real. 

For our purposes, it is enough to consider
the nucleon tensor defined in Eq.~(\ref{eq:hadronTensor2}) associated to 
$\widetilde W^{\mu\nu}$ (independent of $\phi^*_\pi$), that can be written as 
the trace\footnote{The discussion runs totally in parallel 
if one makes instead reference to  $W^{\mu\nu}$, where the pion 
three-momentum, $ \vec k^*_\pi$, conserves its full $\phi^*_\pi$ dependence. 
We choose to use explicitly ${\cal \widetilde H^{\mu\nu}}$ to make direct contact with Eq.~(\ref{eq:abcde}).}
\bea
{\cal \widetilde H}^{\mu\nu}(p^*, p^{\prime *},\hat R^{-1} k^*_\pi)=
&&\frac12{\rm Tr}\left[(/\hspace{-.2cm}  p^{\prime *}+M){\cal J}^\mu( p^*, p^{\prime *},
\hat R^{-1} k^*_\pi)(/\hspace{-.2cm}  p+M)\gamma^0
{\cal J}^{\nu\,\dagger}( p^*, p^{\prime *},\hat R^{-1} k^*_\pi)\gamma^0\right ],
\eea
where here $ p^{\prime *}= q^*+ p^*-\hat R^{-1} k^*_\pi$, and  
${\cal J}^\mu( p^*, p^{\prime *},\hat R^{-1} k^*_\pi)$ is defined from the hadronic current 
operator matrix element
\bea
\langle N'( p^{\prime *},s')\,\pi(\, \hat R^{-1}  k^*_\pi)
|J^\mu_{{\rm CC}+}(0)|N( p^*,s)\rangle=\bar u_{s'}( p^{\prime *})
{\cal J}^\mu( p^*, p^{\prime *},\hat R^{-1} k^*_\pi)\,u_s(p^*).
\label{eq:calJdefi}
\eea
The amputated  ${\cal J} ^\mu( p^*, p^{\prime *},\hat R^{-1}  k^*_\pi)$ current contains a 
vector and an axial contribution that
 one can write as
\bea
{\cal J}^\mu( p^*, p^{\prime *}\,\hat R^{-1} k^*_\pi)=\sum_{j_1} 
\gamma_5e^{i\varphi_{Vj_1}( p^*,
 p^{\prime *}\,\hat R^{-1} k^*_\pi)}
{\cal J}^\mu_{Vj_1}( p^*, p^{\prime *},\hat R^{-1} k^*_\pi)+\sum_{j_2}
e^{i \varphi_{Aj_2}( p^*, p^{\prime *},\hat R^{-1} k^*_\pi)}
{\cal J}^\mu_{Aj_2}( p^*, p^{\prime *},\hat R^{-1} k^*_\pi).\nonumber\\
\label{eq:jdecomp}
\eea
 where the ${\cal J}^\mu_{Vj}$ and ${\cal J}^\mu_{Aj}$  correspond to the
 different Dirac operator structures present in the hadronic current\footnote{ Such an
 expancion can be seen for instance in  Ref.~\cite{Adler:1968tw}, though there
 the hadronic current is already contracted with the leptonic one.}.
 They are built from $\gamma$
matrices (no $\gamma_5$ however) and momenta and,  for each term in the
two  sums,  $e^{i\varphi_{Vj}( p^*, p^{\prime *},\hat R^{-1} k^*_\pi)}$ and 
$ e^{i\varphi_{Aj}( p^*, p^{\prime *},\hat R^{-1} k^*_\pi)}$ stand
for the global phase of all multiplicative factors in that term other than 
$\gamma$
matrices. 
Note that for the HNV model there is a correspondence between the phases 
$\varphi_A$ and $\varphi_V$ and the complex structure of the $\Delta$ 
(corrected by the Olsson phases introduced in Eq.~(\ref{eq:watson}) and the $D_{13}$(1520) resonance). However, for the DCC and SL models, in addition to 
the complex structure of the resonances ($m_R$ and $\Gamma_R$) one should account for loop effects that provide further relative phases between different contributions to the amplitude. 
Simplifying the notation, we will have
\small
\bea
\hspace{-.5cm}{\cal \widetilde H}^{\mu\nu}&=&\frac12{\rm Tr}
\Bigg[(/\hspace{-.2cm} p^{\prime *}+M)\Big[\sum_{j_1}\gamma_5e^{i\varphi_{Vj_1}}
{\cal J}^\mu_{Vj_1}+\sum_{j_2}
e^{i\varphi_{Aj_2}}{\cal J}^\mu_{Aj_2}\Big](/\hspace{-.2cm}  p^*+M)
\gamma^0\Big[\sum_{k_1}e^{-i\varphi_{Vk_1}}{\cal J}^{\nu\,\dagger}_{Vk_1}\gamma_5+
\sum_{k_2}
e^{-i\varphi_{Ak_2}}{\cal J}^{\nu\,\dagger}_{Ak_2}\Big]
\gamma^0\Bigg]
\eea
\normalsize
that can be split into two contributions ${\cal \widetilde H}^{\mu\nu}= {\cal \widetilde H}^{\mu\nu}_{VV+AA}+ {\cal \widetilde H}^{\mu\nu}_{VA+AV}$, given by
\bea
{\cal \widetilde H}^{\mu\nu}_{VV+AA} &=&
\hspace{.35cm}\frac12\sum_{j_1}\sum_{k_1}e^{i(\varphi_{Vj_1}-\varphi_{Vk_1})}\,{\rm Tr}
\left[(/\hspace{-.2cm}  p^{\prime *}-M){\cal J}^\mu_{Vj_1}(/\hspace{-.2cm}  p^*+M)
\gamma^0{\cal J}^{\nu\,\dagger}_{Vk_1}\gamma^0\right]\nonumber\\
&&+\frac12\sum_{j_2}\sum_{k_2}e^{i(\varphi_{Aj_2}-\varphi_{Ak_2})}\,{\rm Tr}
\left[(/\hspace{-.2cm}  p^{\prime *}+M){\cal J}^\mu_{Aj_2}(/\hspace{-.2cm}  p^*+M)
\gamma^0{\cal J}^{\nu\,\dagger}_{Ak_2}\gamma^0\right  ]\nonumber\\
{\cal \widetilde H}^{\mu\nu}_{VA+AV} &=&-\frac12\sum_{j_1}\sum_{j_2}e^{i(\varphi_{Vj_1}-\varphi_{Aj_2})}\,{\rm Tr}
\left[(/\hspace{-.2cm}  p^{\prime *}-M){\cal J}^\mu_{Vj_1}(/\hspace{-.2cm}  p^*+M)
\gamma^0{\cal J}^{\nu\,\dagger}_{Aj_2}\gamma^0\gamma_5
\right]\nonumber\\
&&-\frac12\sum_{j_1}\sum_{j_2}e^{-i(\varphi_{Vj_1}-\varphi_{Aj_2})}\,{\rm Tr}
\left[(/\hspace{-.2cm} p^{\prime *}+M){\cal J}^\mu_{Aj_2}(/\hspace{-.2cm}  p^*+M)
\gamma^0{\cal J}^{\nu\,\dagger}_{Vj_1}\gamma^0\gamma_5
\right].
\label{eq:pcpv}
\eea
Let us pay attention  first to ${\cal \widetilde H}^{\mu\nu}_{VV+AA}$. Since the two 
traces  are  real\footnote{\ For $\alpha_1,\dots,\alpha_{2n}=0,1,2,3$, 
one has that
${\rm 
Tr}(\gamma^{\alpha_1}\cdots\gamma^{\alpha_{2n}})\in\mathbb R
$
and does not contain any Levi-Civita pseudotensor. Besides the trace of an odd number
of $\gamma$ matrices is always zero.},  we therefore get 
real symmetric contributions to the hadronic tensor, 
${\cal \widetilde H}^{\mu\nu\, (s)}_{VV+AA}$,  given 
by\footnote{$A^{\mu\nu}_{jk}= {\rm Tr}
\big[(/\hspace{-.175cm}  p^{\prime *}\mp M)\Gamma^\mu_j(/\hspace{-.175cm} p^*+M)
\gamma^0 \Gamma^{\nu\,\dagger}_k\gamma^0\big]$ is  real when the vector Dirac matrix $\Gamma^\mu_j$ does not contain an odd number of $\gamma_5$ matrices, this is to say it is built from $\gamma$
matrices (no $\gamma_5$ however) and momenta. Then, it trivially follows that $A^{\mu\nu}_{jk}=\left(A^{\mu\nu}_{jk}\right)^* = A^{\nu\mu}_{kj}$. Hence making use of the fact that  the  cosine is an even function, we conclude 
\begin{equation}
T^{\mu\nu}=\sum_{j,k}  \cos(\varphi_j-\varphi_k)A^{\mu\nu}_{jk}= \sum_{j,k}  \cos(\varphi_j-\varphi_k)A^{\nu\mu}_{kj}= \sum_{j,k} \cos(\varphi_k-\varphi_j)A^{\nu\mu}_{kj}= T^{\nu\mu}. 
\end{equation}}
\bea
\underbrace{{\cal \widetilde H}^{\mu\nu\, (s)}_{VV+AA}}_{\rm PC}&=&\hspace{0.35cm}\frac{1}{2}\sum_{j_1}\sum_{k_1}\cos(\varphi_{Vj_1}-\varphi_{Vk_1}){\rm Tr}
\left[(/\hspace{-.2cm}  p^{\prime *}-M){\cal J}^\mu_{Vj_1}(/\hspace{-.2cm}  p^*+M)
\gamma^0{\cal J}^{\nu\,\dagger}_{Vk_1}\gamma^0\right]\nonumber\\
&&+\frac{1}{2}\sum_{j_2}\sum_{k_2}\cos(\varphi_{Aj_2}-\varphi_{Ak_2}){\rm Tr}
\left[(/\hspace{-.2cm}  p^{\prime *}+M){\cal J}^\mu_{Aj_2}(/\hspace{-.2cm}  p^*+M)
\gamma^0{\cal J}^{\nu\,\dagger}_{Ak_2}\gamma^0\right],
\label{eq:pc1}
\eea
and purely imaginary antisymmetric contributions, 
$i {\cal \widetilde H}^{\mu\nu\, (a)}_{VV+AA}$, given by 
(in this case we have a  sine which is an odd function)
\bea
\underbrace{i {\cal \widetilde H}^{\mu\nu\, (a)}_{VV+AA}}_{\rm PV}&=&
\hspace{0.35cm} \frac{i}{2}\sum_{j_1\neq k_1}\sin(\varphi_{Vj_1}- \varphi_{Vk_1}){\rm Tr}
\left[(/\hspace{-.2cm} p^{\prime *}-M){\cal J}^\mu_{Vj_1}(/\hspace{-.2cm}  p^*+M)
\gamma^0{\cal J}^{\nu\,\dagger}_{Vk_1}\gamma^0\right]\nonumber\\
&&+\frac{i}{2}\sum_{j_2\neq k_2}\sin(\varphi_{Aj_2}-\varphi_{Ak_2}){\rm Tr}
\left[(/\hspace{-.2cm} p^{\prime *}+M){\cal J}^\mu_{Aj_2}(/\hspace{-.2cm}  p^*+M)
\gamma^0{\cal J}^{\nu\,\dagger}_{Ak_2}\gamma^0\right].
\label{eq:pv1}
\eea
The symmetric part, ${\cal \widetilde H}^{\mu\nu\, (s)}_{VV+AA}$,  does not 
contain a Levi-Civita pseudotensor and it is thus  PC  since when it is contracted 
with the symmetric part of the leptonic tensor 
it will give rise to
a true scalar. On the other hand, the 
antisymmetric part, ${\cal \widetilde H}^{\mu\nu\, (a)}_{VV+AA}$,  does not 
contain a Levi-Civita pseudotensor either; it is thus PV since when it is contracted 
with the antisymmetric part of 
the leptonic tensor it will give rise to a pseudoscalar.

With respect to ${\cal \widetilde H}^{\mu\nu}_{VA+AV}$, we see that in 
this
 case the traces involved are purely imaginary
and contain a Levi-Civita pseudotensor\footnote{In this case, for 
$\alpha_1,\dots,\alpha_{2n}=0,1,2,3$, 
one has that $
i{\rm
Tr}(\gamma_5\gamma^{\alpha_1}\cdots\gamma^{\alpha_{2n}})\in\mathbb R$.
Besides, all the contributions to the above trace are
 proportional to the Levi-Civita pseudotensor.}. Then, it gives rise to purely imaginary 
 and antisymmetric contributions, ${\cal \widetilde H}^{\mu\nu\, (a)}_{VA+AV}$, to 
 the hadronic tensor given by\footnote{This now follows from the fact 
 that by construction, 
 the purely imaginary $E^{\mu\nu}_{Vj_1; Aj_2}$ tensor defined as
\begin{equation}
E^{\mu\nu}_{Vj_1; Aj_2}={\rm Tr} \big[(/\hspace{-.175cm}  p^{\prime *}- M){\cal J}^\mu_{Vj_1}
(/\hspace{-.175cm}  p^*+M) 
\gamma^0{\cal J}^{\nu\,\dagger}_{Aj_2}\gamma^0\gamma_5\big] +{\rm Tr}
\big[(/\hspace{-.175cm}  p^{\prime *}+M){\cal J}^\mu_{Aj_2}(/\hspace{-.175cm}  p^*+M)
\gamma^0{\cal J}^{\nu\,\dagger}_{Vj_1}\gamma^0\gamma_5\big] \label{eq:Emunu}
\end{equation}
satisfies $-E^{\mu\nu}_{Vj_1; Aj_2}=(E^{\mu\nu}_{Vj_1; Aj_2})^*= 
E^{\nu\mu}_{Vj_1; Aj_2}$ (note that under the complex-conjugate operation in Eq.~(\ref{eq:Emunu}), implemented by taking $\dagger$ inside of the traces, the first (second) term is reduced to the second (first) one, with the exchange of $\mu$ by $\nu$.) }
\bea
\underbrace{i {\cal \widetilde H}^{\mu\nu\, (a)}_{VA+AV}}_{\rm PC}&=&-\frac{1}{2}\sum_{j_1}\sum_{j_2}\cos(\varphi_{Vj_1}-\varphi_{Aj_2})\Bigg\{{\rm Tr}
\left[(/\hspace{-.2cm}  p^{\prime *}-M){\cal J}^\mu_{Vj_1}(/\hspace{-.2cm}  p^*+M) 
\gamma^0{\cal J}^{\nu\,\dagger}_{Aj_2}\gamma^0\gamma_5\right] \nonumber \\
&& +{\rm Tr}
\left[(/\hspace{-.2cm}  p^{\prime *}+M){\cal J}^\mu_{Aj_2}(/\hspace{-.2cm}  p^*+M)
\gamma^0{\cal J}^{\nu\,\dagger}_{Vj_1}\gamma^0\gamma_5\right]\Bigg\}.
\label{eq:pc2}
\eea
and to real symmetric contributions, 
${\cal \widetilde H}^{\mu\nu\, (s)}_{VA+AV}$ given 
by\footnote{In this case $-F^{\mu\nu}_{Vj_1; Aj_2}=(F^{\mu\nu}_{Vj_1; Aj_2})^*=
 -F^{\nu\mu}_{Vj_1; Aj_2}$, where $F$ 
is the tensor between the curly brackets in Eq.~(\ref{eq:pv2}); 
the minus sign appears in the last identity because $ F $ 
is defined as the difference between two terms.}
\bea
\underbrace{{\cal \widetilde H}^{\mu\nu\, (s)}_{VA+AV}}_{\rm PV}&=&-\frac{i}{2}\sum_{j_1}\sum_{j_2}\sin(\varphi_{Vj_1}-\varphi_{Aj_2})\Bigg\{{\rm Tr}
\left[(/\hspace{-.2cm} p^{\prime *}-M){\cal J}^\mu_{Vj_1}(/\hspace{-.2cm}  p^*+M) 
\gamma^0{\cal J}^{\nu\,\dagger}_{Aj_2}\gamma^0\gamma_5\right] \nonumber \\
&& -{\rm Tr}
\left[(/\hspace{-.2cm} p^{\prime *}+M){\cal J}^\mu_{Aj_2}(/\hspace{-.2cm}  p^*+M)
\gamma^0{\cal J}^{\nu\,\dagger}_{Vj_1}\gamma^0\gamma_5\right]\Bigg\},
\label{eq:pv2}
\eea
The symmetric part, ${\cal \widetilde H}^{\mu\nu\, (s)}_{VA+AV}$, is now PV since it contains a Levi-Civita
pseudotensor coming from the trace, whereas the antisymmetric part, ${\cal \widetilde H}^{\mu\nu\, (a)}_{VA+AV}$, is PC 
for the same reason. Note that Eqs~(\ref{eq:pc1}), (\ref{eq:pv1}), (\ref{eq:pc2}) and (\ref{eq:pv2}) show explicitly the decomposition 
\begin{equation}
{\cal \widetilde H}_{\mu\nu} = {\cal \widetilde H}_{\mu\nu}^{(s)} + i\, {\cal \widetilde H}_{\mu\nu}^{(a)}, \qquad {\cal \widetilde H}_{\mu\nu}^{(s,a)} \in \mathbb R
\end{equation}
which trivially leads to that of the tensor ${\widetilde W}_{\mu\nu}$ in Eq.~(\ref{eq:prop}).

As we have just shown, the PV terms are always proportional 
to the sine of phase differences
and they would cancel exactly if all contributions to the hadronic current 
were relatively real. These PV terms give rise to the
$\sin\phi^*_\pi$ and $\sin2\phi^*_\pi$ terms in the differential cross
sections in Eq.~(\ref{eq:dcsabcde}). As seen in Eq.~(\ref{eq:abcde}),  $E^*$
is given in terms of a symmetric contribution to the hadronic tensor 
($\widetilde W^{(s)}_{12}$) that involves Levi-Civita tensors, and thus 
the  $\sin2\phi^*_\pi$
dependence in the differential cross section must come necessarily from the 
symmetric ${\cal \widetilde H}^{\mu\nu\, (s)}_{VA+AV}$ PV term. The latter is generated from
 vector-axial interference
and then it will be absent in the case of photo- or electro-production. On the
other hand,  the  $\sin\phi^*_\pi$ dependence
in the differential cross section gets contributions from both PV
terms: the symmetric ${\cal \widetilde H}^{\mu\nu\, (s)}_{VA+AV}$ and  
the antisymmetric ${\cal \widetilde H}^{\mu\nu\, (a)}_{VV+AA}$ tensors, which 
give rise to $\widetilde W^{(s)}_{02,23}$ and 
$\widetilde W^{(a)}_{01,13}$, respectively. The former (symmetric) ones 
contain  Levi-Civita tensors, while the latter (antisymmetric) ones do not. We remark that 
${\cal \widetilde H}^{\mu\nu\, (a)}_{VV+AA}$ is generated from vector-vector and axial-axial interferences, and the $VV$ part will also appear in polarized electron scattering.
The PV hadron tensors also lead to time-reversal odd correlations in the amplitudes 
(see discussion in Refs.~\cite{Hernandez:2006yg, Hernandez:2007qq}).

In the case of the HNV model, neglecting for simplicity in the discussion the 
$D_{13}(1520)$ contribution, and in the absence of Olsson phases,
those PV terms can only be generated by the 
interference between the part of the $\Delta P$ contribution that is 
proportional to the
$\Delta$ propagator and all other non-resonant terms~\cite{Hernandez:2007qq}. They are not relatively real
 due to the presence of  a nonzero $\Delta$ width in the $\Delta$ propagator. Once the Olsson
phases are included,  there are other sources of parity violation in the model like
  the interference  
 between the contact term  generated from the $\Delta P$ amplitude by the
 $c\, \delta P_{\mu\nu}(p_\Delta)$ term
in Eq.~(\ref{eq:modifipropa})  and the background, or the interference between the vector and axial
 parts  of the contact term in
 $\Delta P$. 
 
 In the case of the SL and DCC models, the unitarization procedure 
 guarantees that, for energies below the two pion production threshold, 
 each amplitude ${\cal M}$ corresponding to a given isospin,
 total angular momentum  and pion orbital angular momentum,   is given by
 $e^{i\delta}|{\cal M}|$, with $\delta$ the corresponding 
 $\pi N$ phase
 shift for the given quantum numbers.  In this case it is better to work in a multipole language.
 For that,  we can rewrite 
 \bea
{\cal \widetilde H}^{\mu\nu}(p^*, p^{\prime *},\hat R^{-1} k^*_\pi)=
&&\frac12{\rm Tr}\left[h^\mu( p^*, p^{\prime *},
\hat R^{-1} k^*_\pi)
h^{\nu\,\dagger}( p^*, p^{\prime *},\hat R^{-1} k^*_\pi)\right ],
\eea
 with $h^\mu( p^*, p^{\prime *},
\hat R^{-1} k^*_\pi)$ and the ${\cal J}^\mu( p^*, p^{\prime *},
\hat R^{-1} k^*_\pi)$ of Eq.~(\ref{eq:calJdefi}) related via
\bea
\bar u_{s'}( p^{\prime *})
{\cal J}^\mu( p^*, p^{\prime *},\hat R^{-1} k^*_\pi)\,u_s(p^*)=\chi^\dagger_{s'}
h^\mu( p^*, p^{\prime *},\hat R^{-1} k^*_\pi)\chi_s,
\eea
with $\chi_{s,s'}$ Pauli bispinors. Since the main objective is to see the
origin of the PV terms, we use in what follows a very simplified  notation. 
Corresponding full expressions can be found for instance in 
Refs.~\cite{Sato:2009de,Adler:1968tw}.  One can expand
\bea
h^\mu=\sum_{j_1}e^{i\delta_{j_1}}(|M_{Vj_1}| {\cal O}_{Vj_1}^\mu-|M_{Aj_1}|
 {\cal O}_{Aj_1}^\mu)
\eea 
where the sum is over all possible multipoles and the ${\cal O}_{V,A}^\mu$
operators are constructed from Pauli matrices and momenta. The ${\cal
O}_{V}^\mu$ operators violate parity while the ${\cal
O}_{A}^\mu$ ones do not. Then,
\beas
{\cal \widetilde
H}^{\mu\nu}=\frac12\sum_{j_1}\sum_{j_2}e^{i(\delta_{j_1}-\delta_{j_2})}
{\rm Tr}\left[( |M_{Vj_1}|{\cal O}_{Vj_1}^\mu-
 |M_{Aj_1}|{\cal O}_{Aj_1}^\mu)(|M_{Vj_2}| {\cal O}_{Vj_2}^{\nu\dagger}-
 |M_{Aj_2}|{\cal O}_{Aj_2}^{\nu\dagger})\right]
 \eeas
 Similar to the case before, the traces
 \bea
 {\rm Tr}( {\cal O}_{Vj_1}^\mu{\cal O}_{Vj_2}^{\nu\dagger})\ \ ,\ \
 {\rm Tr}( {\cal O}_{Aj_1}^\mu{\cal O}_{Aj_2}^{\nu\dagger}),
  \eea
  are real and   do not violate parity (they are tensors), while
 \bea
 {\rm Tr}( {\cal O}_{Vj_1}^\mu{\cal O}_{Aj_2}^{\nu\dagger})\ \ ,\ \
 {\rm Tr}( {\cal O}_{Aj_1}^\mu{\cal O}_{Vj_2}^{\nu\dagger}),
  \eea
  are imaginary and  violate parity (they are pseudotensors). Thus, we will have
 \bea
\underbrace{{\cal \widetilde H}^{\mu\nu\, (s)}_{VV+AA}}_{\rm PC}=
\frac{1}{2}\sum_{j_1}\sum_{j_2}\cos(\delta_{j_1}-\delta_{j_2})\left\{
|M_{Vj_1}||M_{Vj_2}|{\rm Tr}\left[{\cal O}_{Vj_1}^\mu
{\cal O}_{Vj_2}^{\nu\dagger}\right]
+|M_{Aj_1}||M_{Aj_2}|{\rm Tr}
\left[{\cal O}_{Aj_1}^\mu{\cal O}_{Aj_2}^{\nu\dagger}\right]\right\},
\label{eq:pc1bis}
\eea
which is real, symmetric and parity conserving since when it is contracted 
with the symmetric part of
the lepton tensor gives rise to a pure scalar, and 
\bea
\underbrace{i {\cal \widetilde H}^{\mu\nu\, (a)}_{VV+AA}}_{\rm PV}= \frac{i}{2}\sum_{j_1\neq j_2}\sin(\delta_{j_1}-
\delta_{j_2})\left\{|M_{Vj_1}||M_{Vj_2}|
{\rm Tr}
\left[{\cal O}_{Vj_1}^\mu{\cal O}_{Vj_2}^{\nu\dagger}\right]
+|M_{Aj_1}||M_{Aj_2}|{\rm Tr}
\left[{\cal O}_{Aj_1}^\mu{\cal O}_{Aj_2}^{\nu\dagger}\right]\right\}.
\label{eq:pv1bis}
\eea
which is imaginary, antisymmetric and parity violating since when it is 
contracted with the antisymmetric part of
the lepton tensor gives rise to a pseudoscalar. We also have
\bea
\underbrace{i {\cal \widetilde H}^{\mu\nu\, (a)}_{VA+AV}}_{\rm PC}&=&-\frac{1}{2}
\sum_{j_1}\sum_{j_2}\cos(\delta_{j_1}-\delta_{j_2})\Bigg\{|M_{Vj_1}||M_{Aj_2}|{\rm Tr}
\left[{\cal O}_{Vj_1}^\mu{\cal O}_{Aj_2}^{\nu\dagger}\right]  +|M_{Vj_2}||M_{Aj_1}|{\rm Tr}
\left[{\cal O}_{Aj_1}^\mu{\cal O}_{Vj_2}^{\nu\dagger}\right]\Bigg\}.
\label{eq:pc2bis}
\eea
which is imaginary, antisymmetric and parity conserving, since when it is 
contracted with the
antisymmetric part of the leptonic tensor it produces a scalar, and
\bea
\underbrace{{\cal \widetilde H}^{\mu\nu\, (s)}_{VA+AV}}_{\rm PV}&=&-\frac{i}{2}\sum_{j_1}
\sum_{j_2}\sin(\delta_{j_1}-\delta_{j_2})\Bigg\{|M_{Vj_1}||M_{Aj_2}|{\rm Tr}
\left[{\cal O}_{Vj_1}^\mu{\cal O}_{Aj_2}^{\nu\dagger}\right] 
 -|M_{Vj_2}||M_{Aj_1}|{\rm Tr}
\left[{\cal O}_{Aj_1}^\mu{\cal O}_{Vj_2}^{\nu\dagger}\right]\Bigg\},
\label{eq:pv2bis}
\eea
which is real, symmetric and parity violating, since when it is contracted with the
symmetric part of the leptonic tensor it produces a pseudoscalar. The conclusion
from this analysis is that, 
 in fully unitarized models, parity violating effects are due to the interference between
 multipoles that have different phases and thus correspond to different sets 
 of isospin, 
 total angular momentum and  pion orbital angular momentum values.
 For example, interference between the
 Delta resonance $P_{33}$ amplitude and other partial waves.
Other conclusions extracted before as to which part contributes to the $D^*$ 
($\sin\phi^*_\pi$) and $E^*$ ($\sin2\phi^*_\pi$) structure
function remain unchanged. 
%
%
%
%
%
%
%

\section{Comparison of the $\nu_e$ and $\bar \nu_e$ induced cross sections}
\label{sec:neutrino_comparison}
In this section we compare the results of the SL and DCC models with those 
from the  HNV approach for pion production cross
sections for both CC and NC processes. As mentioned before, since we want
the kinematics to be very similar to the case of pion electroproduction, we will
show mainly results for processes induced by electron (anti)neutrinos, though
 we will also compare to the scarce available data obtained from neutrino and 
 antineutrino muon beams.

%
%
\subsection{Total cross sections}
\begin{figure}[h]
\includegraphics[scale=0.6]{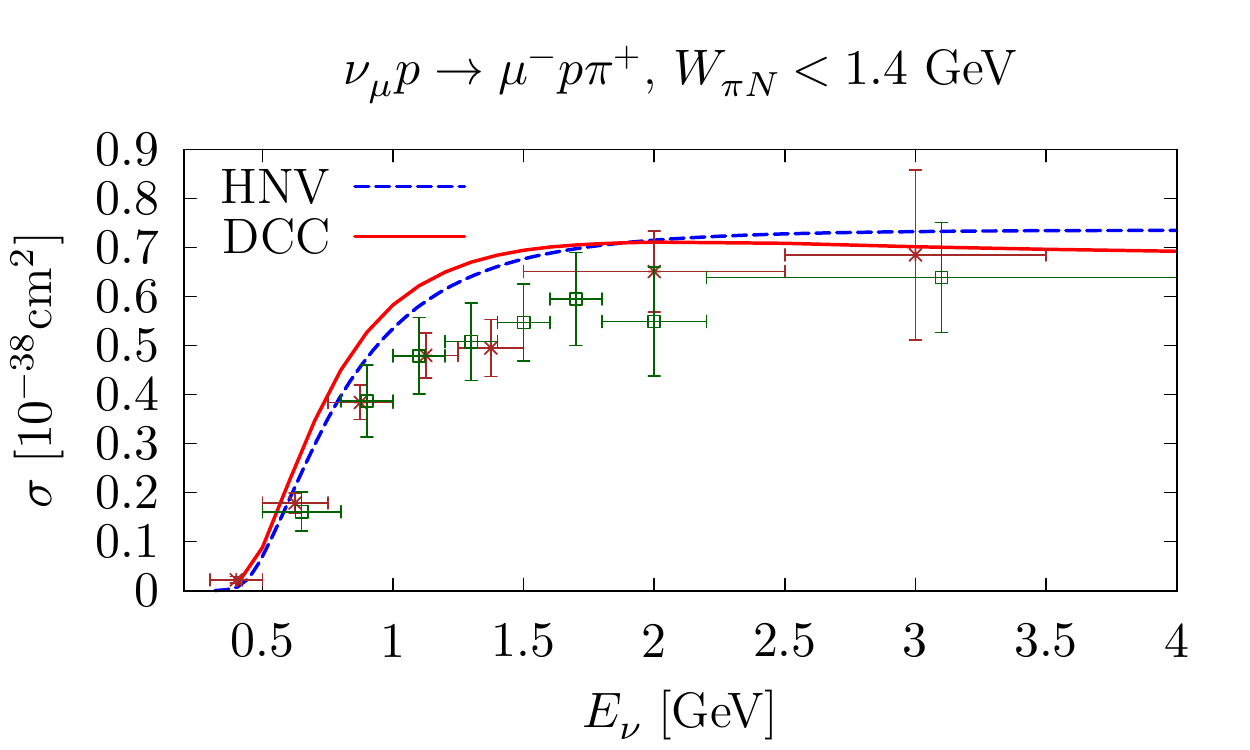}
\includegraphics[scale=0.6]{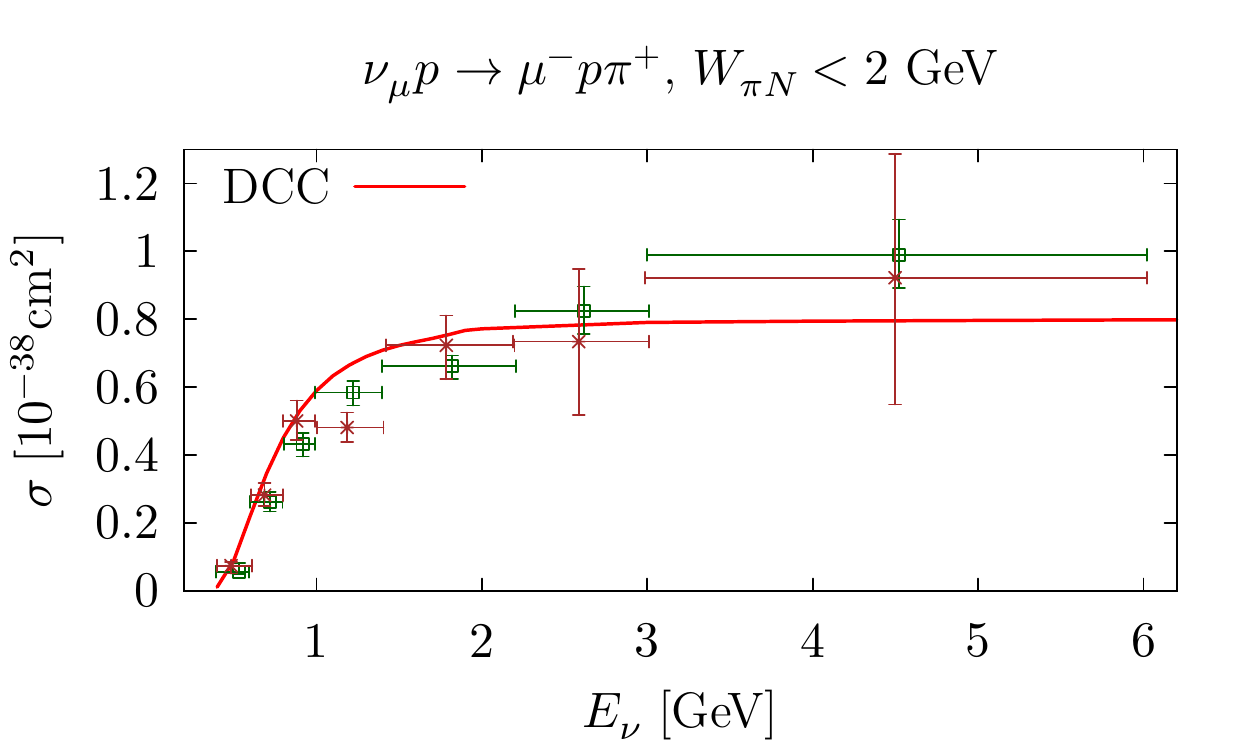}
\caption{ $\nu_\mu p\to \mu^-p\pi^+$ total cross section as a function 
of the neutrino energy. In the left panel a  kinematical 
cut $W_{\pi N}<1.4\, \text{GeV}$  has been included. The corresponding experimental data  has been taken from the
reanalysis done in Ref.~\cite{Rodrigues:2016xjj} of old ANL (crosses) and BNL
(open squares) data, where  the $W_{\pi N}<1.4\,\text{GeV}$ cut is also implemented. In the right panel a kinematical 
cut $W_{\pi N}<2\, \text{GeV}$ has been also applied to the theoretical calculation. The data  has been taken from the
reanalysis done in Ref.~\cite{Wilkinson:2014yfa} of old ANL (crosses) and 
BNL (open squares) data, without any cut on  $W_{\pi N}$.}
  \label{fig:CCtotnumuppi+}
\end{figure}

\begin{figure}[h]
\includegraphics[scale=0.6]{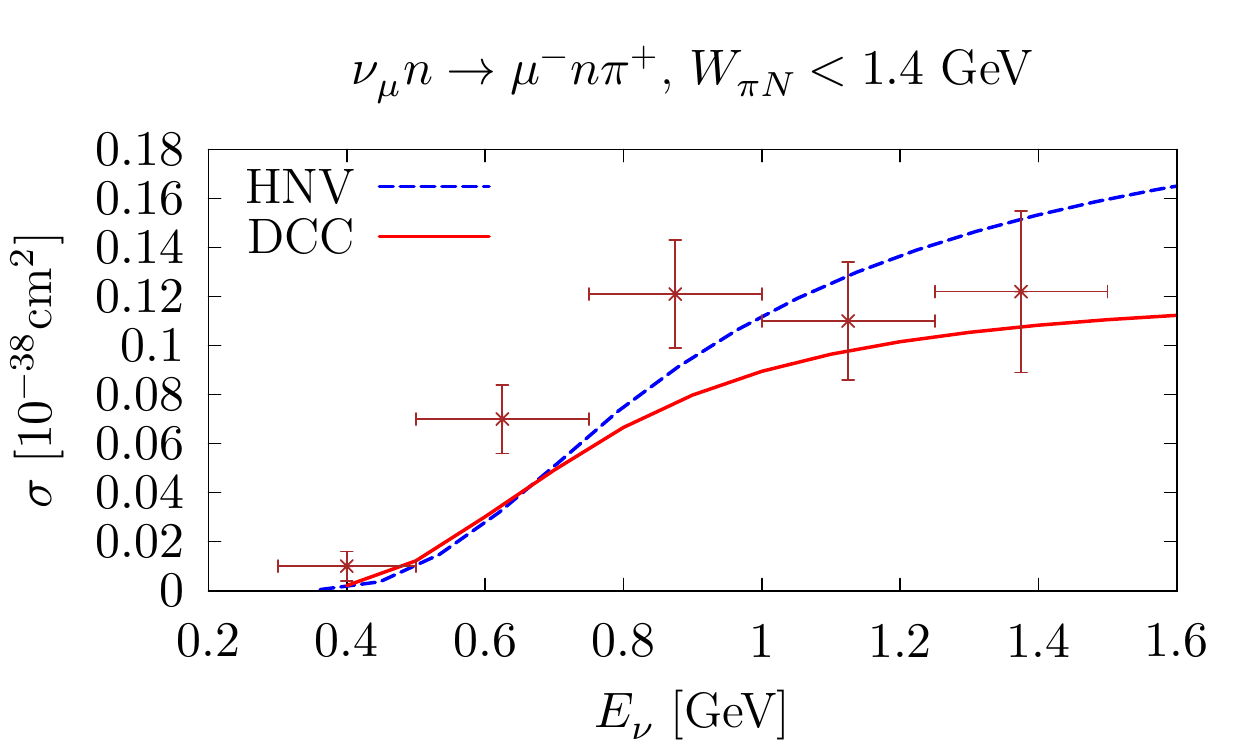}
\includegraphics[scale=0.6]{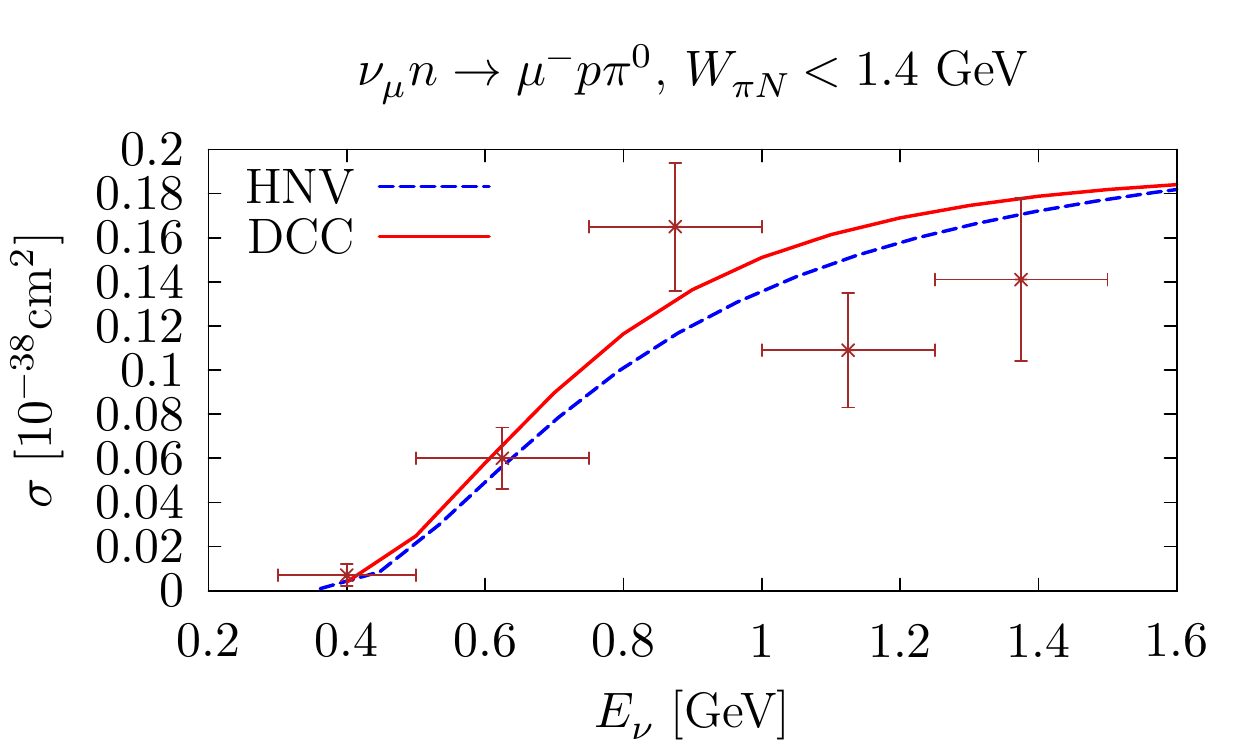}
\includegraphics[scale=0.6]{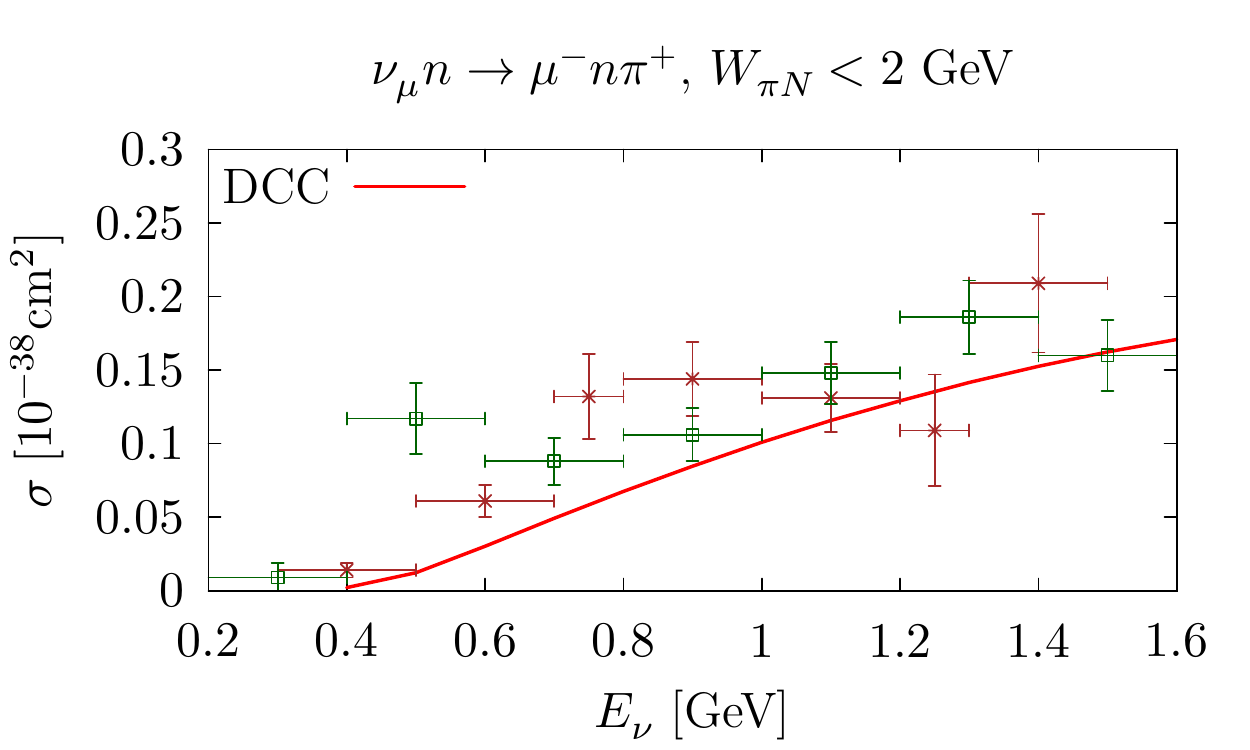}
\includegraphics[scale=0.6]{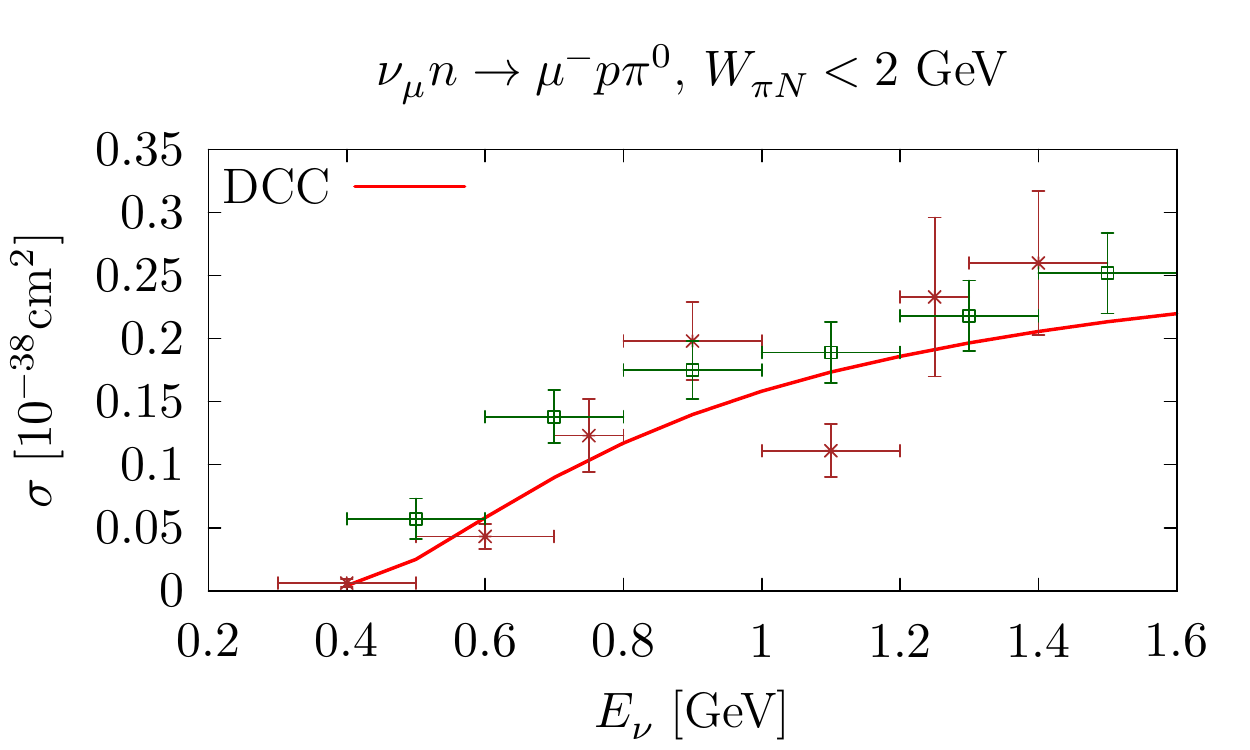}
\caption{ $\nu_\mu n\to \mu^-n\pi^+$ (left panels) and  $\nu_\mu n\to \mu^-p\pi^0$
(right panels) total cross sections as a function 
of the neutrino energy. In the upper panels, the  kinematical 
cut $W_{\pi N}<1.4\,\text{GeV}$  has been included in the data points, 
taken from the reanalysis of the experimental cross sections carried 
out in ~\cite{Rodrigues:2016xjj}, and both in the DCC and HNV theoretical 
predictions. For the  DCC model, we also show the results for 
$W_{\pi N}<2\, \text{GeV}$  in the bottom panels. The corresponding experimental 
data  has been taken also from the
reanalysis carried out in Ref.~\cite{Rodrigues:2016xjj} of the old ANL (crosses) and 
BNL (open squares) data, 
that does not incorporate any cut in the available phase space. }
  \label{fig:CCtotnumun}
\end{figure}

\begin{figure}[h]
\includegraphics[scale=0.7]{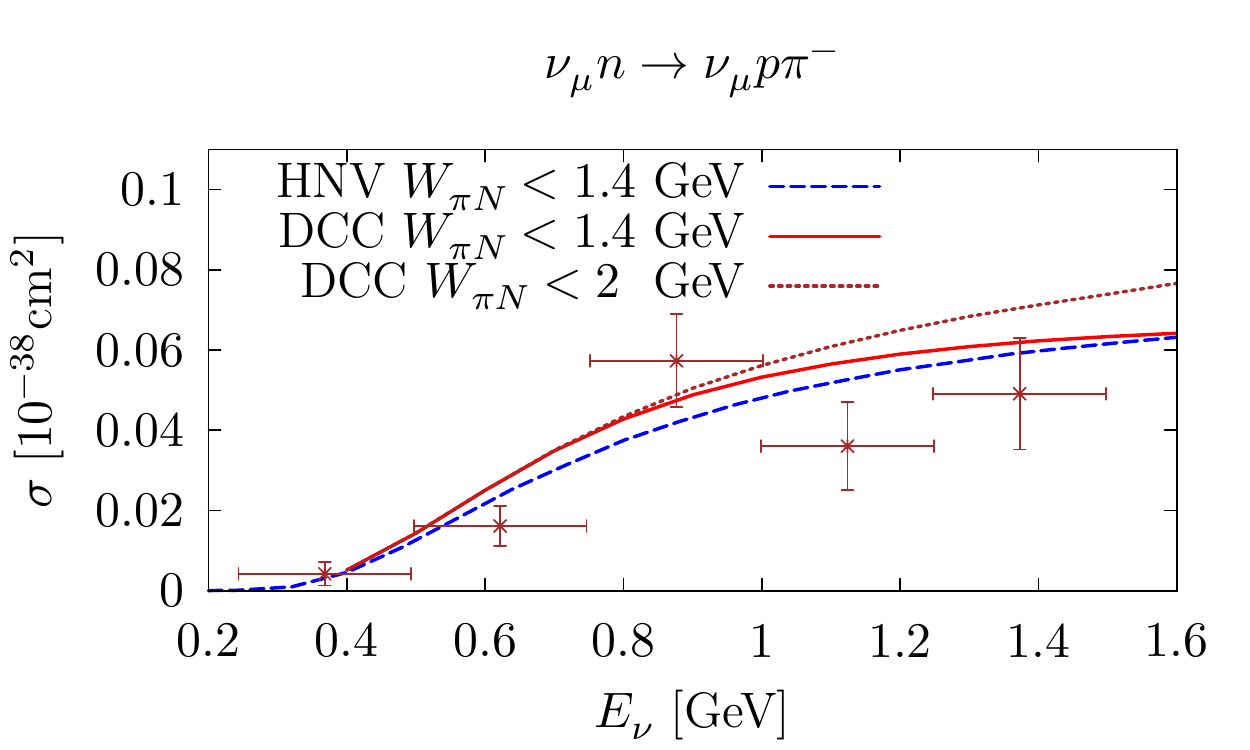}
\caption{ $\nu_\mu n\to \nu_\mu p\pi^-$  total cross section as a function 
of the neutrino energy. The corresponding experimental data  has 
been taken from Ref.~\cite{Derrick:1980nr} where no kinematical cut was
implemented. A  kinematical 
cut $W_{\pi N}<1.4\, \text{GeV}$  has been, however,  imposed for the HNV model (it 
has a moderate effect in this energy-range). We present the DCC results 
both with $W_{\pi N}<1.4\, \text{GeV}$ and  $W_{\pi N}<2\, \text{GeV}$ cuts. }
  \label{fig:NCtotnumuppi-}
\end{figure}

We start by showing in Figs.~\ref{fig:CCtotnumuppi+}, 
\ref{fig:CCtotnumun} and  \ref{fig:NCtotnumuppi-} total cross section results for $\nu_\mu$
induced reactions for which there is experimental data measured in deuterium.
The theoretical results we present have been evaluated, however, at the nucleon
level. Taking into account deuteron wave function effects reduces the cross
section by some 5\% ~\cite{Hernandez:2010bx}. 
For a meaningful   comparison between the HNV and DCC models we impose a  $W_{\pi
N}<1.4\, \text{GeV}$ cut.
This  is done to 
minimize the effect of higher order contributions in the chiral
expansion not taken into account in the  evaluation of  the nonresonant background
within the HNV model and,  also, the 
possible unphysical behavior of the $\Delta$ 
amplitudes  far from the $\Delta$ 
peak  that would affect the HNV model (this unphysical behavior is discussed in 
Ref.~\cite{Gonzalez-Jimenez:2016qqq}). 
Besides, below this $W_{\pi N}$ cut, contributions from higher mass resonances,
not taken into account in the HNV model, should be negligible. 

For the $\nu_\mu p\to \mu^- p\pi^+$ channel we see that  the DCC and HNV models produce 
similar results that lie above  experimental data in the $1-2$ GeV
neutrino energy region. To a lesser extent, this seems to also  be the case for the DCC model
evaluated with $W_{\pi N}<2\, \text{GeV}$ and shown in comparison with data in the right
 panel of
Fig.~\ref{fig:CCtotnumuppi+}. Note, however, that for the latter data no cut in 
$W_{\pi N}$ has been
applied.
For the $\nu_\mu n\to \mu^- n\pi^+$ channel  the discrepancies between the two
models are larger in the high
neutrino energy region  (see the top left panel of Fig.~\ref{fig:CCtotnumun}). The fact that the HNV model gives larger cross sections
for that channel is a direct consequence of the $\Delta$ propagator modification in
Eq.~(\ref{eq:mod}). The HNV predictions for this channel, without 
including the additional terms generated by the latter modification, can be 
seen (black dashed line) in the bottom panel of Fig.~3 in 
Ref.~\cite{Hernandez:2016yfb}, and they were smaller than those obtained 
in the DCC model and shown here. 
For the $\nu_\mu n\to \mu^- p\pi^0$ and the NC
 $\nu_\mu p\to \nu_\mu p\pi^-$ channels, both the HNV and DCC models give again similar
 results  that are in a good global agreement with data, as can be appreciated in the
  right upper panel of Fig.~\ref{fig:CCtotnumun} and in Fig.~\ref{fig:NCtotnumuppi-}.

\begin{figure}[h]
\includegraphics[scale=0.5]{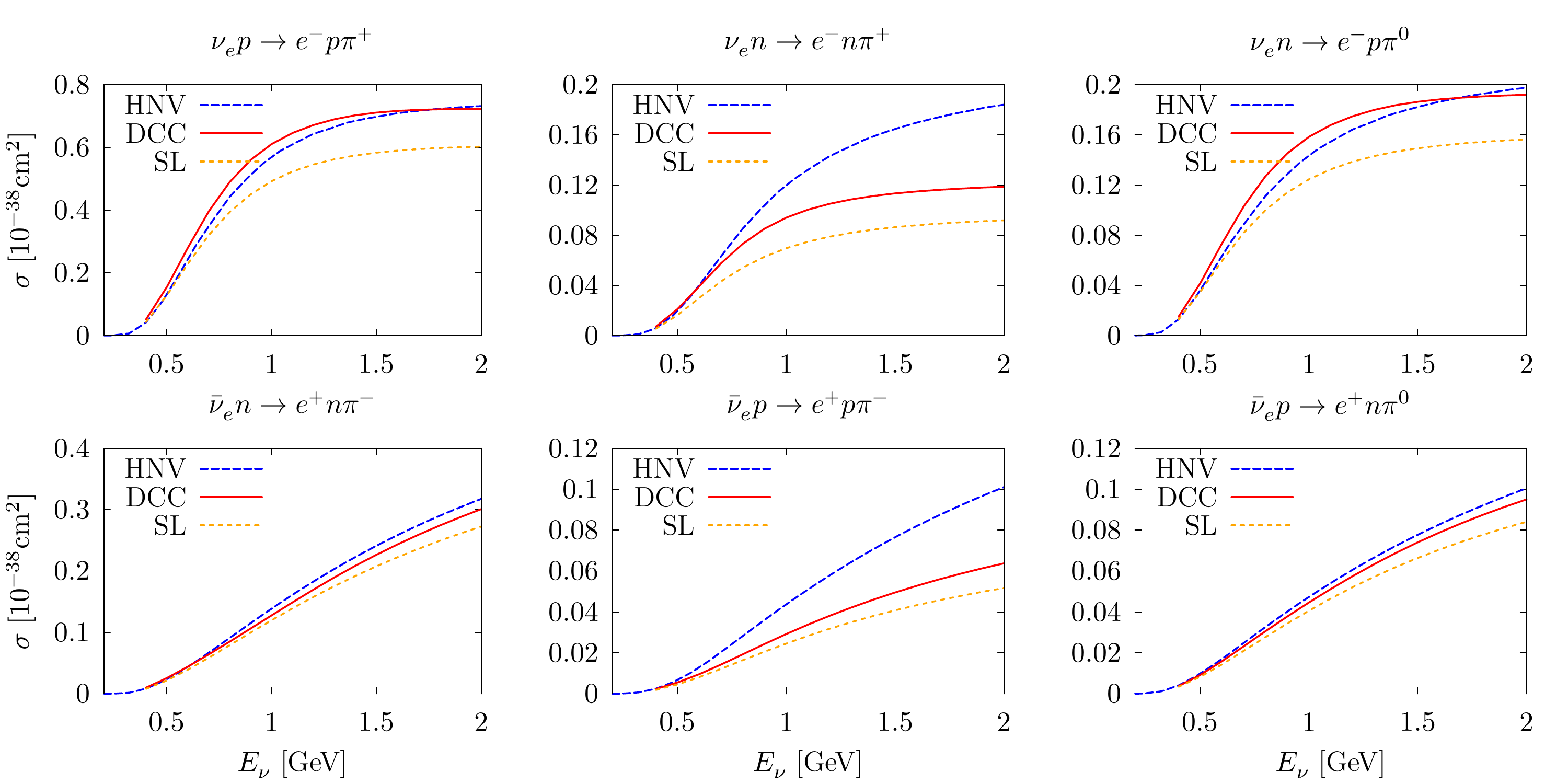}
\caption{  CC total cross sections as a function of the neutrino energy from different theoretical models. A kinematical 
cut $W_{\pi N}<1.4\,\text{GeV}$ on the final pion-nucleon invariant mass has been included.  }
  \label{fig:CCtot}
\end{figure}

\begin{figure}[h]
\center
{\includegraphics[scale=0.45]{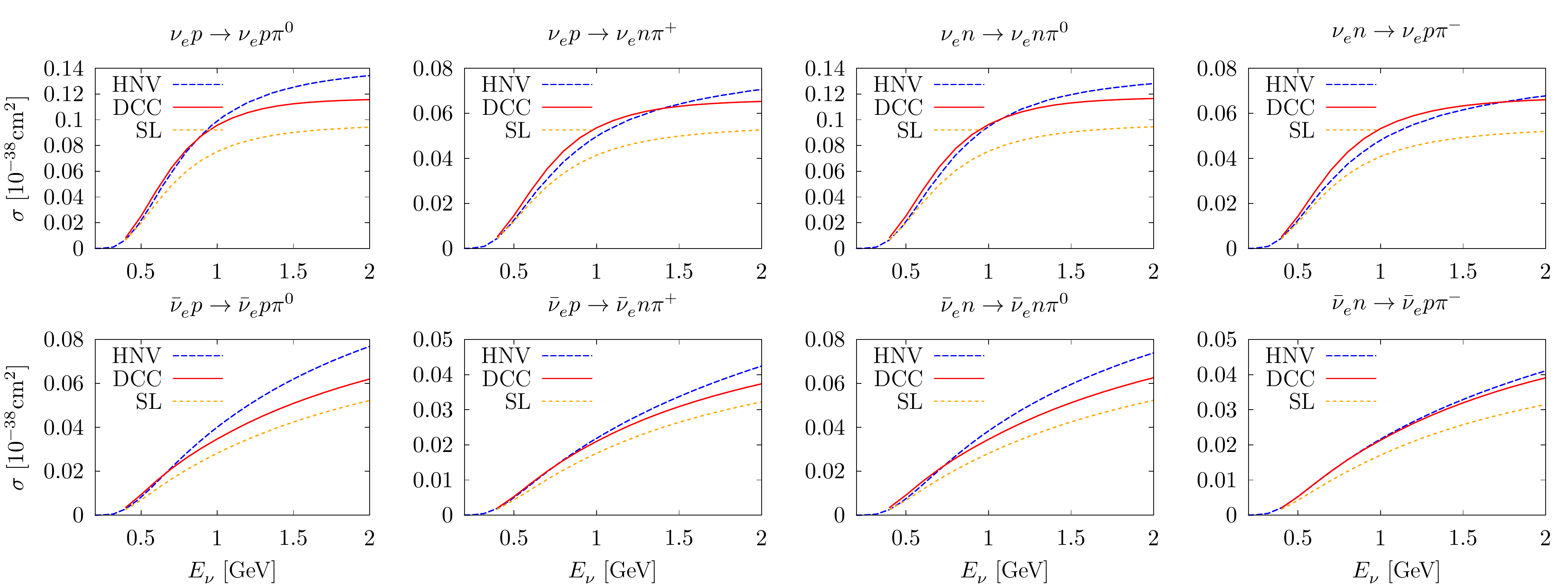}}
\caption{NC total 
cross sections as a function of the neutrino energy from different theoretical models. A kinematical cut $W_{\pi N}<1.4\,\text{GeV}$ on the final pion-nucleon
invariant mass has been included. }
  \label{fig:NCtotn}
\end{figure}

Moving now to reactions induced by electron (anti-)neutrinos, in 
Figs.~\ref{fig:CCtot} and \ref{fig:NCtotn} we compare 
the HNV, SL and DCC total cross section predictions for  all possible
 channels. We show results up to  2\,GeV neutrino energy but imposing the cut 
$W_{\pi N}<1.4$\,GeV.  First, in Fig.~\ref{fig:CCtot}, we display the CC channels, where we observe  that 
the HNV and DCC models produce always larger
cross section  than the SL approach. This is mainly because the SL model uses
 the axial
$N \to \Delta$ coupling predicted by a constituent quark model, while the DCC 
and HNV models use somewhat stronger
couplings close to the PCAC prediction. The difference is particularly large in neutrino $n\pi^+$ and antineutrino $p \pi^-$ channels\footnote{Note that isospin invariance tells us that
$\langle p\pi^-|J^\mu_{cc-}(0)|p\rangle=\langle n\pi^+|J^\mu_{cc+}(0)|n\rangle$ so
that the $\nu_e n\to e^-n\pi^+$ and $\bar \nu_e p\to e^+p\pi^-$ channels share
the same hadronic tensor and they only differ in
the antisymmetry part of the lepton tensor that changes sign.}, for which the HNV cross sections are also significantly bigger than those predicted using the DCC model.  
As mentioned, this latter enhancement in the HNV predictions is due to the new contact term
  resulting from  the $\Delta$ propagator modification of Eq.~(\ref{eq:mod}), 
  and as discussed in Ref.~\cite{Hernandez:2016yfb}, 
  it seems to be supported by the old ANL and BNL bubble chamber 
  $\nu_\mu n\to \mu^- n\pi^+$  data (see also upper left panel of 
  Fig.~\ref{fig:CCtotnumun}). In these two channels, the strength of the 
  crossed $\Delta$ term is enhanced by spin and isospin factors and it greatly 
  cancels with  the rest of the background. The modification of the $\Delta$
   propagator significantly reduces the crossed $\Delta$ contribution, 
   leading to a smaller cancellation with the background and, as net result, 
   to an increase of the cross section.   For the rest of the channels, 
   the crossed $\Delta$ term is much smaller, and  the DCC and HNV models 
   produce similar results. 

Next in Fig.~\ref{fig:NCtotn} we compare the results for the NC channels. The pattern 
is similar to that outlined above for the CC cross sections. DCC and HNV predictions 
agree reasonably well in general, while those obtained from the  SL model are systematically 
lower for the reason mentioned in the previous paragraph. Here  the modification in the $\Delta$ propagator  of Eq.~(\ref{eq:mod}) implemented 
in the HNV model produces significantly smaller effects, because the isovector contribution 
to the amplitudes in all NC channels always involves both the $p\pi^+$ and $n \pi^+$ 
CC amplitudes, and there are no NC channels determined  only by the latter 
of the two~\cite{Hernandez:2007qq}.   

%
%
%

\subsection{Differential cross sections}\label{sec:diff_cross-sec}

\begin{figure}[h]
\centerline{\includegraphics[scale=0.55]{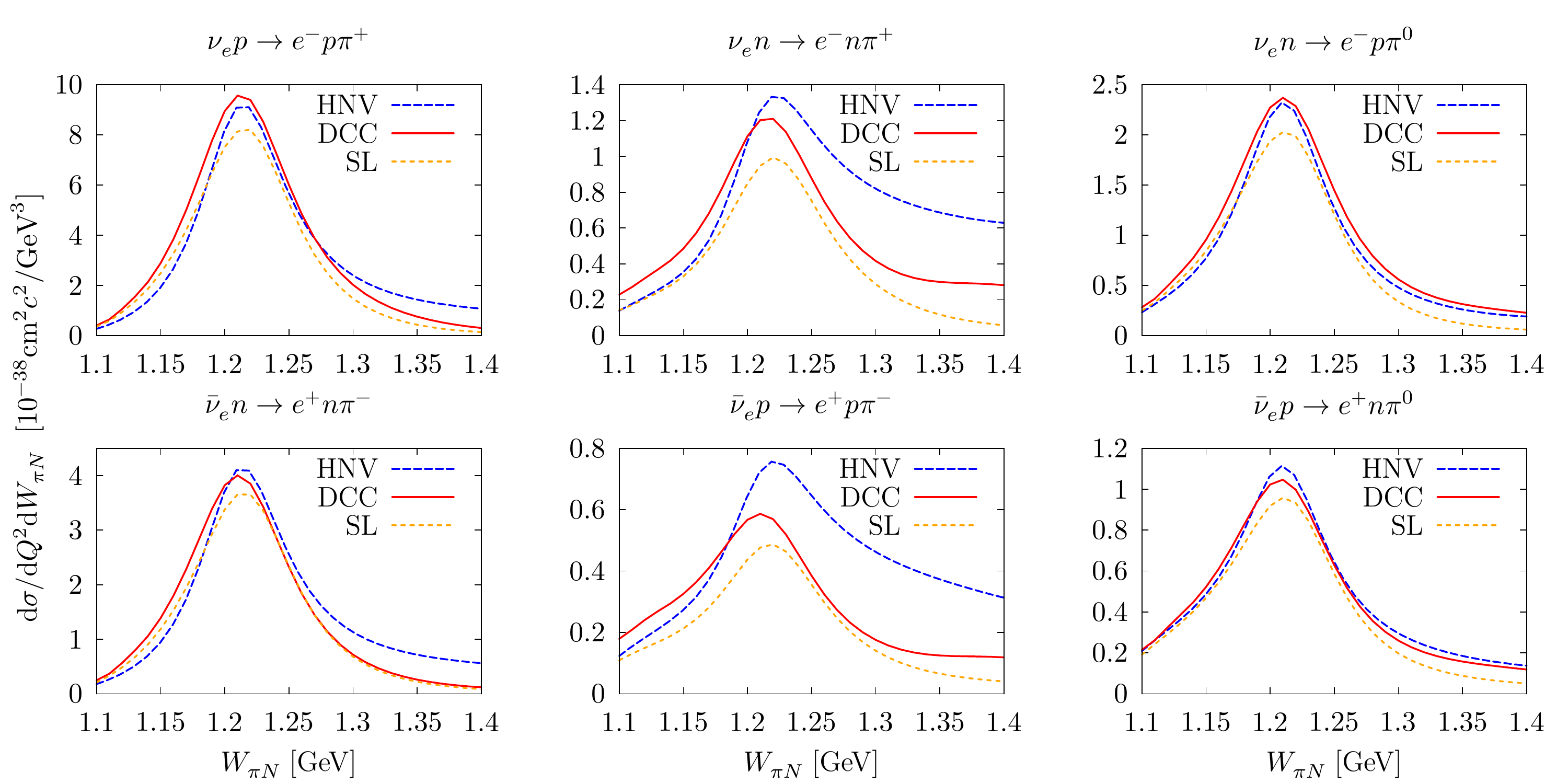}}
\caption{CC--$d\sigma/(dQ^2dW_{\pi N})$ differential  cross
sections   as a function of $W_{\pi N}$, for  fixed $E_\nu= 1\,\text{GeV}$ and
$Q^2=0.1\,\text{GeV}^2/c^2$.   }
  \label{fig:CCdWdQ2}
\end{figure}

\begin{figure}[h]
{\includegraphics[scale=0.45]{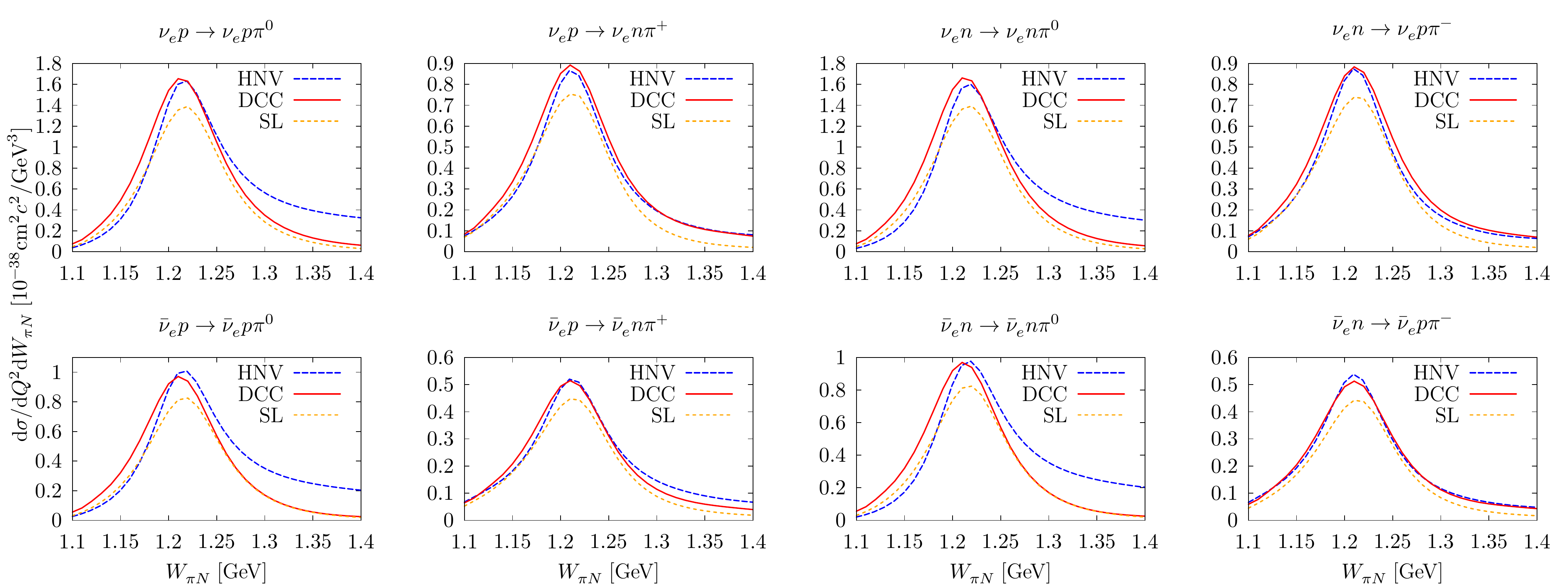}}
\caption{NC--$d\sigma/(dQ^2dW_{\pi N})$ differential  cross
sections as a function of $W_{\pi N}$, for fixed $E_\nu= 1\,\text{GeV}$ and
$Q^2=0.1\,\text{GeV}^2/c^2$.   }
  \label{fig:NCdWdQ2}
\end{figure}

In Figs.~\ref{fig:CCdWdQ2} and  \ref{fig:NCdWdQ2}, we now show CC and NC  
$d\sigma/(dQ^2dW_{\pi N})$
differential cross sections as a function of $W_{\pi N}$, for 
fixed  $E_\nu= 1\,\text{GeV}$ and $Q^2=0.1\,\text{GeV}^2/c^2 $ values. 
The $Q^2$ value is in the 
range where the $d\sigma/dQ^2$ differential cross section is 
maximum. Very similar results (not shown) are obtained when $Q^2$ is
varied in the interval $(0.05,0.15)\,\text{GeV}^2/c^2 $. 
All the distributions show the characteristic peak at the $\Delta$ pole.  
 Apart from the differences in normalization, stemming from 
 different total cross section predictions, we see that, in general, the
  SL and DCC
models show more strength at lower $W_{\pi N}$ values, whereas the opposite 
happens
for the HNV model. Again, this is more pronounced for the 
$\nu_e n\to e^-n\pi^+$ and $\bar \nu_e p\to e^+p \pi^-$ channels where the
 effects of the changes in the $\Delta$ propagator in Eq.~(\ref{eq:mod}) are
  more relevant. Nevertheless, and with the exception of these two latter 
  reactions,  we observe a reasonable agreement between the 
  HNV and DCC models, in spite of the relative simplicity of the former as 
  compared to the latter.  
\begin{figure}[h!]
\centerline{\includegraphics[scale=0.5]{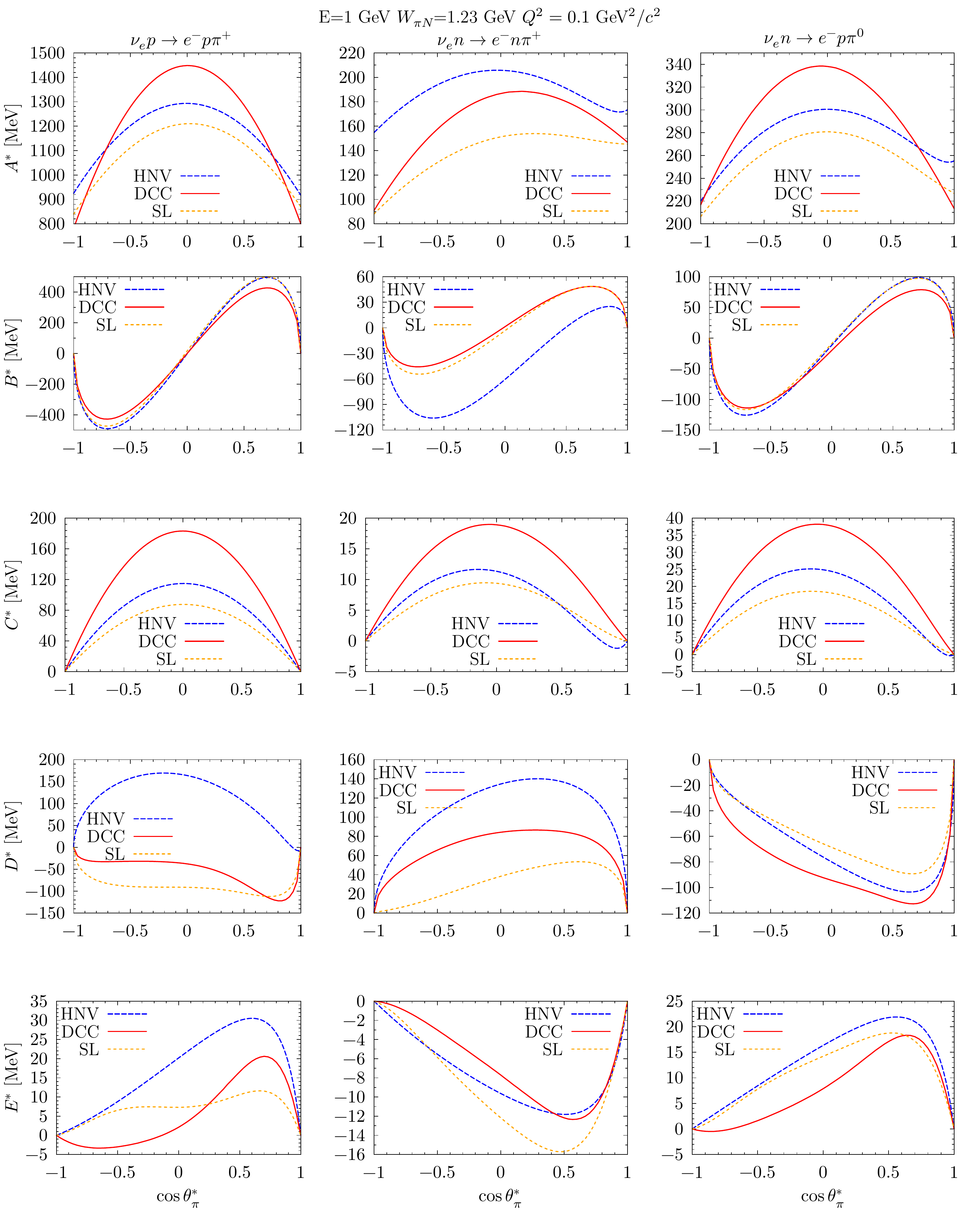}}
\caption{ CC-neutrino $A^*,B^*,C^*,D^*$ and $E^*$ structure functions, as defined in
Eq.~(\ref{eq:abcde}), represented as a function of $\cos\theta^*_\pi$ for 
fixed 
$E_\nu=1\,\text{GeV}$, $W_{\pi N}=1.23\,\text{GeV}$ and $Q^2=0.1\,
\text{GeV}^2/c^2$. }
  \label{fig:ABCDE}
\end{figure}
%
\begin{figure}[h!]
\centerline{\includegraphics[scale=0.5]{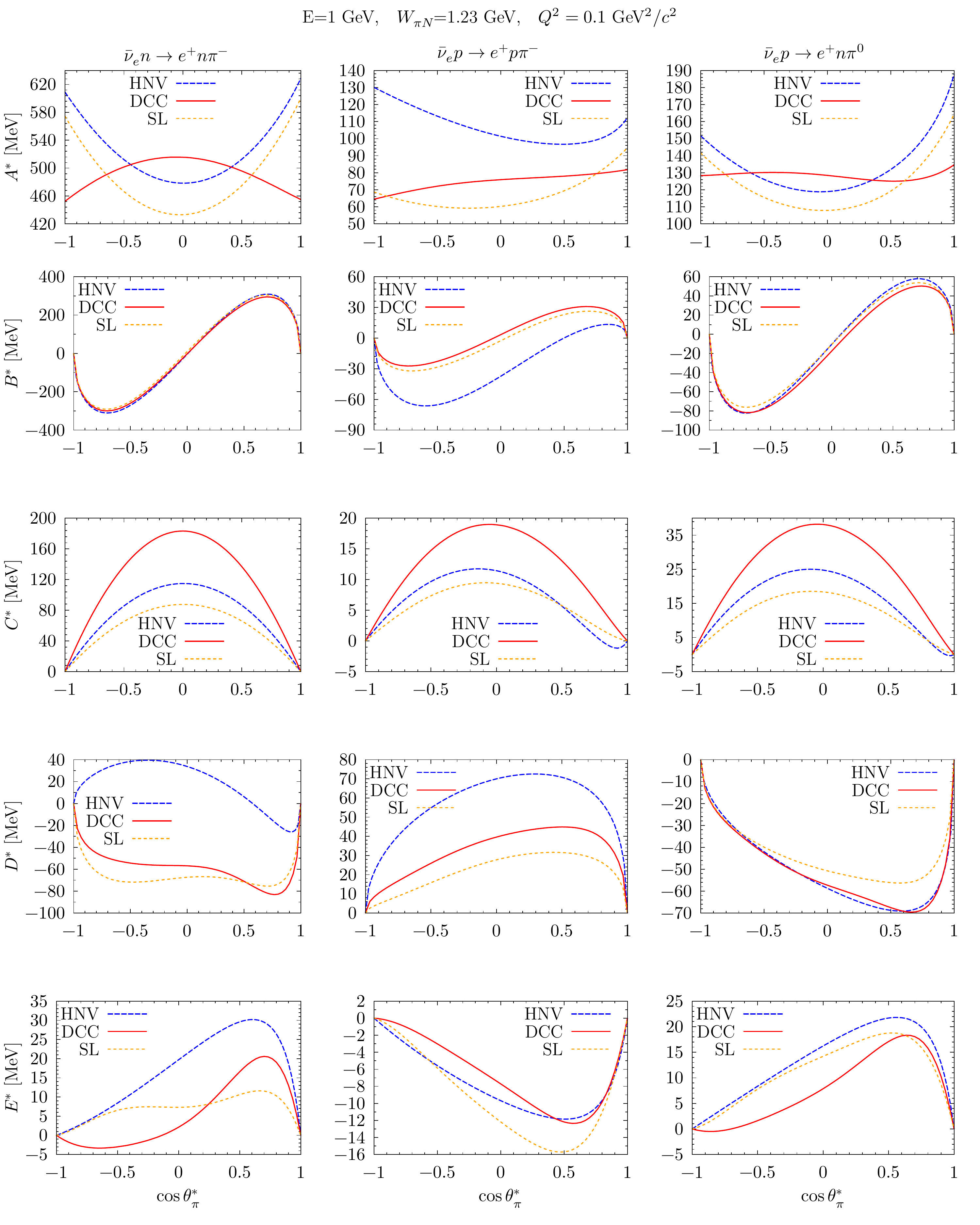}}
\caption{ Same as Fig.~\ref{fig:ABCDE} for CC-antineutrino reactions. }
  \label{fig:ABCDE_anti}
\end{figure}
\begin{figure}[h!]
\centerline{\includegraphics[scale=0.75]{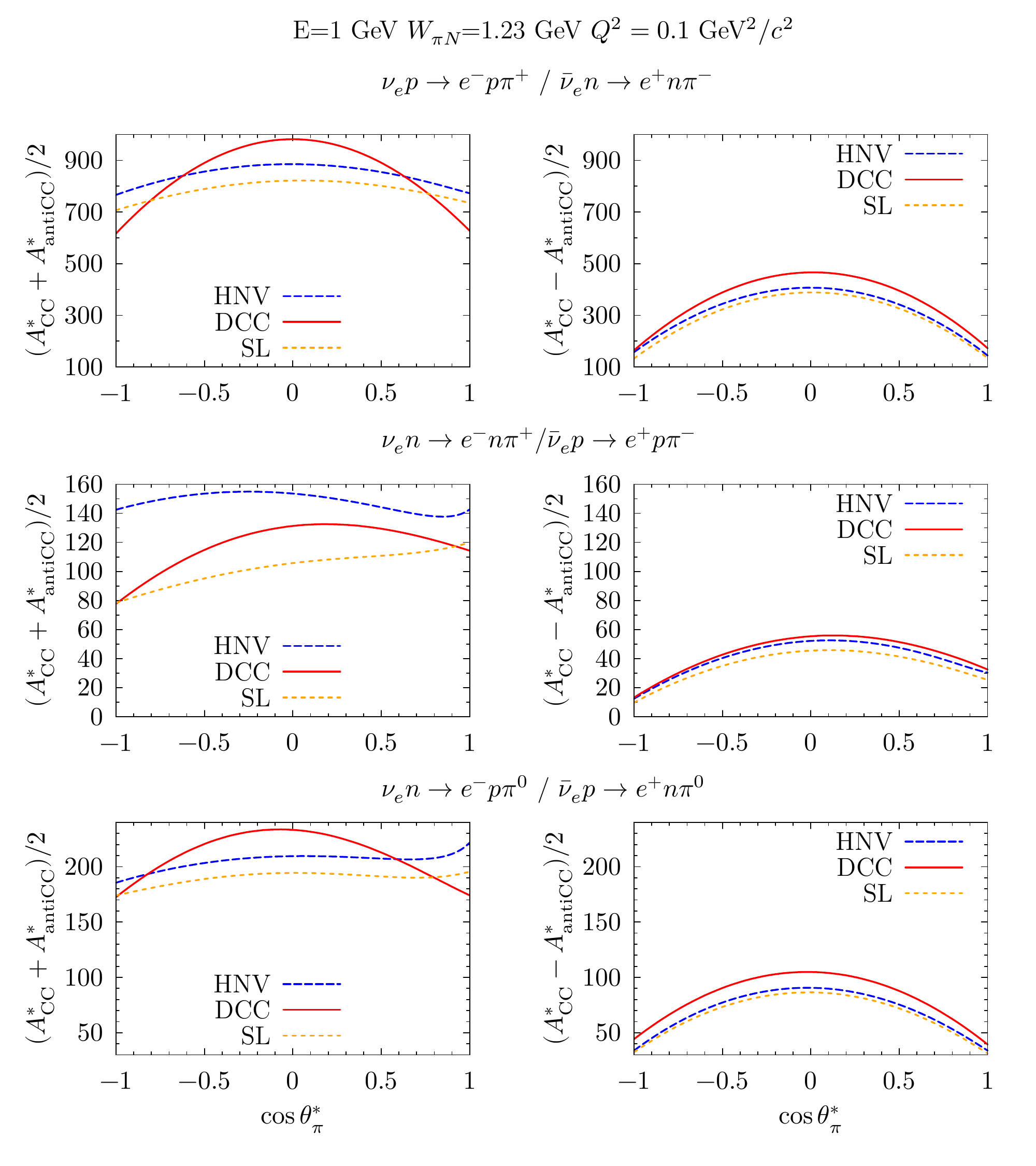}}
\caption{ Semi-sums and semi-differences  (in MeV units) of the neutrino and antineutrino $A^*$ structure functions displayed in Figs.~\ref{fig:ABCDE} and \ref{fig:ABCDE_anti}. }
  \label{fig:Astar}
\end{figure}

%

 Further fixing $W_{\pi N}=1.23\,\text{GeV}$,
 we show,  in Figs.~\ref{fig:ABCDE} (neutrinos) and \ref{fig:ABCDE_anti} 
(antineutrinos), the $\cos\theta_\pi^*$  dependence of the  $A^*,B^*,C^*,D^*$ and $E^*$  
CC structure functions introduced in Eq.~(\ref{eq:abcde}). Some gross features of the 
shapes of these functions can be understood from the expressions given in the latter set 
of equations,  bearing in mind that not only the second but also  the first spatial 
components of $ q^*$ and $ p^* $ are zero, and that only $\hat R^{-1} k^*_\pi$ 
has a non-vanishing $ X^*$ component, which is proportional to $\sin\theta_\pi^*$. Thus, 
we immediately see that $C^*$ and $E^*$ must be proportional to  $\sin^2\theta_\pi^*$, since 
$(\widetilde W_{11}^{(s)}-\widetilde W_{22}^{(s)})$ and $\widetilde W_{12}^{(s)}$ are 
necessarily proportional to the square of $[\hat R^{-1} k^*_\pi]_1$.  In addition, 
there might exist some additional dependence on $\theta_\pi^*$, because all the nucleon 
structure responses could be a function of the Lorentz scalar $q^*\cdot k^*_\pi$. These corrections
 look small for $C^*$ and more sizable for $E^*$, for which the DCC model, for example,  
 predicts a change of 
sign 
in the $\nu_e p\to e^-p\pi^+$ 
and $\bar \nu_e n\to e^+ n \pi^-$ channels. If one uses a multipole expansion 
 of the hadronic amplitude, the deviation of  $E^*$ from a pure 
$\sin^2\theta_\pi^*$ dependence originates from the interference with multipoles
corresponding to a pion orbital angular momentum higher than
one~\cite{Sato:2009de,Adler:1968tw}.

Using the same type of argument, 
one can also factorize the $\sin\theta_\pi^*$ function in $B^*$ and $D^*$, 
which explains why these structure functions vanish at the extremes ($\cos\theta_\pi^*=\pm 1$). 
The additional $\theta_\pi^*$ dependencies, generated by $q^*\cdot k^*_\pi$ and by some other 
tensor terms in $B^*$ and $D^*$, give rise to large deviations from the $\sin\theta_\pi^*$  
shape for these response functions.

Let us focus now on the neutrino processes. For the $\nu_e n\to e^-p\pi^0$ and $\nu_e n\to e^-n\pi^+$
 channels, the three  models produce  structure functions with a similar shape. For the
 $\nu_e p\to e^-p\pi^+$, the   $D^*$ structure function, and to a lesser extent
 the $E^*$ structure function, show larger differences in shape.  These are 
 precisely the two PV contributions to the
 differential cross section. As discussed in Sec.~\ref{sec:originpv}, PV terms in the hadronic tensor derive from the interference between
 different contributions to the hadronic current that are not relatively real. 
 The origin of these discrepancies should be found in the  different pattern 
 of relative phases in the three models. 
 As seen from Eqs.~(\ref{eq:pv1bis}) and (\ref{eq:pv2bis}), $D^*$ and $E^*$
are sensitive to the difference in phase of the multipole amplitudes. 
Below the two-pion production threshold, Watson
theorem tells us that those phases
 are determined by the
corresponding $\pi N$ phase shifts. The latter requirement  is fully satisfied by 
the DCC and SL models, whereas this is not true for the HNV model where only 
a partial unitarization of the $P_{33}$ amplitude is
implemented through the use of the Olsson phases. In the case of the $E^*$ 
structure function for the 
 $\nu p \rightarrow  e^- \pi^+ p$ reaction, and keeping only $s$ and $p$ pion 
 partial
 waves,
one can explicitly observe that  its value is given by the interference between
 the $P_{33}$  (dominated by the $\Delta$) and the  $P_{31}$
 (non-resonant) amplitudes
 \begin{eqnarray}
    E^* \propto \sin^2\theta_\pi \sin(\delta_{P_{33}} -
 \delta_{P_{31}})[|M_{1+}^V| |E_{1-}^A|
              + |M_{1-}^V|(4|M_{1+}^A| + 2|E_{1+}^A|)].
\end{eqnarray}
  Hypothetical future measurements of 
 these structure functions might serve to further constrain the pion production 
 models.    Let us notice, however, that for the $\nu_e p\to e^-p\pi^+$ channel the magnitude of $D^*$ and $E^*$ is much  smaller than $A^*$, getting at most $10\%$ of its value, whereas for the other two channels it reaches $\sim 30\%$.

  For all structure functions, the various predictions differ not only in 
  shape but also in  size. This shows how demanding the test carried out 
  in this work is. This is even more evident when the antineutrino structure 
  functions, shown in Fig.~\ref{fig:ABCDE_anti}, are  examined. Isospin 
  symmetry~\cite{Hernandez:2007qq} implies that the hadron tensors of the  
   $\nu_e p \to e^- p \pi^+$ and $\bar\nu_e n \to e^+ n \pi^-$ reactions
    are equal. The same happens for the  $\nu_e n \to e^- n \pi^+$ 
    and $\bar\nu_e p \to e^+ p \pi^-$ processes, and for the 
    $\nu_e n \to e^- p \pi^0$ and 
    $\bar\nu_e p \to e^+ n \pi^0$ processes. Therefore, the structure 
    functions depicted in the first, second and third columns of 
    Figs.~\ref{fig:ABCDE} (neutrino) and ~\ref{fig:ABCDE_anti} (antineutrino) 
    should differ only in the terms proportional to the antisymmetric part
     of the lepton tensor, that changes sign. From the explicit expressions
      given in Eqs.~(\ref{eq:abcde}), we immediately realize that neutrino 
      and antineutrino 
$C^*$ and $E^*$ structure functions are identical, when looking at the 
appropriate channels. For the $A^*$  response function one sees significant
 differences between the DCC and HNV  predictions  for the antineutrino case. Thus, for instance in the $\bar\nu_e n \to e^+ n \pi^-$ 
 channel, we see that, compared to the HNV and SL models, the DCC model predicts 
    a different shape, in contrast to the situation  discussed above for the related
  neutrino  $\nu_e p \to e^- p \pi^+$ channel. For the antineutrino reaction, 
  the first two approaches lead to concave-up profiles,  as a function of
   $\cos\theta^*_\pi$, while the latter one gives rise to  a concave-down 
   shape. However, DCC and HNV  integrated $A^*$  structure functions  
   differ by less than 5\%, as can be inferred from the 
   $d\sigma/dQ^2dW_{\pi N}$ differential  cross
sections depicted in the left bottom panel of Fig.~\ref{fig:CCdWdQ2}. To shed
 light on this different behavior, we show  in Fig.~\ref{fig:Astar} the 
 symmetric and antisymmetric contributions~\footnote{The antisymmetric
  contribution, whose sign is different for neutrinos and for antineutrinos, 
   is given by
\begin{equation}
 A^*_{\rm antisymmetric}=\int\frac{|\vec k^*_\pi|^2d|\vec
k^*_\pi|}{E^*_\pi}\,2i L^{12}\,\widetilde W_{12}^{(a)},
\end{equation}
while the rest of the terms in the expression of $A^*$ in Eq.~(\ref{eq:abcde}) 
is the same for neutrino and antineutrino reactions, and it is driven by  
the symmetric lepton tensor.} 
to $A^*$ for the $\nu_e p \to e^- p \pi^+/\bar\nu_e n \to e^+ n \pi^-$ 
(first row), 
$\nu_e n \to e^- n \pi^+/\bar\nu_e p \to e^+ p \pi^-$ (second row) and 
$\nu_e n \to e^- p \pi^0/\bar\nu_e p \to e^+ n \pi^0$ (third row) 
isospin related channels, at $W_{\pi N}=1.23\,\text{GeV}$ and $Q^2=0.1\, 
\text{GeV}^2/c^2$ as in Figs.~\ref{fig:ABCDE} and ~\ref{fig:ABCDE_anti}.
 DCC antisymmetric contributions to 
$A^*$ are larger than those obtained within the HNV and SL models. If we 
focus on the results found for $\nu_e p \to e^- p \pi^+/\bar\nu_e n \to e^+ n 
\pi^-$, we see that all models predict similar $\cos\theta^*_\pi$ shapes 
(concave-down) for both the symmetric and antisymmetric terms of $A^*$, but
 when they are subtracted to obtain the antineutrino structure functions, 
 they  give rise to concave-up shapes in the HNV and SL approaches. This
  illustrates the importance of carrying out a thorough test of the different
   model results at the level of outgoing pion angular distributions, 
   going beyond comparisons done for partially integrated cross sections, 
   where the differences tend to cancel.  In addition, we
    can conclude  from Fig.~\ref{fig:Astar} that  
the  inclusion in the HNV model of a local term, induced by the 
$\Delta$ propagator modification discussed in Eq.~(\ref{eq:mod}), 
 produces visible effects in the  symmetric 
contribution to $A^*$ in the $\nu_e n \to e^- n \pi^+$ and 
$\bar\nu_e p \to e^+ p \pi^-$ reactions. 
 
Returning to the discussion of Figs.~\ref{fig:ABCDE} and \ref{fig:ABCDE_anti},  we see that, 
in general, $|D^*|$ is greater than $|E^*|$, and thus PV effects are dominated by the 
$\sin\phi^*_\pi$ dependence of the differential cross section. Comparing the relative sizes 
of $A^*$ and $|D^*|$, we expect the largest parity violations in the $\nu_e n \to e^- n \pi^+$, 
$\bar\nu_e p \to e^+ p \pi^-$,  $\nu_e n \to e^- p \pi^0$ and $\bar\nu_e p \to e^+ n \pi^0$ 
reactions, while the smallest ones should occur in the isospin 3/2 $\nu_e p \to e^- p \pi^+$ 
and $\bar\nu_e n \to e^+ n \pi^-$ channels, that are dominated by the direct $\Delta$ mechanism. 
In addition, in this latter case, we observe that PV effects are greatly reduced  
 for $\bar\nu_e n \to e^+ n \pi^-$, since the relative size of the $|D^*|/A^*$ ratio for this 
 reaction is significantly smaller than for the isospin related one  $\nu_e p \to e^- p \pi^+$.
 
\begin{figure}[h!]
\centerline{\includegraphics[scale=0.6]{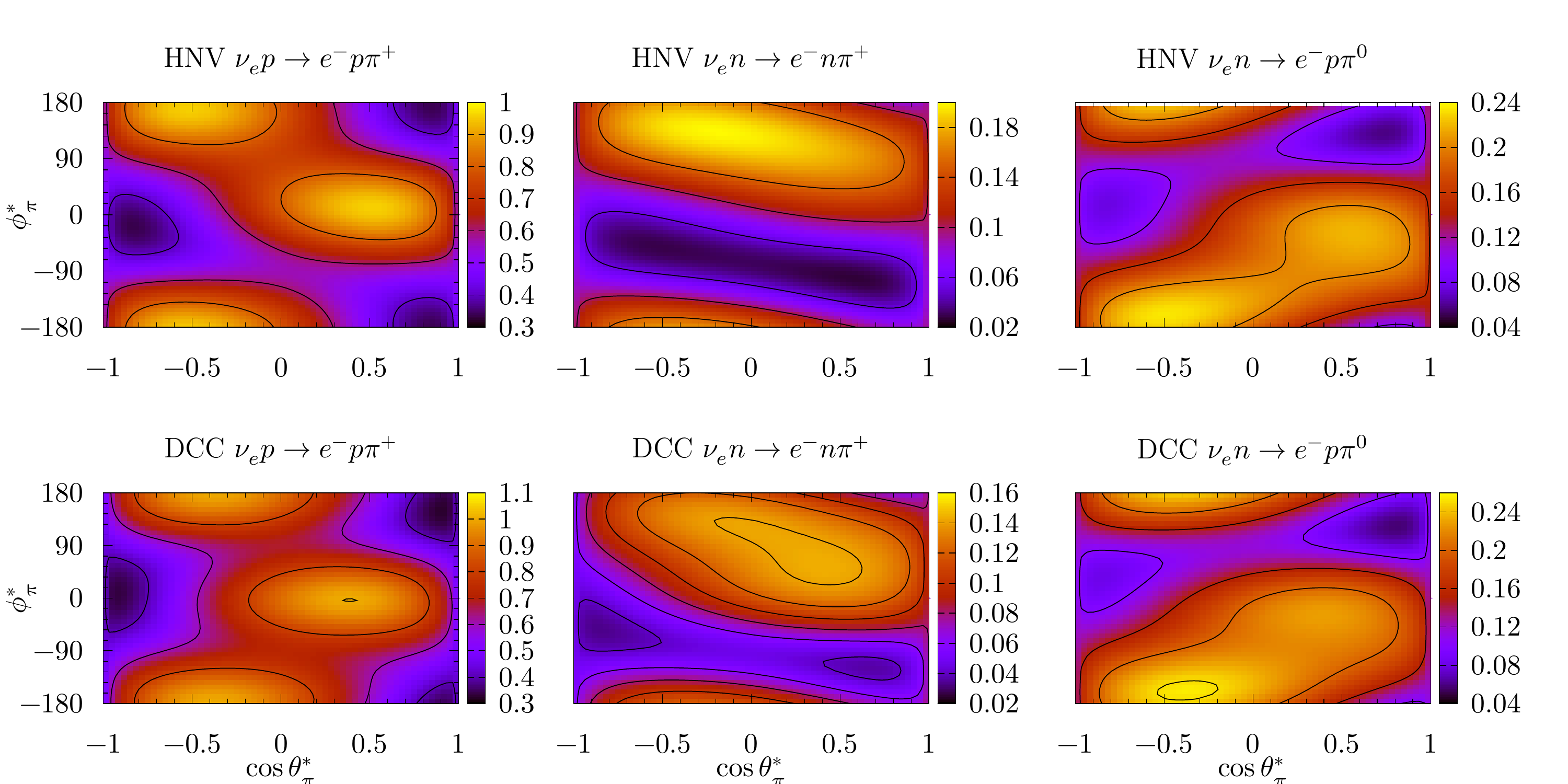}}
\caption{ Neutrino CC--$d\sigma/(dQ^2dW_{\pi N}d\Omega^*_\pi)$  
differential cross section in units of $10^{-38}\text{cm}^2c^2/\text{GeV}^3$, as a
function of $\phi_\pi^*$ and $\theta_\pi^*$, evaluated for fixed 
$E_\nu=1\,\text{GeV}$, $Q^2=0.1\,\text{GeV}^2/c^2$ and $W_{\pi N}=1.23\,
\text{GeV}$ values. }
  \label{fig:ang_CC}
\end{figure}

\begin{figure}[h!]
\centerline{\includegraphics[scale=0.6]{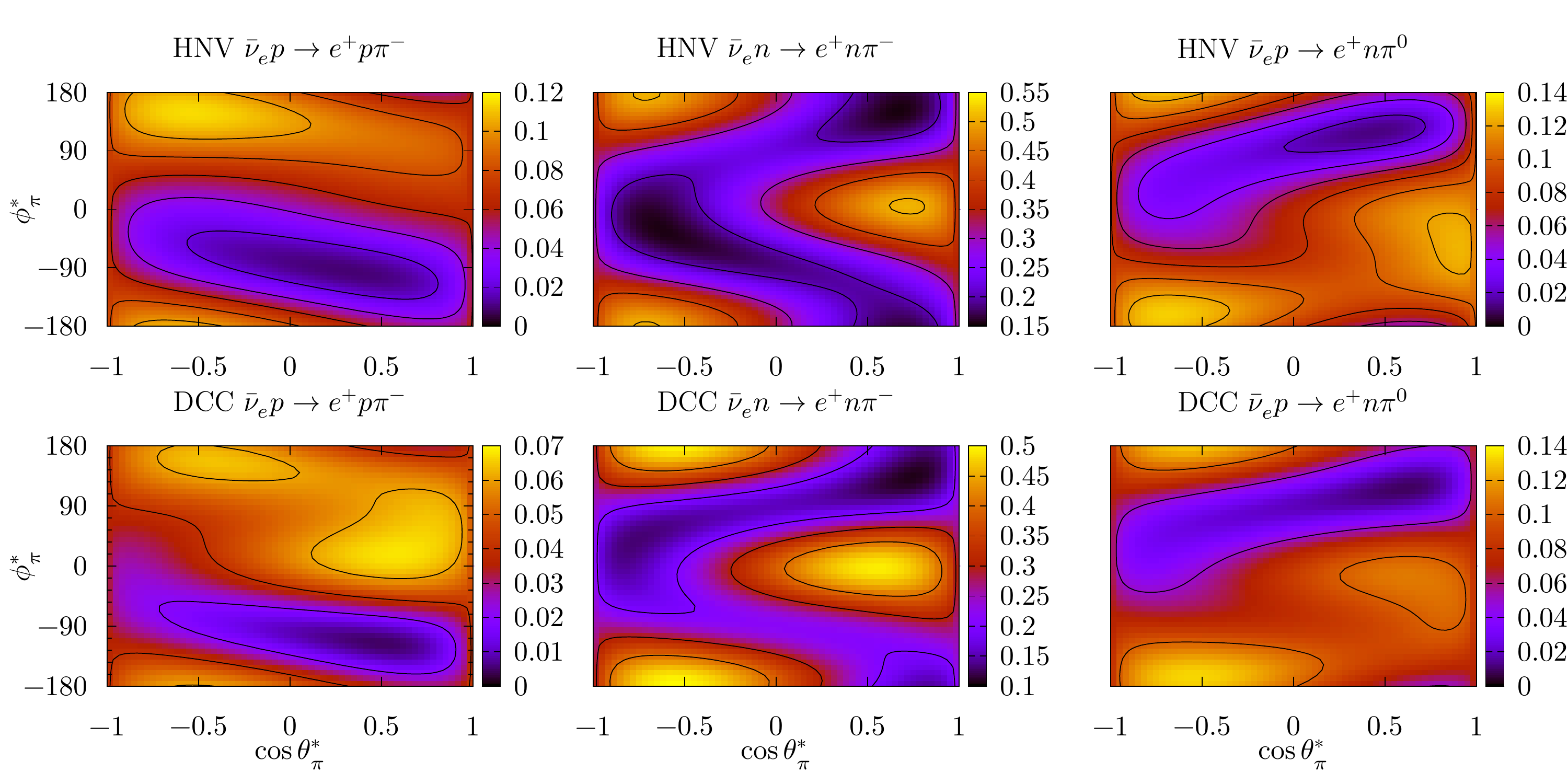}}
\caption{ The same as in Fig.~\ref{fig:ang_CC} for antineutrino CC--$d\sigma/(dQ^2dW_{\pi N}d\Omega^*_\pi)$.}
  \label{fig:ang_CC_anti}
\end{figure}

\begin{figure}[h!]
\centerline{\includegraphics[scale=0.495]{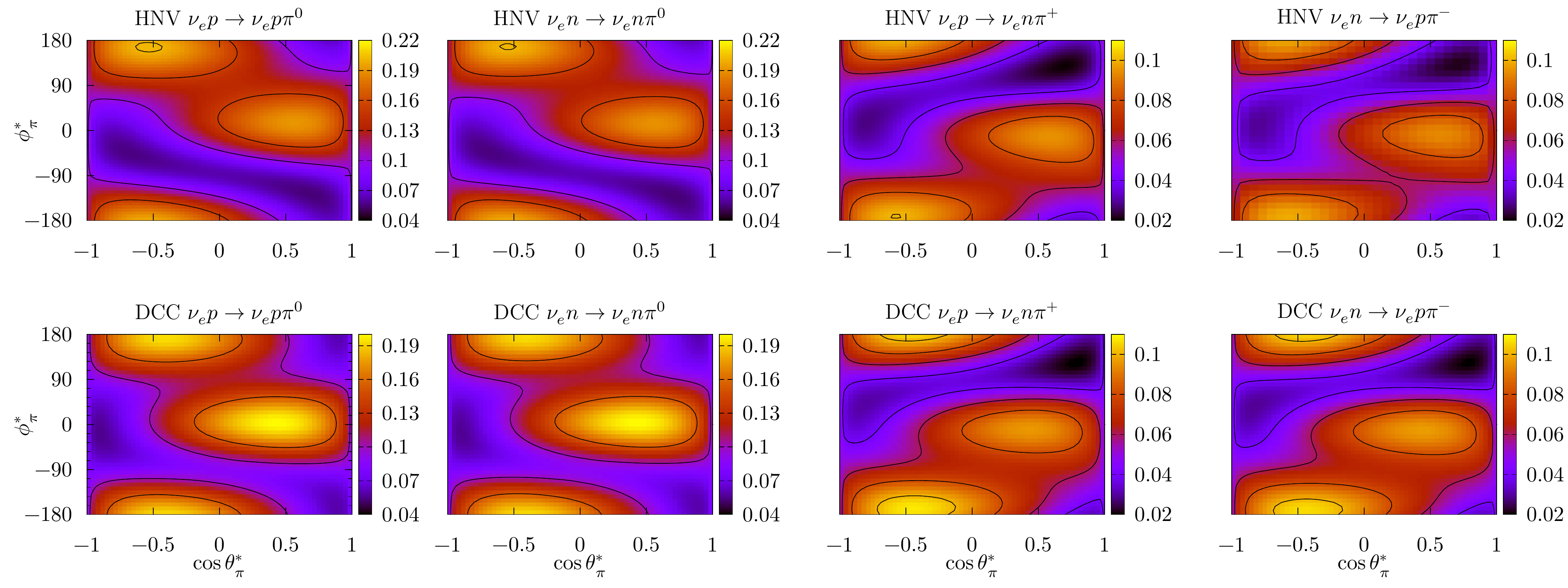}}
\caption{  The same as in Fig.~\ref{fig:ang_CC} for neutrino NC--$d\sigma/(dQ^2dW_{\pi N}d\Omega^*_\pi)$.}
  \label{fig:ang_NC}
\end{figure}

\begin{figure}[h!]
\centerline{\includegraphics[scale=0.495]{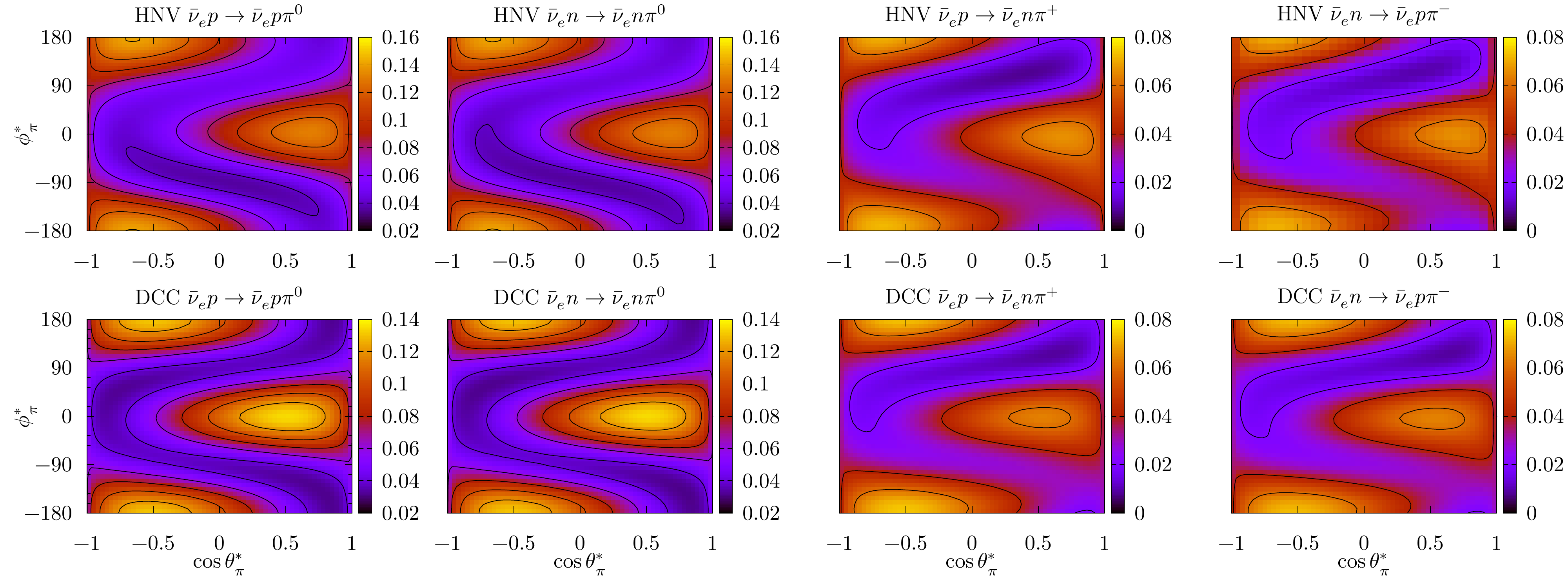}}
\caption{  The same as in Fig.~\ref{fig:ang_CC} for antineutrino NC--$d\sigma/(dQ^2dW_{\pi N}d\Omega^*_\pi)$,  }
  \label{fig:ang_NC_anti}
\end{figure}

 All of the above features are visible in the 
  neutrino and antineutrino CC $d\sigma/(dQ^2dW_{\pi N}d\Omega^*_\pi)$ 
 differential cross sections that  are displayed as contour 
 plots in  Figs.~\ref{fig:ang_CC} and \ref{fig:ang_CC_anti} for the DCC and 
 HNV models. 
 They are given  as a
function of $\phi_\pi^*$ and $\theta_\pi^*$, and have been evaluated for fixed 
$E_\nu=1\,\text{GeV}$, $Q^2=0.1\,\text{GeV}^2/c^2$ and $W_{\pi N}=1.23\,
\text{GeV}$ values.
Despite the differences, we find a good qualitative agreement between the 
two models that predict
similar regions where the pion angular distribution reaches its maximum and 
minimum. 
The same applies to the case of NC processes that are shown in
Figs.~\ref{fig:ang_NC} and \ref{fig:ang_NC_anti}. Note that the $n\pi^0$ 
and $p\pi^0$ NC reactions are driven by the same isovector amplitude, and 
they differ only in the sign of the interference of the latter   with the 
isoscalar part of the amplitude, which is also the same in both 
reactions~\cite{Hernandez:2007qq}. This is the reason why, as long as 
these processes are largely dominated by the isovector excitation of 
the $\Delta-$resonance, the cross sections are similar. The same occurs 
in the case of the $p\pi^-$ and $n\pi^+$ NC reactions. Let us note, 
in addition,  that the isoscalar contributions for these two latter 
processes are a factor of two bigger  than for the two previous 
NC reactions where neutral pions are  produced~\cite{Hernandez:2007qq}.

Since in Figs.~\ref{fig:ang_CC}--\ref{fig:ang_NC_anti} we take
$\phi^*_\pi$ in the interval $[-\pi,\pi]$, parity violation for $d\sigma/(dQ^2dW_{\pi N}d\Omega^*_\pi)$  is clearly seen
in most cases by the lack of reflection symmetry with respect to the
$\phi^*_\pi=0$ line.  It is  significant for CC $\nu_e$
 scattering off neutron ($\bar{\nu}_e$ off proton), where the direct $\Delta$ excitation term is not so dominant,  and for neutrino NC 
 reactions producing charged pions\footnote{Remember that background 
 isoscalar contributions in this case are twice as high as  for NC 
 production of neutral pions.}. 
  It means that for these channels, the $D^*$ and/or $E^*$ terms should have sizes 
  comparable in magnitude to those of the $A^*$, $B^*$ and $C^*$  parity-conserving structure functions.
Parity violation  is less prominent for the antineutrino NC processes
for which both models predict rather symmetric distributions. By looking at the
NC channels with a final charged pion, one sees a transition between a clear
asymmetry for neutrino reactions to a fairly symmetric distribution for the
antineutrino case.
Since the NC hadronic tensor is the same for neutrinos and antineutrinos, 
the different behavior seen in the figures originates from the change of sign of the 
antisymmetric part of
the leptonic tensor. From the general discussion in Sec.~\ref{sec:originpv}, 
there are two types of PV terms in $D^*$, that correspond to those induced by the antisymmetric 
${\cal \widetilde H}^{\mu\nu\, (a)}_{VV+AA}$ and the symmetric 
${\cal \widetilde H}^{\mu\nu\, (s)}_{VA+AV}$  nucleon tensors, discussed in
Eqs.~(\ref{eq:pv1}) and (\ref{eq:pv2}), respectively. When contracted with the leptonic tensor, these two contributions 
tend to cancel each other 
 for the NC antineutrino case, implying that both PV contributions must be  similar in magnitude for NC 
 processes\footnote{Note that in the antineutrino $A^*$ and $B^*$  PC terms, 
 there exist also some cancellations between symmetric and antisymmetric contributions, 
 which explain why they are smaller than those found for neutrinos. However, the point 
 is that these latter cancellations should be less effective than those produced in  $D^*$, 
 and this imbalance gives rise to smaller PV effects in antineutrino NC driven processes. 
 In addition, one might also have to consider possible modifications in the interference 
 pattern between the PV $D^*\sin\phi^*_\pi$ and $E^*\sin2\phi^*_\pi$ contributions. However,  
 in general $|E^*|$ is significantly smaller than $|D^*|$, though details depend on the 
 particular kinematics under study.}. A similar behavior is seen in the HNV model  for 
 NC channels with a final $\pi^0$. For this latter case,  the DCC model produces almost 
 perfect symmetric distributions for antineutrinos, and though some asymmetries can be 
 seen for neutrinos, they are not as pronounced as in the HNV case.

Another feature worth noticing, easily deduced from
Figs.~\ref{fig:ang_CC}-\ref{fig:ang_NC_anti},
is that  the $\phi^*_\pi$
dependence of the differential cross section is very different for 
$\cos\theta_\pi^*<0$ and $\cos\theta_\pi^*>0$.

\begin{figure}[h!]
\centerline{\includegraphics[scale=0.495]{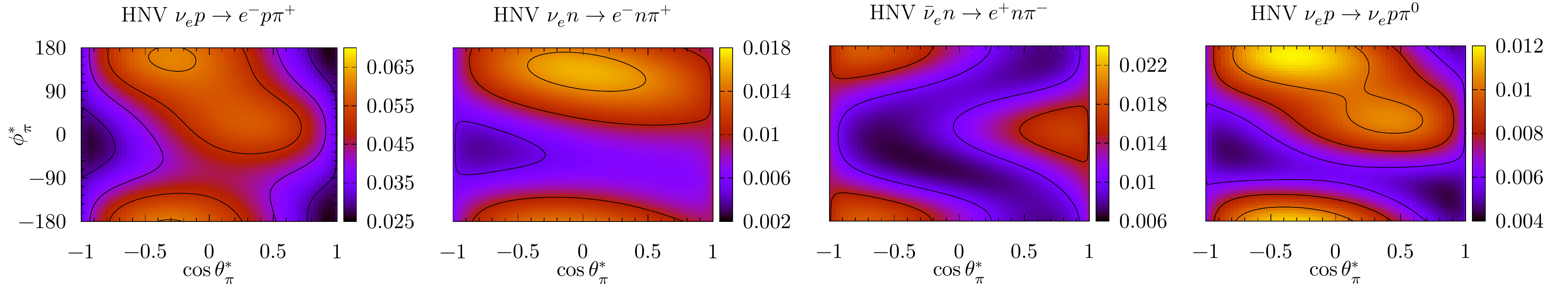}}
\centerline{\includegraphics[scale=0.495]{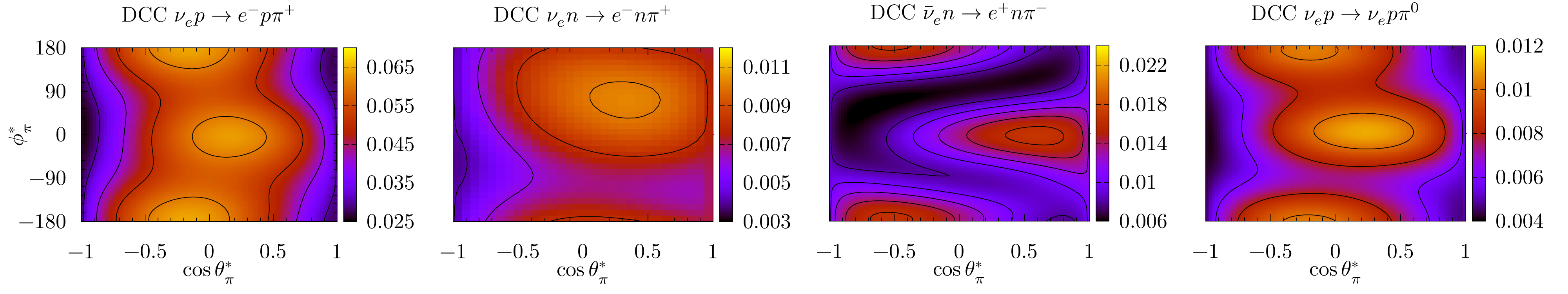}}
\caption{  $d\sigma/d\Omega^*_\pi$ 
differential cross section in units of $10^{-38}\text{cm}^2$, as a
function of $\phi_\pi^*$ and $\theta_\pi^*$, evaluated at
$E_\nu=1\,\text{GeV}$ and with a $W_{\pi N}<1.4$\,GeV cut.}
  \label{fig:ang_inte}
\end{figure}

In Fig.~\ref{fig:ang_inte} we show now the $d\sigma/d\Omega^*_\pi$ differential
cross
section  for
the $\nu_e p\to e^-p\pi^+, \nu_e n\to e^-n\pi^+, \bar \nu_e n\to e^+n\pi^-$ and 
$\nu_e p\to\nu_e p\pi^0$ channels evaluated at $E_{\nu}=1\, \text{GeV}$ and with a cut 
$W_{\pi N}<1.4$\,GeV. 
Parity violation is 
 seen in both models in the case of the $\nu_e n\to e^-n\pi^+$ reaction, 
 while for $\nu_e p\to e^-p\pi^+$ and  $\nu_e p\to\nu_e p\pi^0$ a PV 
 pattern is only
clearly appreciable in the HNV
model.  Both models predict very small PV effects in the case of the 
$\bar \nu_e n\to e^+n\pi^-$ reaction. The three latter processes are largely
 dominated by the excitation of the $\Delta$ and its subsequent $\pi N$
  decay, and thus finding small PV signatures is not surprising.
   Moreover, we see once more that PV effects get substantially reduced 
   in the antineutrino $\bar \nu_e n\to e^+n\pi^-$ reaction as compared to 
   those found in the isospin related neutrino $\nu_e p\to e^-p\pi^+$ process 
   (see discussion of Figs.~\ref{fig:ABCDE} and \ref{fig:ABCDE_anti}).

In any case, all distributions show a clear anisotropy. This
means that using  an isotropic distribution for the 
pions in the center
of mass of the final pion-nucleon system, as assumed in some Monte Carlo event generators, is not supported by the results
in   Fig.~\ref{fig:ang_inte}.  Moreover,  different 
channels have different angular distributions.

\begin{figure}[h!]
\includegraphics[scale=0.45]{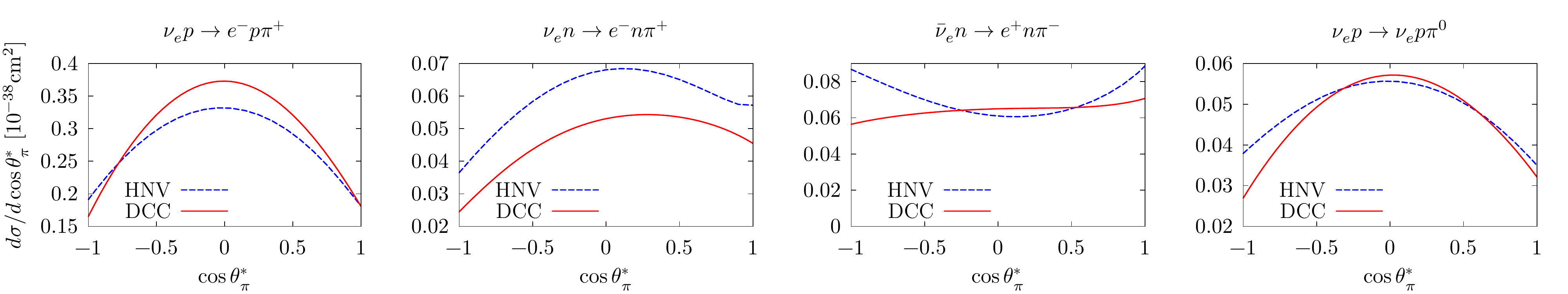}
\caption{ $d\sigma/d\cos\theta^*_\pi$ differential cross section in units of $10^{-38}\text{cm}^2$ for 
$E_\nu=1\,\text{GeV}$,  and with a $W_{\pi N}<1.4$\,GeV cut.}
  \label{fig:costeta_inte}
\end{figure}

\begin{figure}[h!]
\includegraphics[scale=0.45]{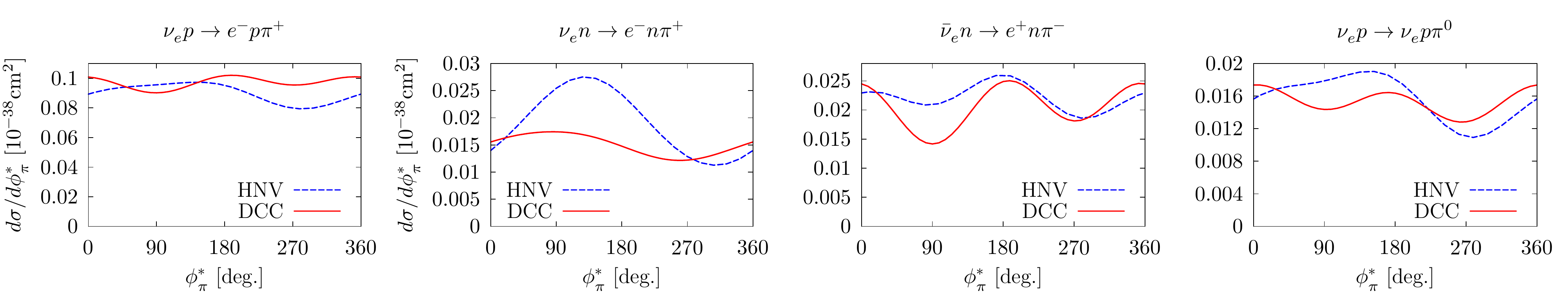}
\caption{ $d\sigma/d\phi^*_\pi$  differential cross section in units of $10^{-38}\text{cm}^2$ for 
$E_\nu=1\,\text{GeV}$,  and with a $W_{\pi N}<1.4$\,GeV cut.}
  \label{fig:phi_inte}
\end{figure}

In Figs.~\ref{fig:costeta_inte} and \ref{fig:phi_inte} we display the 
$d\sigma/d\cos\theta^*_\pi$ and $d\sigma/d\phi^*_\pi$ differential
cross sections, respectively,  for the same channels and incoming neutrino energy as the ones shown in  Fig.~\ref{fig:ang_inte},  and with
the same $W_{\pi N}<1.4$\,GeV cut applied. They are not flat and again  different channels show 
different behaviors. Looking at the $d\sigma/d\cos\theta^*_\pi$ differential
cross section one sees that the two models predict distributions similar in shape  and size  for the 
$\nu_e p\to e^- p\pi^+$ and $\nu_e p\to\nu_e p\pi^0$ channels. The 
discrepancies are more visible for  $\bar\nu_e n\to e^+ n\pi^-$. Note that 
isospin invariance guarantees that the hadron tensors of the $\nu_e p\to e^- 
p\pi^+$ and the $\bar\nu_e n\to e^+ n\pi^-$ processes should be identical, 
and therefore the  differences in the cross sections should only be produced
 by the change of sign of the interference between vector and axial 
 contributions. The largest differences between DCC and HNV 
 predictions are found, however, for the $\nu_e n\to e^- n\pi^+$ channel, as we have 
 already seen in Figs.~\ref{fig:CCtot}, \ref{fig:CCdWdQ2} and \ref{fig:ABCDE}. 
 They are mainly due to the  inclusion  in the 
 HNV model of a local term induced by the 
 $\Delta$ propagator modification discussed in Eq.~(\ref{eq:mod}). This term  
notably improves the description of the  $\nu_\mu n\to \mu^-n\pi^+$ total 
ANL cross section data~\cite{Hernandez:2016yfb} (see also 
Fig.~\ref{fig:CCtotnumun}  here).

As for  the $d\sigma/d\phi^*_\pi$ differential
cross section, first we observe that the distributions are not symmetric 
around $\phi^*_\pi=\pi$, implying certain violations of parity, which are 
quite significant for the $\nu_e n\to e^- n\pi^+$ reaction. Both, the HNV  
and the DCC models predict more pions to be produced above
the scattering plane, i. e. $\phi^*_\pi\in[0,\pi[$, for the $\nu_\mu n\to 
\mu^-n\pi^+$  and $\nu_e p\to\nu_e p\pi^0$ reactions.
The asymmetry 
for the  $\nu_e p\to e^- p\pi^+$ channel is predicted to be small in both 
models
but with a different sign. 
For $\bar \nu_e n\to e^+n\pi^-$, PV effects are larger in the 
DCC predictions than in the HNV ones, since in the former, the number of 
pions produced above the scattering plane is clearly smaller than that below 
that plane.

\begin{figure}[h!]
\includegraphics[scale=0.5]{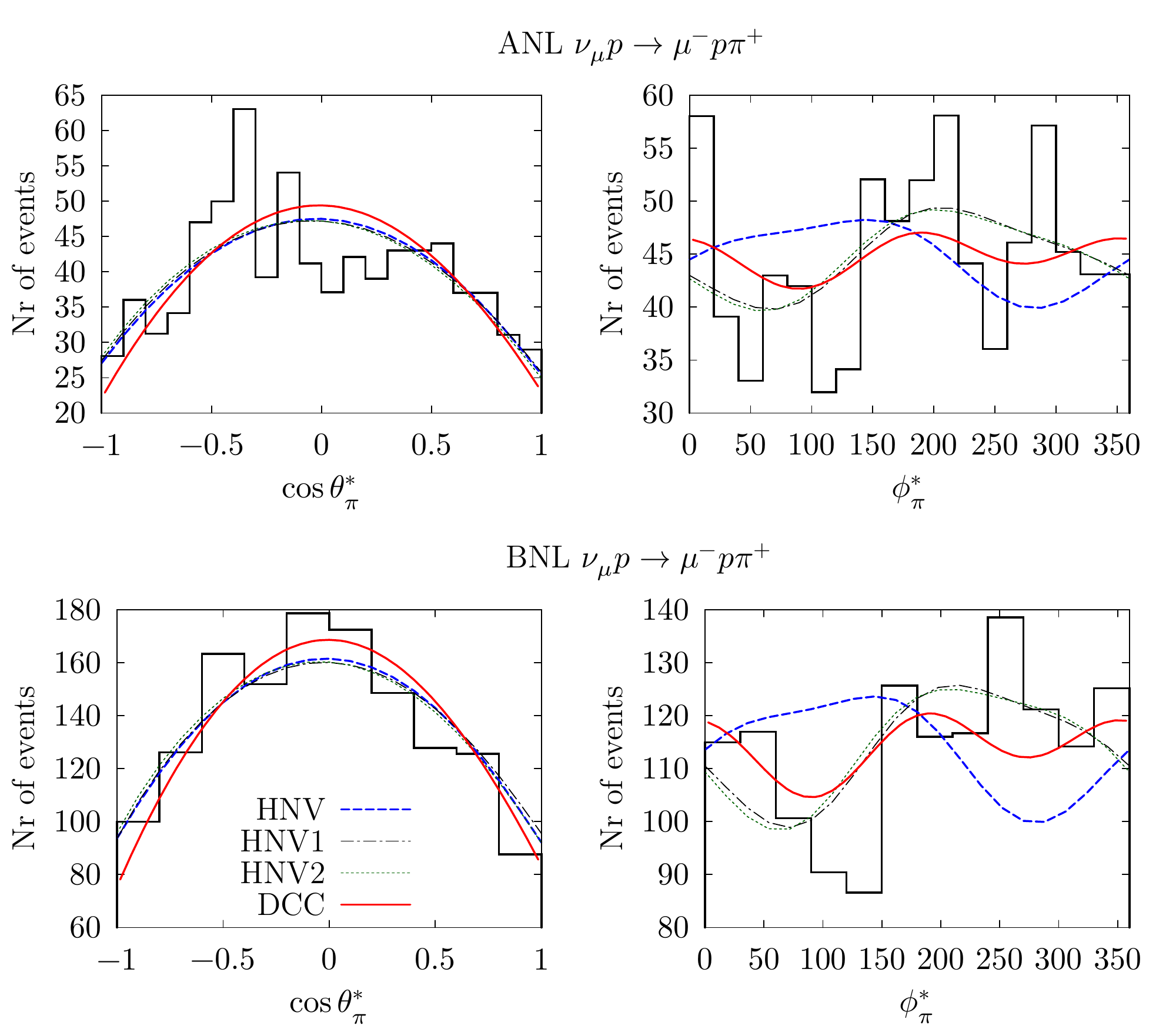}
\caption{ Shape comparison of the theoretical $d\sigma/d\cos\theta^*_\pi$ 
(left panels) and 
$d\sigma/d\phi^*_\pi$ (right panels) differential cross sections
with unnormalized ANL~\cite{PhysRevD.25.1161} and BNL~\cite{PhysRevD.34.2554} 
data.
A cut  $W_{\pi N}<1.4$\,GeV, in the
final pion-nucleon  invariant mass is imposed in both data and the theoretical
 results. Theoretical results  have been obtained averaging
over the neutrino flux for neutrino energies in the 
$[0.5,6]$\,GeV interval, setting their overall size to reproduce the areas 
under the experimental data. Predictions from two previous versions of the HNV model are also displayed: HNV1 stands for the HNV model  without  the $\Delta$ propagator modification of Eq.~(\ref{eq:mod}), while to compute the HNV2 results, the
implementation of Watson theorem has been further suppressed. }
  \label{fig:cosphi_fluxavg}
\end{figure}

Finally, in Fig.~\ref{fig:cosphi_fluxavg}, we make a shape-only comparison of 
our theoretical results for the  $d\sigma/d\cos\theta^*_\pi$ 
 and $d\sigma/d\phi^*_\pi$  differential cross sections for the 
$\nu_\mu p\to \mu^- p\pi^+$ reaction 
with unnormalized ANL~\cite{PhysRevD.25.1161} and BNL~\cite{PhysRevD.34.2554} old bubble chamber data.
Both in the data and in the theoretical calculations,  the  cut 
$W_{\pi N}<1.4$\,GeV in the final pion-nucleon  invariant mass is imposed, and the theoretical distributions  have been obtained averaging
over the neutrino flux for neutrino energies in the 
$[0.5,6]$\,GeV interval. The theoretical results have been
area-normalized to the data. Predictions from two previous versions of the HNV model are also shown to elucidate how the local terms discussed in Eq.~(\ref{eq:mod}) \cite{Hernandez:2016yfb} and  
the implementation of Watson theorem~\cite{Alvarez-Ruso:2015eva} affect this channel, dominated by the direct excitation of the $\Delta$ resonance.

All the models give similar
predictions for the flux averaged $d\sigma/d\cos\theta^*_\pi$
differential cross section, and show a good  agreement with BNL data. This means that the corrections for the HNV model proposed in Refs.~\cite{Hernandez:2016yfb} and  \cite{Alvarez-Ruso:2015eva} have 
little effect not only on the integrated, but also on the $\cos\theta^*_\pi$ differential cross section for the  $\nu_\mu p\to \mu^- p\pi^+$ reaction, that we remind again, it is largely dominated by the direct $\Delta$
excitation term. For the flux averaged  $d\sigma/d\phi^*_\pi$ differential cross section, the DCC model exhibits a global better agreement 
with data. As expected, the PV effects, both in the data and theoretical predictions, are small, being largely obscured  by the uncertainties
in the experimental distribution.  HNV models predict larger asymmetries, though still small in absolute value, around 10\% maximum. On the other hand,  the inclusion of  the local terms discussed in Eq.~(\ref{eq:mod}) \cite{Hernandez:2016yfb} increases the differences with the DCC results, and it also seems that the induced changes in the shape of the distribution do not receive data support.
%
%
%
\section{Study of pion electroproduction as a test of the vector part of the
DCC, SL and HNV models}
\label{sec:electropi}
\begin{figure}[h!]
\includegraphics[scale=0.6]{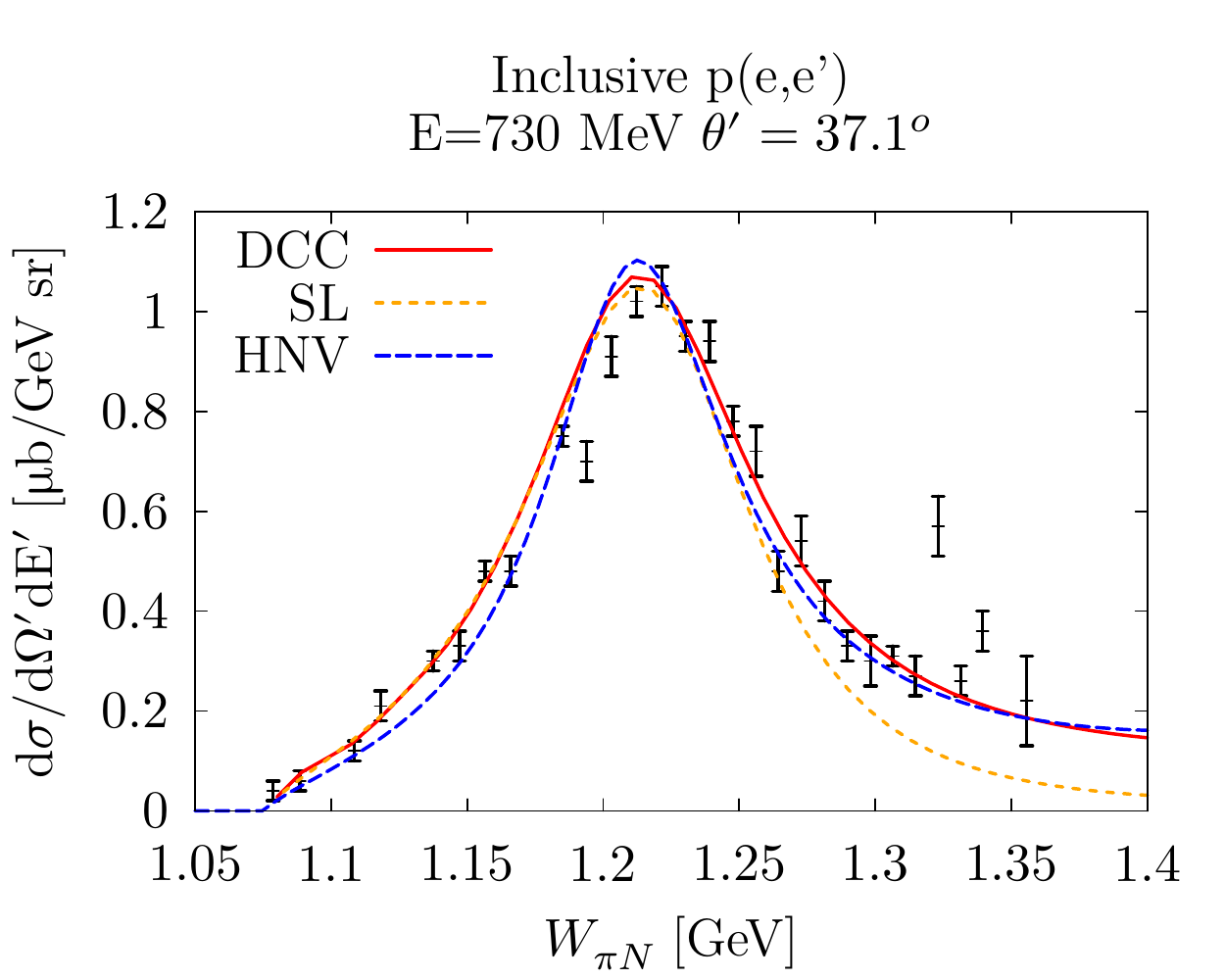}
\includegraphics[scale=0.6]{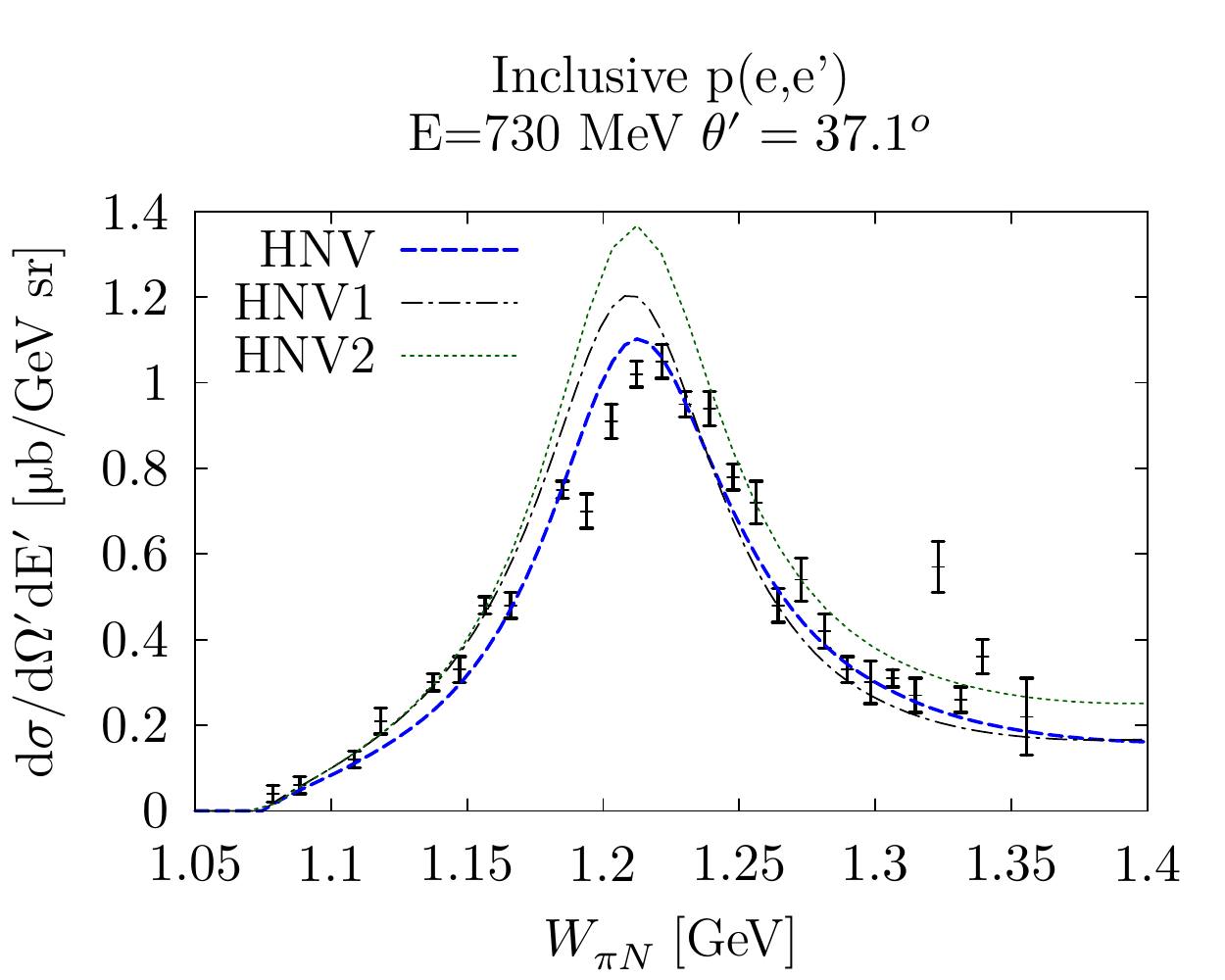}

\caption{ Inclusive $d\sigma/(d\Omega' dE')$ cross section off a proton 
(sum of the differential distributions for the $e^- p\to e^- p\pi^0$ and 
$e^- p\to e^- n\pi^+$ reactions), as a function of the invariant mass $W_{\pi N}$ and for fixed
$\theta'=37.1^{\rm o}$. 
The four-momentum transfer square $Q^2$ varies in the interval $[0.18,0.04]\,\text{GeV}^2/c^2$, when
$W_{\pi N} \in [1, 1.4]$ GeV. Data  taken from Ref.~\cite{OConnell:1984qim}. 
In the right panel, predictions from the HNV and two previous versions 
of that model are displayed: HNV1 stands for the HNV model  without  the 
$\Delta$ propagator modification of Eq.~(\ref{eq:mod}), while to compute 
the HNV2 results, the implementation of Watson theorem has been further 
suppressed.}
  \label{fig:electro_730}
\end{figure}
\begin{figure}[h!]
\includegraphics[scale=0.6]{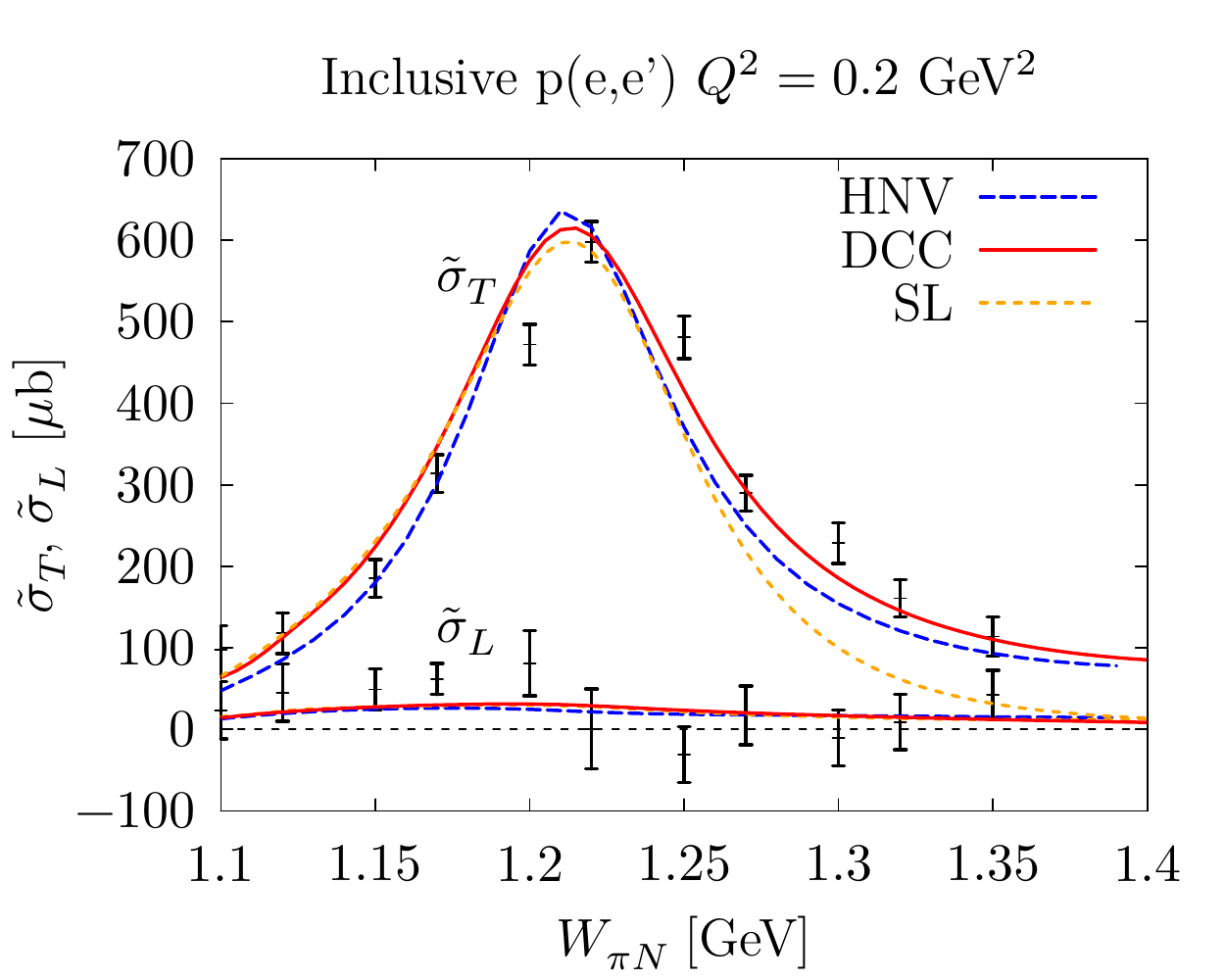}
\includegraphics[scale=0.6]{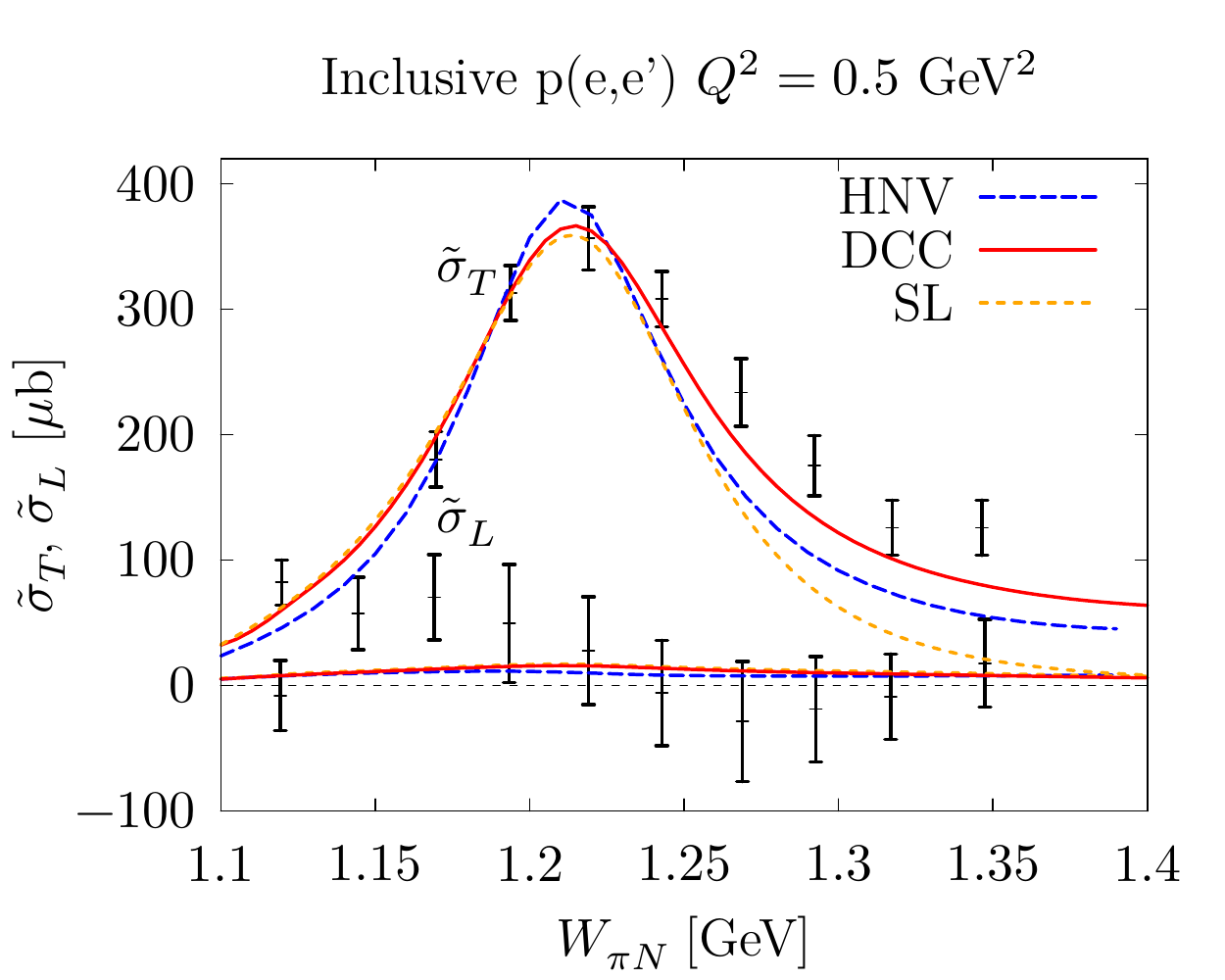}
\caption{Data and theoretical predictions for the 
$\tilde \sigma_T = \int \sigma_T d\Omega_\pi^*$ and $\tilde \sigma_L = 
\int \sigma_L d\Omega_\pi^*$ inclusive cross sections off protons 
($p\pi^0+n\pi^+$), as a function of the $\pi N$ invariant mass, and  
for two fixed values of $Q^2=0.2\ \rm GeV^2$ (left panel) and 
$Q^2=0.5 \rm GeV^2$ (right panel). Data taken from Ref.~\cite{Baetzner:1972bg}.}
  \label{fig:batzner}
\end{figure}
Pion electroproduction provides a testing ground for the vector part of the
pion production models. We do not aim here to perform an exhaustive comparison 
with the abundant  data that are available. In fact, such a test 
has already been done for the SL and DCC models \cite{Kamano:2013iva, Sato:1996gk,Sato:2000jf}. Here we just 
want to show  the observables which are described in a similar way by the HNV, SL and DCC models, as well as those that differ, trying to understand the origin of the discrepancies.  
This should help us  to better understand the differences observed in the weak 
pion production. 

In Sec.~\ref{sec:neutrino_comparison} we have compared the three models 
for CC and NC reactions induced by neutrinos in the vicinity of the $\Delta$ 
peak, and for a relatively 
low $Q^2$ value  in the region where the $d\sigma/dQ^2$ cross section is maximum. For a similar kinematical setup,  
we now show  results 
 for  pion electroproduction differential cross sections integrated over the 
 outgoing pion variables. 
 In Fig.~\ref{fig:electro_730}, we show results for the
 $d\sigma/(dE'd\Omega')$ differential cross section off protons evaluated for an
 incoming electron energy of $E=0.73\,$ GeV and for fixed $\theta'=37.1^{\rm o}$.
 The results are plotted as a function of $W_{\pi N}$ and we compare them with experimental data
  taken from Ref.~\cite{OConnell:1984qim}.
In its left panel  we see that the HNV and DCC models give very similar
predictions which, in turn, are in a good agreement with the data. The HNV 
model predicts less strength for low $W_{\pi N}$, something that
 has also been observed for the neutrino 
  induced reactions, see Figs.~\ref{fig:CCdWdQ2} and \ref{fig:NCdWdQ2}. 
 At the $\Delta$ resonance peak and below, the SL and DCC give very
similar results, since the $N\to\Delta$ vector  form factors were adjusted 
to reproduce the pion electroproduction data.  Above resonance, the SL model
gives smaller cross sections. 
In the case of neutrino cross sections, the differences seen between the SL and the DCC models
are, however, mainly due to the difference in   strength 
in the  axial current in those two models.
  In the right panel  we  show the predictions of the HNV model when 
  the modification of the
  $\Delta$ propagator in Eq.~(\ref{eq:mod}) is not taken into account (HNV1),
  and when we further suppress the implementation of Watson theorem (HNV2). One
  sees that the results significantly improve when  going from HNV2 to HNV1 and from HNV1 to the
  full HNV model, leading to an excellent description of the experimental distribution. This is particularly reassuring because, though the HNV model uses vector form-factors that have been in principle fitted to data, its latest refinement~\cite{Hernandez:2016yfb} (modification of the $\Delta$ propagator, motivated by the use of the so called consistent couplings~\cite{Pascalutsa:2000kd}) was derived only from  neutrino pion production data. Note that the final $p\pi^0$ and $n\pi^+$ states in the electron induced reactions are not purely isospin 3/2 states, and thus they receive sizable contributions from non-resonant mechanisms, in particular from the $\Delta$ crossed term which is corrected by the use of consistent couplings.

For electrons we have access to very precise experimental measurements of 
the
 pion angular distributions. It is common to write the differential cross 
 section as (see Eq.~(\ref{eq:finem1}))
\bea
&&\hspace*{-1cm}\frac{d\sigma_{em}}{d\Omega'dE'd\Omega^*_\pi}=\Gamma_{em}\Big\{
\sigma_{T}+\varepsilon\,
\sigma_{L}+
\sqrt{2\varepsilon(1+\varepsilon)}\
\sigma_{LT}
\,\cos\phi^*_\pi
+h\sqrt{2\varepsilon(1-\varepsilon)}\
\sigma_{LT'}
\,\sin\phi^*_\pi+\varepsilon\,
\sigma_{TT}
\,\cos2\phi^*_\pi
\Big\}
\label{eq:sigmas}
\eea
where the  different quantities have been introduced in 
Appendix~\ref{app:helicity2}. 
It is a
valid expression when both  electrons are ultrarelativistic and
the initial electron is polarized with well defined helicity $h$. As also
 mentioned
in Subsec.~\ref{sec:pv} and the Appendix~\ref{app:helicity2}, the presence 
of the $\sin\phi^*_\pi$ term does not imply parity
violation in this case, since the helicity also changes sign under parity. It 
is
 straightforward to see a direct correspondence of the terms $ \sigma_{T}+
 \varepsilon\,\sigma_{L}$, $\sigma_{LT}$, $\sigma_{TT}$ and $\sigma_{LT'}$ 
and the $A^*$, $B^*$, $C^*$ and $D^*$ structure functions introduced for 
 neutrinos 
in Eq.~(\ref{eq:abcde}).

After integrating over $\Omega_\pi^*$, only the $\sigma_T$ and $\sigma_L$ terms contribute to the $d\sigma_{em}/(d\Omega'dE')$ differential cross section. These partially integrated distributions
\bea 
\tilde \sigma_T = \int \sigma_T d\Omega_\pi^* \ , \ \ \tilde \sigma_L = \int \sigma_L d\Omega_\pi^*\nonumber
\eea
have been measured for various values of $Q^2$ and $W_{\pi N}$. 
In Fig.~\ref{fig:batzner}, we present the predictions for 
$\tilde \sigma_{T,L}$ obtained from the DCC, SL and HNV models and they 
are compared to the data of Ref.~\cite{Baetzner:1972bg}. Not much can be 
said about the accuracy of the predictions for $\tilde \sigma_L$  because 
of the large experimental uncertainties. For $\tilde \sigma_T$, which largely 
dominates over  $\tilde \sigma_L$, we find an acceptable description of the
 data, and we observe a similar behavior as in the case of $d\sigma_{em}/
 (d\Omega'dE')$ presented in Fig.~\ref{fig:electro_730}: the HNV predicts 
 less strength below the $\Delta$ peak, while the SL model underestimates the 
 experimental points above it.

\begin{figure}[h!]
\centerline{\includegraphics[scale=0.5]{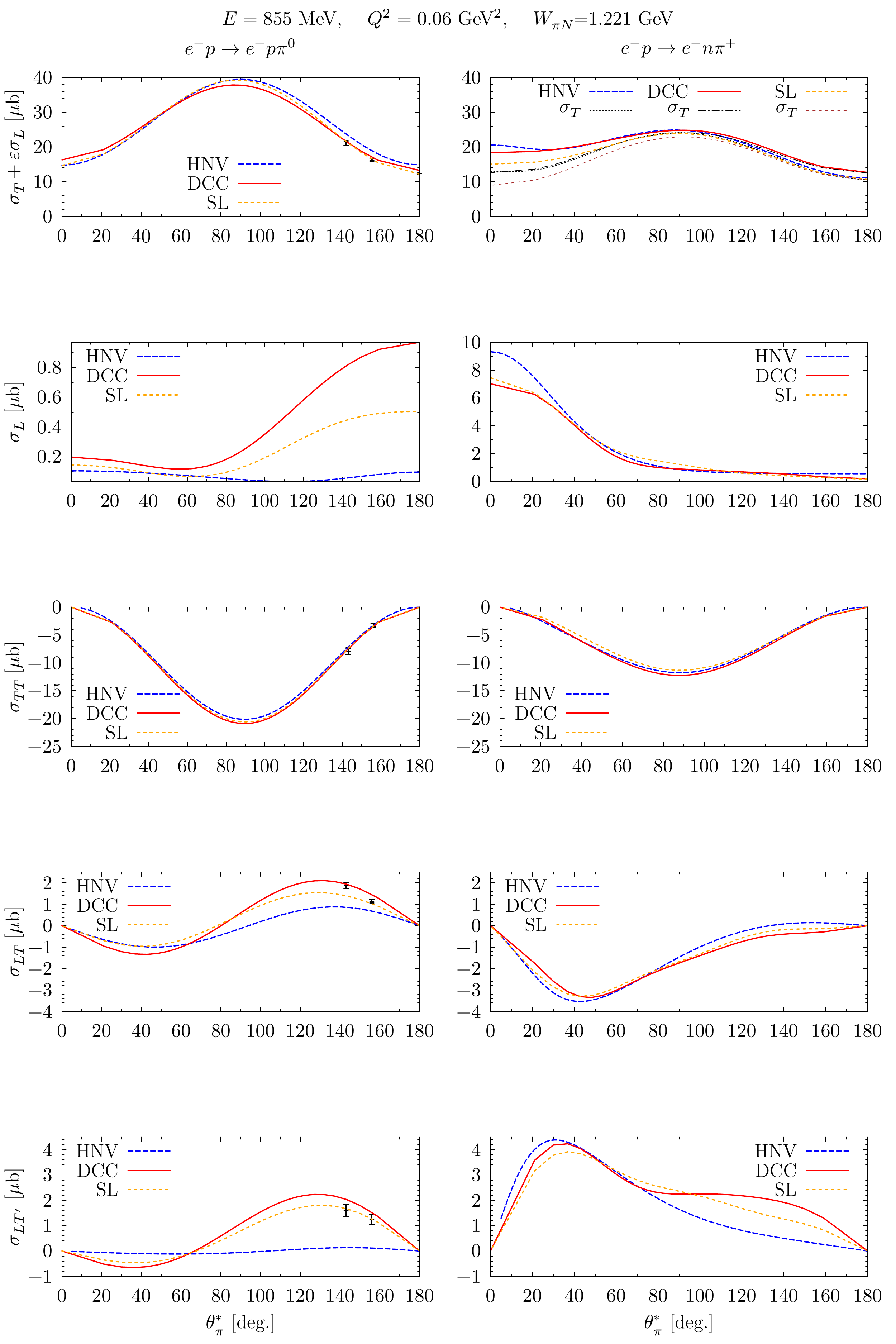}}
\caption{Comparison of the $ \sigma_{T}+\varepsilon\,
\sigma_{L}$, $\sigma_{L}$, $\sigma_{TT}$, $\sigma_{LT}$, $\sigma_{LT'}$ 
pion polar angular distributions obtained using the DCC, SL and HNV models for the  $e^-p\rightarrow e^- p\pi^0$ 
(left panels) and $e^-p\rightarrow e^- n\pi^+$ (right panels) channels. The kinematics correspond to $Q^2=0.06\ \text{GeV}^2/c^2$, $W_{\pi N}=1.221$ GeV and an incoming electron energy of 0.855 GeV. For $e^-p\rightarrow e^- p\pi^0$, the  $\sigma_L$ contribution is negligible so that $ \sigma_{T}+\varepsilon\,
\sigma_{L} \approx \sigma_T$,  while for $e^-p\rightarrow e^- n\pi^+$ 
we also show  $\sigma_T$ in the first panel. Data from Ref.~\cite{Stave:2006ea} are available only for the $p\pi^0$ channel. }
  \label{fig:MAMI}
\end{figure}
\begin{figure}[h!]
\centerline{\includegraphics[scale=0.5]{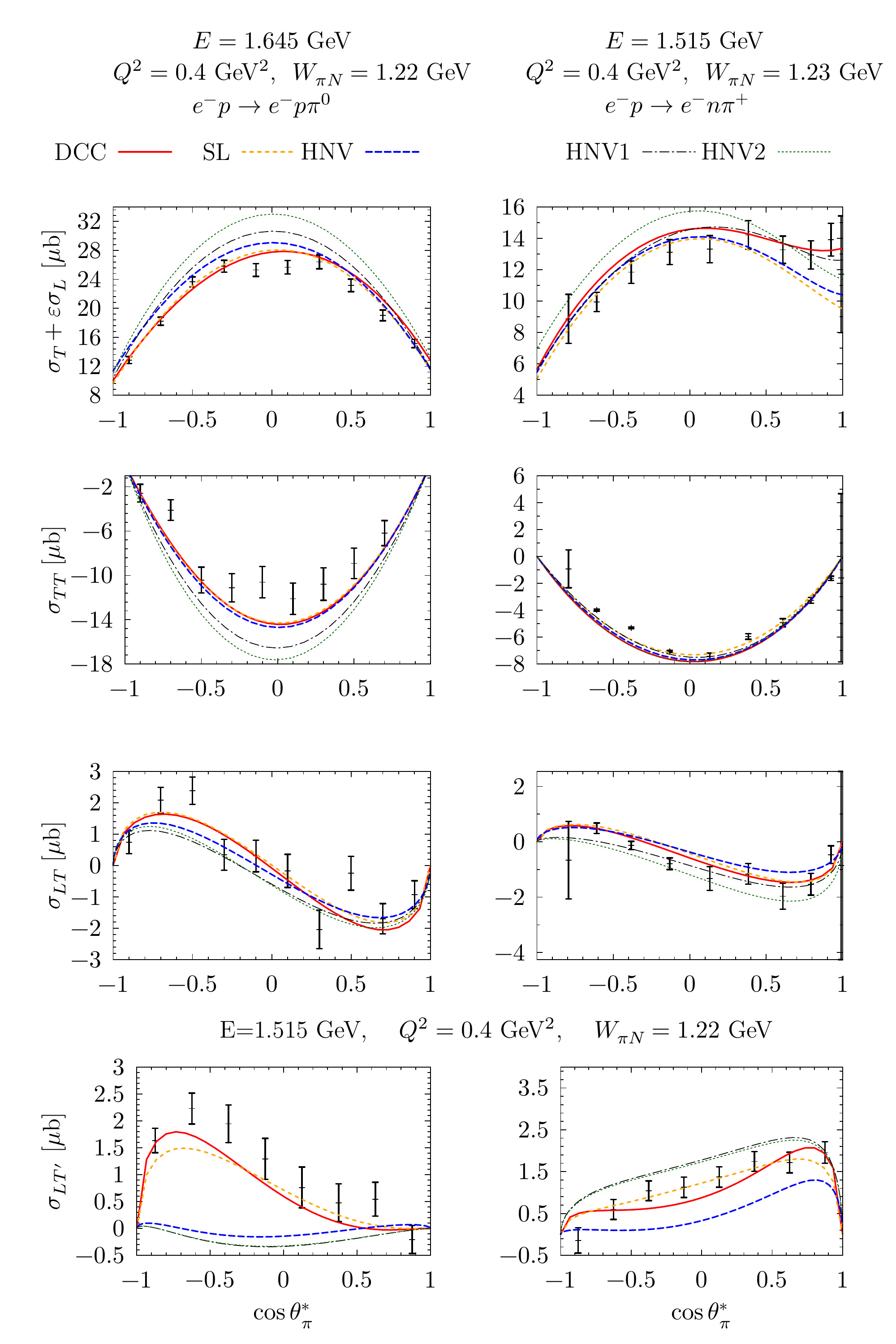}}
\caption{ Comparison of the  $ \sigma_{T}+\varepsilon\,
\sigma_{L}$, $\sigma_{TT}$ and  $\sigma_{LT}$ pion polar angular distributions 
obtained using the DCC, SL and HNV models for $e^-p\rightarrow e^- p\pi^0$ 
(left panels) and $e^-p\rightarrow e^- n\pi^+$ (right panels) at 
$Q^2=0.4\,\text{GeV}^2/c^2$ and $W_{\pi N}=1.22$ GeV or $1.23$ GeV, respectively. 
 Data from Refs.~\cite{Joo:2001tw} and ~\cite{Egiyan:2006ks} for 
 $e^-p\rightarrow e^- p\pi^0$ and $e^-p\rightarrow e^- n\pi^+$, respectively, 
 are displayed as well.
In the two bottom panels, we also show the $p\pi^0$ and  $n\pi^+$ measurements 
of Refs.~\cite{Joo:2003uc} (left) and ~\cite{Joo:2005gs} (right)  at 
$Q^2=0.4\,\text{GeV}^2/c^2$ and $W_{\pi N}=1.22$ GeV, together with the 
theoretical predictions, of the 
$\sigma_{LT'}$ distribution.  Finally in all panels,  the HNV1 curves stand for the 
results obtained within the HNV model, when  the 
propagator
modification of Eq.~(\ref{eq:mod}) is not considered, while to obtain 
the HNV2 predictions, the implementation of Watson theorem is further 
suppressed. }
  \label{fig:CLAS}
\end{figure}
\begin{figure}[h!]
\centerline{\includegraphics[scale=0.5]{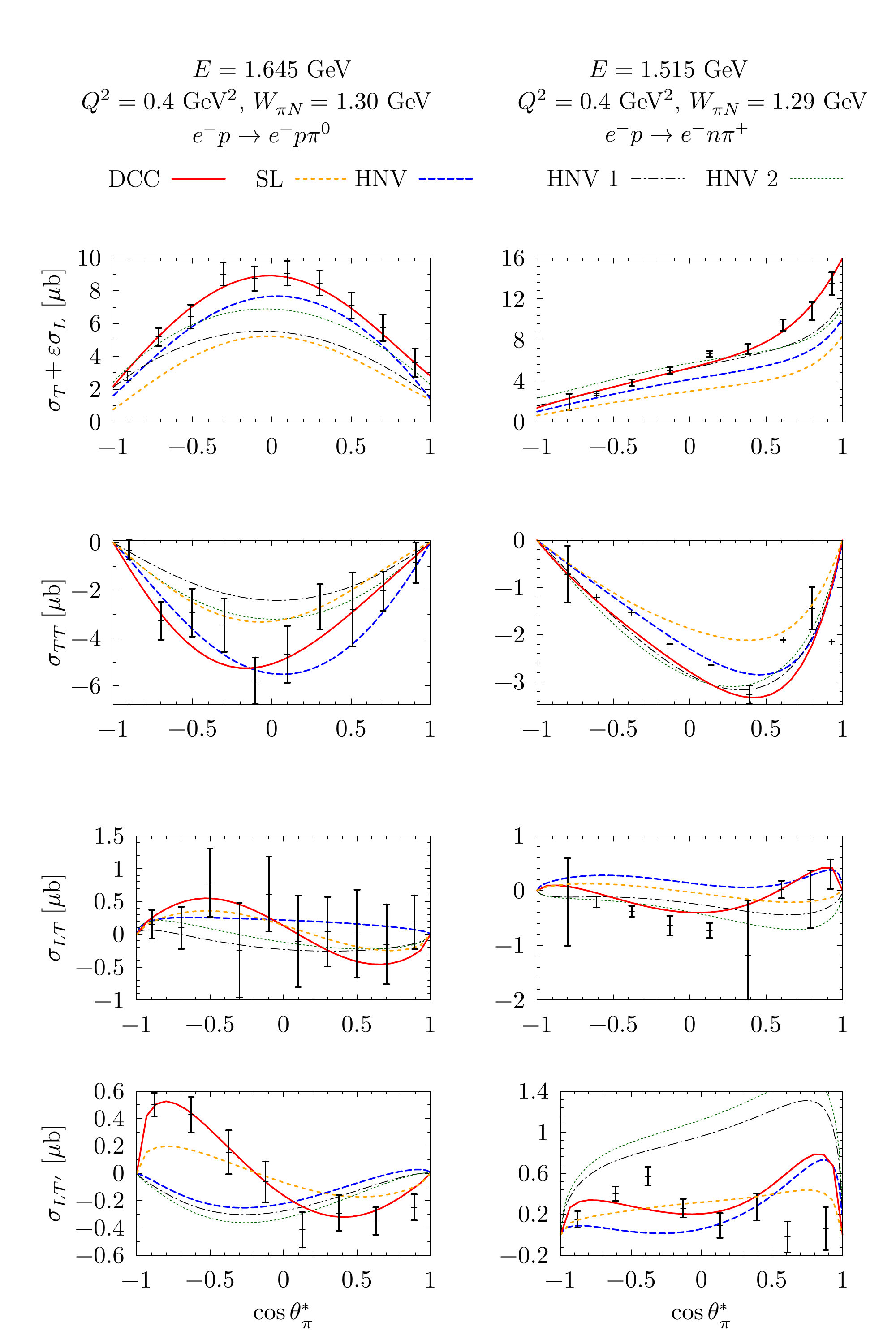}}
\caption{ Same as Fig.~\ref{fig:CLAS} ($Q^2=0.4\,\text{GeV}^2/c^2$), but for higher $\pi N$ invariant masses, $W_{\pi N}=1.30$ GeV and  $1.29$ GeV for $e^-p\rightarrow e^- p\pi^0$ and 
$e^-p\rightarrow e^- n\pi^+$, respectively.}
\label{fig:CLAS-13}
\end{figure}

In the following, we shall further compare the theoretical pion angular distributions
 for the $e^-  p\rightarrow e^-p\pi^0$ and 
$e^-  p\rightarrow e^-n\pi^+$ channels, for $W_{\pi N}$ invariant masses in 
the vicinity of the $\Delta$ peak and  for  two  $Q^2$ values for which  precise 
data are available. In Fig. \ref{fig:MAMI}, we show  results for  $W_{\pi N}=1.221$
 GeV and a very low 
$Q^2=0.06\ \text{GeV}^2/c^2$ value and compare them to data taken from 
Ref.~\cite{Stave:2006ea}. The latter  
correspond 
 to the lowest  $Q^2$ measurement of these observables that has been performed so 
 far. They cover a small $\theta_\pi^*$ range, above $140^\circ$, and  only for the 
 $e^-  p\rightarrow e^-p\pi^0$ channel. We show  
 results from the three models, for both $p\pi^0$ and $n\pi^+$ final states, 
 and the full $\theta_\pi^*$ range.
For the  $e^-  p\rightarrow e^-n\pi^+$ channel 
(right panels in Fig.~\ref{fig:MAMI}) all models 
 give very similar results for all the structure functions. For 
 $e^-  p\rightarrow e^-p\pi^0$ (left panels in Fig.~\ref{fig:MAMI}), the 
 theoretical predictions 
  differ for the transverse-longitudinal interference 
 terms, $\sigma_{LT'}$ and $\sigma_{LT}$, and also for the
 longitudinal $\sigma_L$ differential cross section. These contributions are much smaller than $\sigma_T$ ($\le 5\%$),  in particular $\sigma_L$,  so that all models would predict similar $d\sigma/dQ^2dW_{\pi N}$
 cross sections.  As it has been discussed at 
 the end of Sec.~\ref{sec:originpv}, 
 in the case of the HNV model, $\sigma_{LT'}$ (or correspondingly the $D^*$ 
 function for neutrinos) appears as a consequence 
 of  interference between the $\Delta P$ term and the background contributions 
 (which 
 have  different phases mainly because of the nonzero imaginary part of the 
 $\Delta$
 propagator). Background terms  in the $e^-  p\rightarrow e^-p\pi^0$ channel are 
 small within the HNV model 
 (isospin symmetry  forbids the CT and the PF contributions), and thus that  
 interference is necessarily small. 
The 
situation is entirely different for the $e^- p\rightarrow e^-n\pi^+$ channel, for which the
background contribution is sizable. This is the reason why 
 the three models give very similar predictions in this case.
  By looking at Eq.~(\ref{eq:stletc}), one realizes that $\sigma_{L}$, 
 $\sigma_{LT}$ and  $\sigma_{LT'}$ depend on the third component of the hadronic electromagnetic
 current. The above discussion tells us that for the 
 $e^-  p\rightarrow e^-p\pi^0$ channel, this component may not  be correct  
 within the HNV model. 
 
 In a multipole language, the main features of the $\sigma_{LT'}$ angular distribution in the $\Delta$
 region can be understood using $s$ and $p$ wave pion production multipoles, as
 done for instance in Ref.~\cite{Joo:2005gs}
 \bea
 \sigma_{LT'}\approx \frac{|\vec k^*_\pi|_0}{k_\gamma}
 \sqrt{\frac{Q^2}{|\vec q^*|^2}}\sin\theta^*_\pi({\cal A}+{\cal
 B}\cos\theta_\pi^*),
 \eea
with
\bea
{\cal A}={\rm Im}[S^*_{0+}(-M_{1+}+3E_{1+}+M_{1-})+(S_{1-}-2S_{1+})^*E_{0+}]\ \
,\ \ {\cal B}=6{\rm Im}[S^*_{1-}E_{1+}+S^*_{1+}(E_{1+}-M_{1+}+M_{1-})],
\eea
and $k_\gamma$ defined in Appendix~\ref{app:helicity}.
${\cal A}$ comes from the interference between  $s$ and $p$ wave multipoles while 
${\cal B}$ is generated from the interference among $p$ wave multipoles. Since the
direct Delta contributes only to $P_{33}$ multipoles, all with the same phase,
$\sigma_{LT'}$ is very sensitive to background contributions. In the DCC and SL
models, the main contributions to ${\cal A}$ and  ${\cal B}$ are
respectively $-{\rm Im}[S^*_{0+}M_{1+}]$ and $-{\rm Im}[S^*_{1+}M_{1+}]$. The
latter can only come from isospin 3/2 and isospin 1/2 interference and it changes
sign when going from $\pi^0$ to $\pi^+$ production.
This change of sign of ${\cal B}$ explains the difference in shape for 
$\sigma_{LT'}$ seen when going from $\pi^0$ to $\pi^+$ production. It is also 
clear that the relative phases between the multipoles have to be well under
control to get $\sigma_{LT'}$ right. This is achieved in the DCC and SL 
models below the two-pion production threshold.

 Next,  we show $e^-  p\rightarrow e^-p\pi^0$ and $e^-  p\rightarrow e^-n\pi^+$ 
 results evaluated at a  higher 
$Q^2=0.4\ \text{GeV}^2/c^2$ value, and for $\pi N$ invariant masses located at the 
$\Delta$ peak (Fig.~\ref{fig:CLAS}) or slightly above (Fig.~\ref{fig:CLAS-13}). The 
three models give similar
results in good agreement with data, with the exceptions of the SL $\sigma_{T}+\varepsilon\sigma_L$ and $\sigma_{TT}$ distributions above the $\Delta$, and the  HNV $\sigma_{LT'}$ structure function, particularly  for the $e^-  p\rightarrow e^-p\pi^0$ channel,  for which the
background contribution within the HNV model is small. We also show in these 
two figures results obtained when we eliminate from the HNV model
 the $\Delta$ propagator modification of Eq.~(\ref{eq:mod}), and when we 
 further
  suppress the partial unitarization of the amplitudes, implemented 
 by imposing  Watson theorem for the multipoles dominated by the 
 $\Delta$ resonance. In the $p\pi^0$ case, one 
 sees a clear improvement  in the $\sigma_T+\varepsilon\sigma_L$ and $\sigma_{TT}$ observables when going
 from HNV2 to HNV1 and from the latter to the full HNV calculation. 
 For $\sigma_{LT}$ the quality of the data does not allow us to be very conclusive, while
  the three versions of the HNV model fail to reproduce the data of the small $\sigma_{LT'}$. 
 For the $n\pi^+$ reaction, though in general the modifications proposed in Refs.~\cite{Alvarez-Ruso:2015eva, Hernandez:2016yfb} improve the global agreement with data, the effects   are not as pronounced as those found in the $p\pi^0$ case.

The conclusion to be drawn from this comparison of electromagnetic results is
that one needs a full unitarization procedure,  like the one implemented in the complex DCC model 
in order to get a good reproduction of all scattering observables. Its effect 
seems to be crucial to explain the $\sigma_{LT}$  and $\sigma_{LT'}$ data  for the  
$e^-  p\rightarrow e^-p\pi^0$ reaction, where
background contributions are small. In the 
$e^-  p\rightarrow e^-n\pi^+$ channel, the non-resonant contributions are much more
 important, and the simple HNV model predictions  agree reasonably well with those 
 obtained within the sophisticated DCC approach.
All that notwithstanding,  it is important to stress  that  the $\sigma_{L}$, 
$\sigma_{LT}$  and $\sigma_{LT'}$ structure functions, where the HNV model shows larger discrepancies with the DCC results,  are much smaller in magnitude than $\sigma_T$ and 
$\sigma_{TT}$ and thus their effects are  not so relevant when looking at pion angular
distributions. Besides, if one integrates on the outgoing pion $\phi_\pi^*$ variable,  the contributions from  $\sigma_{LT}$ 
and $\sigma_{LT'}$ (and $\sigma_{TT}$) cancel exactly and the resulting 
differential cross section is governed by $\sigma_T+\varepsilon\sigma_L$ for
which, both the HNV and DCC models give similar predictions.
%
%
%
%
\section{Summary and conclusions}
\label{sec:conclusions}

We have carried out a careful analysis of the pion angular dependence of the CC and NC neutrino and antineutrino  pion production reaction
 off nucleons. We have shown  that 
the possible dependencies on the azimuthal angle measured in the final pion-nucleon CM system are $1,\cos\phi^*_\pi,\cos 2\phi^*_\pi, \sin\phi^*_\pi$ and $\sin 2\phi^*_\pi$, and that the two latter ones  give rise to parity
violation and time-reversal odd correlations  in the weak $d\sigma/(d\Omega'dE'd\Omega^*_\pi)$ 
and $d\sigma/(dQ^2 dW_{\pi N}d\Omega^*_\pi)$ differential cross sections. These findings were 
already derived in 
Refs.~\cite{Sato:2003rq,Hernandez:2007qq, Hernandez:2006yg}, but here we have made a detailed discussion of 
the origin of the PV  contributions. Hence, we have seen that these are generated 
from the interference between different contributions to the hadronic current that are not 
relatively real. When the hadronic current is further expanded in multipoles, one  sees that the
only PV  contributions that survive are the ones associated to the interference between
multipoles corresponding to different quantum numbers.
In particular, we have shown that the $\sin 2\phi^*_\pi$ term comes from
symmetric contributions to the hadronic tensor 
 generated from
 vector-axial interference (${\cal \widetilde H}^{\mu\nu\, (s)}_{VA+AV}$). Thus, as expected,  
the $\sin 2\phi^*_\pi$ structure function will be absent in the case of photo- or 
electro-production. On the
other hand,  the  $\sin\phi^*_\pi$ dependence
in the differential cross section gets contributions from two different PV tensors. The first 
one, as in the $\sin2\phi^*_\pi$ case, comes from  the  symmetric ${\cal \widetilde H}^{\mu\nu\, (s)}_
{VA+AV}$ tensor,
 while the second one comes from the   antisymmetric 
 ${\cal \widetilde H}^{\mu\nu\, (a)}_{VV+AA}$ 
 tensor  
generated from  vector-vector and axial-axial 
interferences. The pion electroproduction polarized differential cross section contains 
a $\sin\phi^*_\pi$ structure function, $\sigma_{LT'}$, coming only from the vector-vector 
interference.

As a test of the
 vector content of the DCC, SL and HNV models, we have compared their predictions  for pion electroproduction in 
the $\Delta$ region, and we have also confronted these predictions with data. The DCC scheme provides an excellent description of the 
 existing measurements for $ \sigma_{T}+\varepsilon\,
\sigma_{L}$, $\sigma_{TT}$, $\sigma_{LT}$ and $\sigma_{LT'}$ pion polar angular distributions 
and also for $(Q^2,W_{\pi N})$ differential cross sections, obtained after integrating over the 
angles of the outgoing pion. 
Despite its simplicity, the HNV model works also quite well and it leads to a fair description
 of the data and a good reproduction of the DCC predictions, except for 
$\sigma_{LT'}$ in the  $e^-  p\rightarrow e^-\pi^0p$ reaction
where   the background contribution
is small. 

Within the DCC model, the hadronic rescattering processes are taken into account by solving 
coupled channel equations for the $\Delta(1232)$ and higher resonances. In this approach, a 
unified treatment of all
resonance production processes satisfying unitarity is provided, and the predictions extracted 
from the DCC model have been extensively and successfully 
compared to data on $\pi N$ and $\gamma N$ reactions, up to  invariant masses slightly above  
2 GeV. The meson-baryon 
channels included in the calculations are $\pi N$, $\eta N$, $K\Lambda$, $K\Sigma$ and $\pi\pi N$ 
through $\pi N$, $\rho N$ and $\sigma N$ resonant components, and the analysis includes 20 partial 
waves, up to the $H_{19}$ and $H_{39}$ (isospin 1/2 and 3/2, orbital angular momentum $L=5$ and
 total angular momentum $J=9/2$)~\cite{Kamano:2013iva}. The model includes a few tens of bare 
 strangeness-less baryon resonances, whose properties (bare masses and couplings to the 
 different channels and form-factors) need to be fitted to data.  The meson-exchange interactions 
 between different meson-baryon pairs, as well as the ultraviolet cutoffs, needed to make the 
 unitarized couple-channels amplitudes finite, should be phenomenologically determined, as well. 
 There is a total of few hundred parameters that were fitted in  \cite{Kamano:2013iva} to a
  large sample 
($\ge 22300$ data points)  of $\pi N \to \pi N$ and  $\pi^\pm p, \gamma p \to \pi N, \eta N,
 K\Lambda, K\Sigma$ measurements.
Given the high degree of complexity of the DCC approach, it is really remarkable
 that the bulk of its predictions for electroproduction of pions in the $\Delta$ 
 region could be reproduced, 
with a reasonable accuracy, by the simpler HNV model. The latter has the advantage 
that it might be more easily implemented in the Monte Carlo event 
generators used for neutrino oscillation analyses. Electron data also support the latest improvements of the HNV model (approximate 
 unitarization of the amplitudes~\cite{Alvarez-Ruso:2015eva}, implemented by 
 imposing the Watson theorem for the multipoles dominated by the  $\Delta$ 
 resonance, and the 
modification of the $\Delta$ propagator~\cite{Hernandez:2016yfb}, motivated by 
the use of the so called consistent couplings) that 
lead to an accurate reproduction of the bubble chamber ANL and BNL neutrino data, 
including the $\nu_\mu n \to \mu^- n\pi^+$ channel, using amplitudes 
fully consistent with PCAC. 

We have presented an exhaustive comparison of the DCC, SL and HNV model predictions 
for CC and NC neutrino and antineutrino  pion production integrated and differential
 cross sections. DCC and HNV totally integrated and  $d\sigma/dQ^2 dW_{\pi N}$
  differential cross sections  agree reasonably well, except for the channels, 
  like $\nu_e n \to e^- n \pi^+$, where the crossed $\Delta$ mechanism is favored 
  by spin-isospin factors  with respect to the  direct excitation of the $\Delta$
   resonance. This is because the modification of the $\Delta$ propagator,  
   implemented in the HNV model, greatly cancels the crossed $\Delta$ mechanism, 
   leading to larger cross section values than the ones obtained in the DCC model. This
    enhancement 
   allows for a better description of the ANL $\nu_\mu n \to \mu^- n \pi^+$ total 
   cross sections.   In most of the cases, the SL model predictions are smaller,
   the main reason for that being that the SL model uses a smaller $N \to\Delta$
   axial  coupling extracted from a constituent quark model.
   It should also be kept in mind that the old bubble chamber data were
   obtained from neutrino-deuteron reactions and that the effects of the final state
   interaction studied in Ref.~\cite{SXN2018} may modify the current cross 
   section data at the
   nucleon level extracted from deuteron data.
   
With respect to the pion angular dependence of the weak cross sections, we 
have observed, first of  all, that  CC and  NC  distributions show  clear anisotropies. 
This means that using  an isotropic distribution for the pions in the CM
 of the final pion-nucleon system, as assumed by some of the Monte Carlo event 
 generators, is not supported by the results of the DCC and HNV models.  
 In addition, we have seen that different channels show different angular 
 distributions. We want to stress once more the importance of carrying out an 
 exhaustive test of 
the different models at the level of outgoing pion angular distributions, going 
beyond comparisons done for partially integrated cross sections, where 
model differences cancel to a certain extent  (see for instance 
$d\sigma/dQ^2 dW_{\pi N}$ and $A^*$ for $\bar\nu_e n \to e^+ n \pi^-$, depicted 
in Figs.~\ref{fig:CCdWdQ2} and \ref{fig:ABCDE_anti}  respectively).

The $d\sigma/d\phi^*_\pi$ differential cross section is not symmetric around
 $\phi^*_\pi=\pi$, implying certain violations of parity, that are dominated 
by the $\sin\phi^*_\pi$ term. PV effects are quite significant for neutrino NC 
 reactions producing charged pions, but even more for the $\nu_e n\to e^- n\pi^+$ 
 and  $\bar \nu_e p\to e^+ p\pi^-$ CC processes. Both, the HNV  and the DCC models 
 predict more pions to be produced above the scattering plane. However, parity 
 violation  effects are less prominent for the antineutrino NC reactions, implying 
  some cancellations between the PV effects induced by the 
  ${\cal \widetilde H}^{\mu\nu\, (s)}_{VA+AV}$ (vector-axial
   interference) and
  ${\cal \widetilde H}^{\mu\nu\, (a)}_{VV+AA}$ (vector-vector and axial-axial 
  interference)  tensors. These cancellations  are not produced in the case of neutrinos, 
   because  the contribution of the latter tensor to the cross sections changes sign.

 Going into finer details, the terms proportional to $1,\cos\phi^*_\pi,\cos 2\phi^*_\pi, 
 \sin\phi^*_\pi$ and $\sin 2\phi^*_\pi$ for the DCC and HNV models show some moderate 
 differences in size and even in shape, for instance for the $\sin\phi^*_\pi$ 
 structure function in the $\nu_e p\to e^- p\pi^+$  reaction. In this latter case, 
 the reason is the same as the one commented above for the $\sigma_{LT'}$ 
 differential cross section in the  $e^-  p\rightarrow e^-\pi^0p$ reaction. This 
 channel is largely dominated by the direct $\Delta$ mechanism, and thus PV 
 effects are notably smaller than in other channels for which the interferences 
  between resonant and non-resonant amplitudes are larger. 
 In the channels where the non-resonant background contributions are sizable, for 
 instance $\nu_e n\to e^- p\pi^0$ or $\nu_e n\to e^- n\pi^+$
%
both DCC and HNV models predict qualitatively similar results. The same occurs in  the
  case of the $\sin 2\phi^*_\pi$ structure function, suggesting that the PV effects 
  encoded in the vector-axial interference are similar in both models. 
 
Given the safety restrictions in current and future experiments,  
 presumably, we will be bound to extract the pion angular dependence from nuclear 
 cross sections, rather than from reactions with nucleon targets. In that 
 case,  the particles produced
in the primary interaction should travel across the high-density nuclear medium 
which alters the particle composition of the event.
Experimentally, the picture is confused even further by the typically broad neutrino
 energy spectrum and by beam flux
uncertainties. The viability of measuring the pion  angular distribution associated 
with the production off nucleons from  neutrino interactions with nuclei was
 analyzed in Ref.~\cite{Sanchez:2015yvw}. The results based on the NEUT Monte 
 Carlo~\cite{Hayato:2009zz} showed that this angular distribution can be determined, 
 with certain accuracy, because the
information is reasonably well maintained despite the FSI  and the need to 
reconstruct the energy of the incoming
neutrino from the experimental data. Nevertheless, further studies are needed 
to reliably estimate the distortion induced in the angular distributions by the FSI. 
 
Since pion production becomes one of the main reaction mechanisms for neutrinos with
energies of a few GeV, the theoretical knowledge of the nuclear cross sections is an 
 important and necessary ingredient to reduce systematic errors affecting present 
 and future neutrino
oscillation experiments. The first requirement for putting neutrino
 induced pion production on nuclear targets on a firm ground   
 is, however, to have a realistic model at the 
 nucleon level. 
 This work, where we have presented a 
 detailed comparison of three, state of the art, microscopic models for electroweak
  pion production off nucleons is, in our understanding, a first step forward in 
  that direction. Moreover, we are firmly convinced that the physics content of 
  the Monte Carlo event generators used in the analysis of neutrino oscillation experiments should 
  necessarily be confronted with the predictions of the three models discussed in 
  this work. A last remark we want to make is the following. Even the realistic models 
  described in this work rely on old data obtained in
  deuterium, so that any improvement   requires to have pion 
 production experiments by neutrinos carried out at the nucleon level. We  strongly support
 any  experimental effort on that line.

%
%
%
%
%
\begin{acknowledgments}
 This research
 has been supported by the Funda\c{c}\~ao de Amparo \`a Pesquisa do Estado de S\~ao Paulo
(FAPESP), Process No.~2016/15618-8, by the JSPS KAKENHI No. 25105010 and 
16K05354, by the Spanish Ministerio de Econom\'\i a y
 Competitividad and European FEDER funds under  Contracts 
  No. FIS2014-51948-C2-1-P,  No. FPA2016-77177-C2-2-P, FIS2017-84038-C2-1-P
 and  No. SEV-2014-0398, and by Junta de
 Castilla y Le\'on under Contract No. SA041U16.
 
 \end{acknowledgments}

\appendix
\section{Lorentz transformation to the CM mass of the final pion-nucleon
system}
\label{app:lt}
The Lorentz transformation will be constructed as the product of a rotation and a boost to the
CM system of the final pion-nucleon.
\beas
\Lambda=BR
\eeas
The
rotation matrix is chosen in a way that, when seen as a passive rotation, it takes the $Z$ axis over
$\vec q$ and the $Y$ axis over $\vec k\wedge \vec k\,'$. It can be written as
\begin{eqnarray}
 R^{\mu}_{\ \nu}=\left(\begin{array}{cccc}1&0&0&0\\0&\cos\theta&0&\sin\theta\\
0&0&1&0\\
0&-\sin\theta&0&\cos\theta\end{array}\right)\times
\left(\begin{array}{cccc} 1&0&0&0\\0&-\cos\phi&-\sin\phi&0\\
0&\sin\phi&-\cos\phi&0\\
0&0&0&1\end{array}\right),
\label{eq:rot1}
\end{eqnarray}
where $\theta,\phi$ are the $\vec q=\vec k-\vec k'$  polar angles in the
original fixed reference frame (the LAB frame that we chose to be oriented such that 
$\vec k=(0,0,|\vec k\,|)$\ ) and they are given by
\begin{eqnarray}
\cos\theta=\frac{|\vec k|-|\vec k'|\cos\theta'}{|\vec q\,|}, \quad \sin\theta=\frac{|\vec k'|}{|\vec q\,|}\sin\theta' ,\quad  |\vec q\,|=\sqrt{|\vec k|^2+|\vec k'|^2-2|\vec k||\vec k'|\cos\theta'}, \quad \phi=\phi'+\pi.
\end{eqnarray}
with $\theta',\phi'$ the final lepton polar and azimuthal angles measured in the same fixed
reference frame. The rotated vector components are given by\footnote{We illustrate here the
general case. For a CC reaction $E=|\vec k\,|$, while for a NC reaction one will further have 
$E'=|\vec k\,'|$.}
\begin{eqnarray}
&& (Rq)^\mu=\left(q^0, 0,0, |\vec q\,|\right) ,\ 
(Rk)^\mu=\left(E, |\vec k|\sin\theta,0, |\vec k|\cos\theta\right)
 ,\ 
(Rk')^{\mu}=\left( E',|\vec k|\sin\theta,0, 
|\vec k|\cos\theta-|\vec q\,|\right) ,\nonumber\\  && (Rp)^\mu=p^\mu=(M,0,0,0),\ 
(Rk)^\mu_{\pi }=(E_\pi,R^1_{\ j} k^j_\pi,R^2_{\ j} k^j_\pi,R^3_{\ j} k^j_\pi).
\end{eqnarray}

Now the boost to the CM mass is given by
\bea
B=\left(\begin{array}{cccc}\gamma&0&0&-\gamma v\\
0&1&0&0\\
0&0&1&0\\
-\gamma v&0&0&\gamma \end{array}\right),\ \ v=\frac{|\vec q\,|}{q^0+M},\ \ \gamma=\frac1{\sqrt{1-v^2}}=
\frac{q^0+M}{W_{\pi N}}
\label{eq:boost}
\eea
and the fully transformed four-vectors are 
\bea
&&q^{*\mu}=(\Lambda q)^\mu=\left(\gamma(q^0-v|\vec q\,|),\,0\,,0, 
\gamma(-vq^0+|\vec q\,|)\right)\ ,\ \
k^{*\mu}=(\Lambda k)^\mu=\left( \gamma(E-v|\vec k|\cos\theta),|\vec k|\sin\theta,\,0,\,
\gamma(-vE+|\vec k|\cos\theta)\right),\nonumber\\
&&k^{\prime *\mu}=(\Lambda k^{\prime})^\mu=\left(\gamma[E' -v(|\vec k|\cos\theta-|\vec q\,|)],
 |\vec k|\sin\theta,\,0,\, 
\gamma[-vE' +(|\vec k|\cos\theta-|\vec q\,|)]\right)\  ,\ \ 
p^{*\mu}=(\Lambda p)^\mu=\left(\gamma M,\, 0,\,0,\, 
-\gamma v M\right),\nonumber\\
&&(\Lambda k_\pi)^{\mu}=(BRk_\pi)^\mu=B^\mu_{\ \alpha}R^\alpha_{\ \beta}k_\pi^\beta.
\eea
\normalsize
These are the  four-vectors as seen in 
a reference frame $X^*Y^*Z^*$ that moves along with the CM  system of the final pion-nucleon
 and that is oriented such that $Z^{*+}\equiv \vec q$, $Y^{*+}\equiv \vec k\wedge\vec k'$ and
$X^{*+}\equiv (\vec k\wedge\vec k')\wedge\vec q$. 

There are a few things that have to be noticed. First, none of the $q^*,k^*,k^{\prime *}$ and $p^*$
four-vectors depend on $\phi'$. Second, their second spatial  components  are all zero. And third, 
for $q^*$ and $p^*$
 also the first spatial components are zero.

\section{Dependence of the hadron tensor on the pion azimuthal angle}
\label{sec:app-varios}
Performing the rotations of Eq.~(\ref{eq:rotation}), one finds 
\begin{align}
&
\begin{aligned}
W^{00}=&\widetilde W^{00},& W^{03}&=\widetilde W^{03}, & W^{30}&=\widetilde W^{30},&W^{33}&=\widetilde W^{33},\nonumber\\
\end{aligned}\\
&
\begin{aligned}
W^{01}&=\cos\phi^*_\pi\,\widetilde W^{01}-\sin\phi^*_\pi\, \widetilde W^{02}, 
&W^{10}&=\cos\phi^*_\pi\,\widetilde W^{10} -\sin\phi^*_\pi\,\widetilde W^{20},\nonumber\\
W^{02}&=\sin\phi^*_\pi\,\widetilde W^{01}+\cos\phi^*_\pi\,\widetilde  W^{02} ,  
&W^{20}&=\sin\phi^*_\pi\,\widetilde W^{10}+ \cos\phi^*_\pi\,\widetilde W^{20},\nonumber\\
W^{13}&=\cos\phi^*_\pi\,\widetilde W^{13}- \sin\phi^*_\pi \widetilde W^{23}, 
&W^{31}&=\cos\phi^*_\pi\,\widetilde W^{31}- \sin\phi^*_\pi \widetilde W^{32},\nonumber\\
W^{23}&=\sin\phi^*_\pi\,\widetilde W^{13}+ \cos\phi^*_\pi \widetilde W^{23}, 
 &W^{32}&=\sin\phi^*_\pi\,\widetilde W^{31}+ \cos\phi^*_\pi \widetilde W^{32},\nonumber\\
\end{aligned}\\
&
\begin{aligned}
W^{11}&=\frac12\Big((\widetilde W^{11}+\widetilde W^{22})
-\sin 2\phi^*_\pi\ (\widetilde W^{12}+ \widetilde W^{21})
+\cos2\phi^*_\pi\ (\widetilde W^{11}-\widetilde W^{22})\Big),
&&\\
W^{22}&=\frac12\Big((\widetilde W^{11}+
\widetilde W^{22})+\sin 2\phi^*_\pi\ (\widetilde W^{12}+ \widetilde W^{21})-\cos2\phi^*_\pi\ (\widetilde W^{11}-\widetilde W^{22})\Big),&&\\
W^{12}&=\frac12\Big((\widetilde W^{12}-\widetilde W^{21})+\sin 2\phi^*_\pi\ (\widetilde W^{11}- 
\widetilde W^{22})+\cos2\phi^*_\pi\ (\widetilde W^{12}+\widetilde W^{21})\Big),&&\\
W^{21}&=\frac12\Big(-(\widetilde W^{12}-\widetilde W^{21})+\sin 2\phi^*_\pi\ (\widetilde W^{11}- 
\widetilde W^{22})+\cos2\phi^*_\pi\ (\widetilde W^{12}+\widetilde W^{21})\Big).&&
\end{aligned}
\label{eq:th}
\end{align}
%
\section{Differential cross section as a sum over virtual $W$ cross sections
}
\label{app:helicity}
In the case of pion
electroproduction, and in the zero lepton mass limit, it is customary to write the differential cross section 
in terms of the differential cross sections,  $d\sigma(\gamma^* N\to N'\pi)/d\Omega^*_\pi \big|_{\phi_\pi^*=0}$, for virtual photons 
of different polarization.
 A similar thing can be done for the weak process, and the differential cross
 sections can be written in terms of $d\sigma(W^* N\to N'\pi)/d\Omega^*_\pi\big|_{\phi_\pi^*=0}$
differential cross sections for pion production by a virtual $W$ boson (virtual $Z$ in the case of NC
processes) of different polarization.  For that purpose let us rewrite the 
$L^{\mu\nu}( k^*, k^{\prime *})
\,W_{\mu\nu}( q^*, p^*,  k^*_\pi)$ product in terms of the helicity
components of the lepton and hadron tensors
\bea
L^{\mu\nu}(k^*, k^{\prime *})\,W_{\mu\nu}(q^*, p^*,  k^*_\pi)=
g_{rr}g_{ss}\,\epsilon^*_{r\mu}L^{\mu\nu}(k^*, k^{\prime *})
\epsilon_{s\nu}\,
\epsilon_{r\alpha}W^{\alpha\beta}(q^*, p^*,  k^*_\pi)
\epsilon^*_{s\beta}=g_{rr}g_{ss}{\cal L}_{rs}{\cal W}_{rs}
\eea
where, for $r=t,+1,-1,L$,  we have introduced the, orthogonal to $q^*$, polarization vectors ($Q^2=-q^2$)
\bea
\epsilon^\mu_t=\frac{1}{\sqrt{Q^2}}(q ^{*0},0,0,|\vec{q}\,^*|),\ \ \ \epsilon^\mu_{\pm1}=\mp\frac1{\sqrt2}(0,1,\pm i,0),
\ \ \ \ \epsilon^\mu_{L}=\frac{1}{\sqrt{Q^2}}(|\vec{q}\,^*|,0,0, q ^{*0}),
\eea
the quantities
\bea
g_{tt}=g_{+1+1}=g_{-1-1}=-1,\ g_{LL}=1,
\eea
and we have used the identity $g_{rr}\,\epsilon^*_{r\mu}\epsilon_{r\nu}=g_{\mu\nu}$. The helicity components of the lepton and hadron tensors are defined as
\bea
{\cal L}_{rs}=\epsilon^*_{r\mu}L^{\mu\nu} (k^*, k^{\prime *})
\epsilon_{s\nu},\ \ \ \ \ {\cal W}_{rs}=\epsilon_{r\alpha}W^{\alpha\beta}(q^*,p^*,k^*_\pi)
\epsilon^*_{s\beta}.
\label{eq:eps_tensor}
\eea
 From the
fact that for both $L^{\mu\nu}(k^*, k^{\prime *})$ and 
$W^{\alpha\beta}(q^*,p^*,k^*_\pi)$ their symmetric parts are real
while their antisymmetric parts are pure imaginary one derives that
\bea
{\cal L}_{rs}={\cal L}^*_{sr}\ ,\ {\cal W}_{rs}={\cal W}^*_{sr}.
\eea
The values of the different components are given by\footnote{For the case of the lepton tensor 
helicity components, their calculation is simplified if ones uses that
${\cal L}_{rs}=\epsilon^*_{r\mu}L^{\mu\nu} (k^*, k^{\prime *})
\,\epsilon_{s\nu}=\tilde\epsilon^*_{r\mu}L^{\mu\nu} (Rk, Rk')\,
\tilde\epsilon_{s\nu}$, with $\tilde\epsilon_r=B^{-1}\epsilon_r$ the corresponding polarization vectors 
associated to $Rq$ ($\tilde\epsilon^\mu_t=\frac{1}{\sqrt{Q^2}}(q ^{0},0,0,|\vec{q}\,|),\ 
 \tilde\epsilon^\mu_{\pm1}=\mp\frac1{\sqrt2}(0,1,\pm i,0),
\  \tilde\epsilon^\mu_{L}=\frac{1}{\sqrt{Q^2}}(|\vec{q}\,|,0,0, q ^{0})\ )
$. }
\begin{align}
{\cal L}_{LL}&=\frac{2}{Q^2}(|\vec{q}\,| |\vec k\,|- q^0|\vec k\,|\cos\theta)^2 -\frac{Q^2+m_l^2}2, &{\cal L}_{tt}&=m_l^2\frac{Q^2+m_l^2}{2Q^2}, \nonumber\\
{\cal L}_{tL}&={\cal L}_{Lt}=-\frac{m_l^2}{Q^2}\,(\, |\vec{q}\,||\vec k\,|-
 q^0|\vec k\,|\cos\theta),\nonumber\\
 {\cal L}_{\pm1\pm1}&=|\vec k\,|^2\sin^2\theta+\frac{Q^2+m_l^2}2\mp(|\vec{q}\,|
 |\vec k\,|- q^0|\vec k\,|\cos\theta), & {\cal L}_{+1-1}={\cal L}_{-1+1}&=-|\vec k\,|^2\sin^2\theta,\nonumber\\
{\cal L}_{\pm1L}={\cal L}_{L\pm1}&=\frac{|\vec k\,|\sin\theta}{\sqrt2\sqrt{Q^2}}
\Big[{-Q^2} \pm2(|\vec{q}\,| |\vec k\,|- q^0|\vec k\,|\cos\theta)\Big], & {\cal
L}_{t\pm1}={\cal
L}_{\pm1t}&= \mp \frac{m_l^2|\vec k\,|\sin\theta}{\sqrt2\sqrt{Q^2}}.
\end{align}

for the leptonic case and
\begin{align}
{\cal W}_{tt}&=\frac1{Q^2}\Big[( q^{*0})^2W^{00}- q^{*0}|\vec{q}\,^*|
(W^{30}+W^{03})+|\vec{q}\,^*|^2W^{33}\Big],\nonumber\\
{\cal W}_{\pm1\pm1}&=\frac12[W^{11}+W^{22}\mp i(W^{12}-W^{21})],\nonumber\\
{\cal W}_{LL}&=\frac1{Q^2}\Big[|\vec{q}\,^*|^2W^{00}- q^{*0}|\vec{q}\,^*|
(W^{30}+W^{03})+( q^{*0})^2W^{33}\Big],\nonumber\\
{\cal W}_{t\pm1}&=\frac{-1}{\sqrt2\sqrt{Q^2}}\Big[\mp q^{*0}W^{01}+i
 q^{*0}W^{02}\pm|\vec{q}\,^*|W^{31}-i|\vec{q}\,^*|W^{32}
\Big],\nonumber\\
{\cal W}_{\pm1t}&=\frac{-1}{\sqrt2\sqrt{Q^2}}\Big[\mp q^{*0}W^{10}-i
 q^{*0}W^{20}\pm|\vec{q}\,^*|W^{13}+i|\vec{q}\,^*|W^{23}
\Big],\nonumber\\
{\cal W}_{tL}&=\frac1{Q^2}\Big[ q^{*0}|\vec{q}\,^*|\,(W^{00}+W^{33})
-( q^{*0})^2\,W^{03}-|\vec{q}\,^*|^2\,W^{30}\Big],\nonumber\\
{\cal W}_{Lt}&=\frac1{Q^2}\Big[ q^{*0}|\vec{q}\,^*|\,(W^{00}+W^{33})
-|\vec{q}\,^*|^2\,W^{03}-( q^{*0})^2\,W^{30}\Big],\nonumber\\
{\cal W}_{\pm 1\mp1}&=-\frac12[W^{11}-W^{22}\pm i(W^{12}+W^{21})],\nonumber\\
{\cal W}_{\pm1L}&=\frac{-1}{\sqrt2\sqrt{Q^2}}\Big[\mp |\vec{q}\,^* |W^{10}-i
|\vec{q}\,^*|W^{20}\pm q^{*0}W^{13}+i q^{*0}W^{23}\Big],\nonumber\\
{\cal W}_{L\pm1}&=\frac{-1}{\sqrt2\sqrt{Q^2}}\Big[\mp |\vec{q}\,^* |W^{01}+i
|\vec{q}\,^*|W^{02}\pm q^{*0}W^{31}-i q^{*0}W^{32}\Big],
\end{align}
for the hadronic case.

The different contributions to the $
g_{rr}g_{ss}{\cal L}_{rs}{\cal W}_{rs}$ sum
can be separated in the following way
\begin{enumerate}
\item{${\cal S}_{tt}$}
\begin{align}
{\cal S}_{tt}={\cal L}_{tt}{\cal W}_{tt}&=m_l^2\frac{Q^2+m_l^2}{2Q^2}
\frac1{Q^2}\Big[( q^{*0})^2W^{00}- q^{*0}|\vec{q}\,^*|
(W^{30}+W^{03})+|\vec{q}\,^*|^2W^{33}\Big]\nonumber\\
&=m_l^2
\frac{Q^2+m_l^2}{2Q^2}\frac1{Q^2}\Big[( q^{*0})^2\widetilde W^{00}- q^{*0}|\vec{q}\,^*|
(\widetilde W^{30}+\widetilde W^{03})+|\vec{q}\,^*|^2\widetilde W^{33}\Big],
\end{align}
that does not depend on $\phi^*_\pi$.\\
\item{${\cal S}_{tT}+{\cal S}_{tT'}$}
\bea
{\cal L}_{t+1}{\cal W}_{t+1}&+&
{\cal L}_{+1t}{\cal W}_{+1t}+{\cal L}_{t-1}{\cal W}_{t-1}
+{\cal L}_{-1t}{\cal W}_{-1t}\nonumber\\
&=&-\frac{m_l^2|\vec k\,|\sin\theta}{{Q^2}}\Big[
 q^{*0}(W^{01}+W^{10})-|\vec{q}\,^*|(W^{31}+W^{13})\Big]\nonumber\\
&=&-\frac{m_l^2|\vec k\,|\sin\theta}{{Q^2}}\Big\{
\cos\phi^*_\pi\Big[ q^{*0}(\widetilde W^{01}+\widetilde W^{10})
-|\vec{q}\,^*|(\widetilde W^{13}+\widetilde W^{31})\Big]\nonumber\\
&&\hspace{2cm}-
\sin\phi^*_\pi\Big[ q^{*0}(\widetilde W^{02}+\widetilde W^{20})
-|\vec{q}\,^*|(\widetilde W^{23}+\widetilde W^{32})\Big]\
\Big\}\nonumber\\
&\equiv&{\cos\phi^*_\pi {\cal S}_{tT} + \sin\phi^*_\pi} {\cal S}_{tT'}
\eea
where $T$ stands for transverse and $tT$ and $tT'$ refer to the contributions
proportional to $\cos\phi^*_\pi$ and $\sin\phi^*_\pi$ respectively.\\
\item{${\cal S}_{tL}$}
\bea
{\cal S}_{tL}&=&-{\cal L}_{tL}{\cal W}_{tL}-
{\cal L}_{Lt}{\cal W}_{Lt}\nonumber\\
&=&\frac{m_l^2}{(Q^2)^2}
\,(|\vec{q}\,| |\vec k\,|- q^{0}|\vec k\,|\cos\theta)\,
\Big\{2 q^{*0}|\vec{q}\,^*|\,(W^{00}+W^{33})
-\big[( q^{*0})^2+|\vec{q}\,^*|^2\big]\,(W^{30}+W^{03})\Big\}\nonumber\\
&=&
\frac{m_l^2}{(Q^2)^2}
\,(|\vec{q}\,| |\vec k\,|- q^0|\vec k\,|\cos\theta)\,
\Big\{2 q^{*0}|\vec{q}\,^*|\,(\widetilde W^{00}+\widetilde W^{33})
-\big[( q^{*0})^2+|\vec{q}\,^*|^2\big]\,(\widetilde W^{30}+\widetilde W^{03})\Big\}
\eea
that does not depend on $\phi^*_\pi$. $L$ stands for longitudinal.\\
\item{${\cal S}_{T}$}
\bea
{\cal S}_{T}&=&{\cal L}_{+1+1}{\cal W}_{+1+1}+{\cal L}_{-1-1}{\cal
W}_{-1-1}\nonumber\\
&=&\Big[|\vec k\,|^2\sin^2\theta+\frac{Q^2+m_l^2}2\Big]\,(W^{11}+W^{22})+i(|\vec{q}\,| 
|\vec k\,|- q^{0}|\vec k\,|\cos\theta)\,(W^{12}-W^{21})\nonumber\\
&=&\Big[|\vec k\,|^2\sin^2\theta+\frac{Q^2+m_l^2}2\Big]
\,(\widetilde W^{11}+\widetilde W^{22})+i(|\vec{q}\,| 
|\vec k\,|- q^{0}|\vec k\,|\cos\theta)\,(\widetilde W^{12}-\widetilde W^{21})
\eea
which is a pure transverse term that  has no $\phi^*_\pi$ dependence.\\
\item{${\cal S}_{L}$}
\bea
{\cal S}_{L}&=&{\cal L}_{LL}{\cal W}_{LL}\nonumber\\
&=&\Big[\frac{2}{Q^2}(|\vec{q}\,| |\vec k\,|- q^{0}|\vec k\,|\cos\theta)^2
-\frac{Q^2+m_l^2}2\Big]\,\frac1{Q^2}\Big[|\vec{q}\,^*|^2W^{00}- q^{*0}
|\vec{q}\,^*|
\,(W^{30}+W^{03})+( q^{*0})^2W^{33}\Big]\nonumber\\
&=&
\Big[\frac{2}{(Q^2)^2}(|\vec{q}\,| |\vec k\,|- q^{0}|\vec k\,|\cos\theta)^2
-\frac{Q^2+m_l^2}{2Q^2}\Big]\,\Big[|\vec{q}\,^*|^2\widetilde W^{00}- q^{*0}
|\vec{q}\,^*|
\,(\widetilde W^{30}+\widetilde W^{03})+( q^{*0})^2\widetilde W^{33}\Big]
\eea
which is purely longitudinal and has no $\phi^*_\pi$ dependence.\\
\item{${\cal S}_{TT}+{\cal S}_{TT'}$}
\bea
{\cal L}_{+1-1}{\cal W}_{+1-1}&+&{\cal L}_{-1+1}
{\cal W}_{-1+1}\nonumber\\
&=&(|\vec k\,|\sin\theta)^2
\,(W^{11}-W^{22})\nonumber\\&=&(|\vec k\,|\sin\theta)^2\Big[\cos 2\phi^*_\pi
\,(\widetilde W^{11}-\widetilde W^{22})-\sin 2\phi^*_\pi
\,(\widetilde W^{12}+\widetilde W^{21})\Big]\nonumber\\
&\equiv&{\cos 2\phi^*_\pi{\cal S}_{TT}+\sin 2\phi^*_\pi{\cal S}_{TT'}}
\eea
This is also purely transverse but it has a term  in $\cos 2\phi^*_\pi$ ($TT$)
and one in  $\sin 2\phi^*_\pi$ ($TT'$).\\
\item{${\cal S}_{LT}+{\cal S}_{LT'}$}
\bea
-{\cal L}_{+1L}{\cal W}_{+1L}&-&{\cal L}_{L+1}{\cal W}_{L+1}-{\cal L}_{-1L}{\cal
W}_{-1L}-{\cal L}_{L-1}{\cal W}_{L-1}\nonumber\\
&=&-i|\vec k\,|\sin\theta\Big[ q^{*0}\,(W^{23}-W^{32})-
|\vec{q}\,^*|\,(W^{20}-W^{02})\Big]\nonumber\\
&&+\frac{2 |\vec k\,|\sin\theta }{Q^2}(|\vec{q}\,| |\vec k\,|- q^{0} |\vec k\,|\cos\theta)\,
\Big[ q^{*0}\,(W^{13}+W^{31})-|\vec{q}\,^*|\,(W^{10}+
W^{01})\Big]\nonumber\\
&{=}&\sin\phi^*_\pi\,\Big\{-i|\vec k\,|\sin\theta\,\Big[ q^{*0}
\,(\widetilde W^{13}-\widetilde W^{31})-
|\vec{q}\,^*|\,(\widetilde W^{10}-\widetilde W^{01})
\Big]\nonumber\\
&&\hspace{2cm}-\frac{2|\vec k\,|\sin\theta}{Q^2}(|\vec{q}\,| |\vec k\,|- q^0 |\vec k\,|
\cos\theta)\,
\Big[
 q^{*0}\,(\widetilde W^{23}+\widetilde W^{32})-
|\vec{q}\,^*|\,(\widetilde W^{20}+
\widetilde W^{02})\Big]\Big\}\nonumber\\
&&+\cos\phi^*_\pi\Big\{\frac{2|\vec k\,|\sin\theta}{Q^2}(|\vec{q}\,| |\vec k\,|- q^0 |\vec k\,|
\cos\theta)\,
\Big[
 q^{*0}\,(\widetilde W^{13}+\widetilde W^{31})-
|\vec{q}\,^*|\,(\widetilde W^{10}+
\widetilde W^{01})\Big]\nonumber\\
&&\hspace{2cm}-i|\vec k\,|\sin\theta\,\Big[ q^{*0}
\,(\widetilde W^{23}-\widetilde W^{32})-
|\vec{q}\,^*|\,(\widetilde W^{20}-\widetilde W^{02})
\Big]\Big\}\nonumber\\
&\equiv&{\cos\phi^*_\pi{\cal S}_{LT}+\sin\phi^*_\pi{\cal S}_{LT'}}
\eea
which comes from longitudinal-transverse interference and has a term in
$\cos\phi_\pi^*$ ($LT$) and one in $\sin\phi_\pi^*$ ($LT'$).\\

\end{enumerate}
Thus,
\bea
g_{rr}g_{ss}{\cal L}_{rs}{\cal W}_{rs}&=&({\cal S}_{tt}+{\cal S}_{T}+
{\cal S}_{L}+{\cal S}_{tL})+({\cal S}_{tT}+{\cal
S}_{LT})\,\cos\phi^*_\pi+({\cal S}_{tT'}+{\cal
S}_{LT'})\,\sin\phi^*_\pi+{\cal S}_{TT}\,\cos2\phi^*_\pi+{\cal
S}_{TT'}\,\sin2\phi^*_\pi\nonumber\\
\label{eq:weakexact}
\eea

\subsection{Zero lepton mass limit}
In the  zero lepton mass limit,  the lepton
 current is conserved and thus  we will have
 $S_{tt}=S_{tT}=S_{tT'}=S_{tL}=0$, and then
\bea
g_{rr}g_{ss}{\cal L}_{rs}{\cal W}_{rs}&\stackrel{m_l=0}{=}&({\cal S}_{T}+
{\cal S}_{L})+{\cal
S}_{LT}\,\cos\phi^*_\pi+{\cal
S}_{LT'}\,\sin\phi^*_\pi+{\cal S}_{TT}\,\cos2\phi^*_\pi+{\cal
S}_{TT'}\,\sin2\phi^*_\pi.
\label{eq:weakaaprox}
\eea
In that case
\bea
&&q^0=|\vec k\,|-|\vec k\,'|\Longrightarrow
Q^2=4|\vec k\,|\,|\vec k\,'|\sin^2\theta'/2\ \ ,\ \ 
|\vec k\,|+|\vec k\,'|=\frac{\sqrt{Q^2+|\vec q\,|^2\tan^2\theta'/2}}
{\tan\theta'/2},\nonumber\\
&&\hspace{2cm}(|\vec q\,|\,|\vec k\,|-q^0|\vec k\,|\cos\theta)=
\frac{Q^2}{2|\vec q\,|\tan\theta'/2}\,\sqrt{Q^2+|\vec q\,|^2\tan^2\theta'/2},
\eea
and introducing the quantity
\bea
\varepsilon=\frac{Q^2}{Q^2+2|\vec q\,|^2\tan^2\theta'/2}
\Longrightarrow\sqrt{ 1-\varepsilon}=\frac{\sqrt2\,|\vec q\,|\tan\theta'/2} {\sqrt{Q^2+2|\vec q\,|^2\tan^2\theta'/2}}=|\vec q\,|\tan\theta'/2 \frac{\sqrt{2\varepsilon}}{\sqrt{Q^2}}
\eea
one has that
\bea
&&\hspace{2cm}|\vec
k\,|\sin\theta=\frac{Q^2}{1-\varepsilon}\,\sqrt{\varepsilon(1-\varepsilon)}
\frac1{\sqrt2\sqrt{Q^2}}\ \ ,\   \ \
|\vec
k\,|^2\sin^2\theta+\frac{Q^2}2=\frac{Q^2}{1-\varepsilon}\frac{1}2,
\nonumber\\
&&\hspace{-1cm}\frac{2}{Q^2}\,(|\vec q\,|\,|\vec k\,|-q^0|\vec k\,|\cos\theta)^2-
\frac{Q^2}2=\frac{Q^2}{1-\varepsilon}\,\varepsilon\ \ ,\ \ \ 
\frac{2|\vec k\,|\sin\theta}{Q^2}\,(|\vec q\,|\,|\vec k\,|-q^0
|\vec k\,|\cos\theta)=
\frac{Q^2}{1-\varepsilon}\,\,\sqrt{\varepsilon(1+\varepsilon)}
\frac{1}{\sqrt2\sqrt{Q^2}}.
\eea
With the above information we can rewrite
\bea
g_{rr}g_{ss}{\cal L}_{rs}{\cal
W}_{rs}&\stackrel{m_l=0}{=}&\frac{Q^2}{1-\varepsilon}\Big\{
(\hat{\cal S}_{T1}+\sqrt{1-\epsilon^2}\,\hat{\cal S}_{T2}+\varepsilon\,
\hat{\cal S}_{L})+(\sqrt{2\varepsilon(1+\varepsilon)}\,\hat{\cal
S}_{LT1}+\sqrt{2\varepsilon(1-\varepsilon)}\,\hat{\cal
S}_{LT2})\,\cos\phi^*_\pi\nonumber\\
&&\hspace{1.cm}+(\sqrt{2\varepsilon(1-\varepsilon)}\,\hat{\cal
S}_{LT'1}+\sqrt{2\varepsilon(1+\varepsilon)}\,\hat{\cal
S}_{LT'2})\,\sin\phi^*_\pi+\varepsilon\,\hat{\cal S}_{TT}\,\cos2\phi^*_\pi+\varepsilon\,\hat{\cal
S}_{TT'}\,\sin2\phi^*_\pi\Big\},\nonumber\\
\eea
where
\begin{gather}
\begin{aligned}
\hat{\cal S}_{T1}&=\frac12(\widetilde{ W}^{11}+\widetilde{
W}^{22})=\frac12(\widetilde{\cal W}_{+1+1}+\widetilde{\cal W}_{-1-1}), & \hat{\cal S}_{T2}&=\frac{i}2\,(\widetilde{
W}_{12}-\widetilde{ W}_{21})=-\frac12(\widetilde{\cal W}_{+1+1}-\widetilde{\cal W}_{-1-1}),\nonumber\\
\hat{\cal S}_{TT}&=\frac12(\widetilde{ W}^{11}-\widetilde{
W}^{22})=-\frac12(\widetilde{\cal W}_{+1-1}+\widetilde{\cal
W}_{-1+1}), & \hat{\cal S}_{TT'}&=-\frac12(\widetilde{ W}^{12}+\widetilde{
W}^{21})=-\frac{i}2(\widetilde{\cal W}_{+1-1}-\widetilde{\cal W}_{-1+1}),\\
\end{aligned}\\
\begin{aligned}
\hat{\cal S}_{L}&=\frac1{Q^2}\Big[|\vec{q}\,^*|^2\widetilde W^{00}- q^{*0}|\vec{q}\,^*|
(\widetilde W^{30}+\widetilde W^{03})+( q^{*0})^2\widetilde W^{33}\Big] = \widetilde{\cal W}_{LL},\\
\hat{\cal S}_{LT1}&=\frac1{2\sqrt{Q^2}}\Big[
 q^{*0}\,(\widetilde W^{13}+\widetilde W^{31})-
|\vec{q}\,^*|\,(\widetilde W^{10}+
\widetilde W^{01})\Big]=-\frac{1}{2\sqrt2}(\widetilde{\cal W}_{+1L}+
\widetilde{\cal W}_{L+1}-\widetilde{\cal W}_{-1L}-\widetilde{\cal W}_{L-1}),\\
\hat{\cal S}_{LT2}&=-\frac{i}{2\sqrt{Q^2}}\Big[ q^{*0}
\,(\widetilde W^{23}-\widetilde W^{32})-
|\vec{q}\,^*|\,(\widetilde W^{20}-\widetilde W^{02})
\Big]=\frac{1}{2\sqrt2}(\widetilde{\cal W}_{+1L}+
\widetilde{\cal W}_{L+1}+\widetilde{\cal W}_{-1L}+\widetilde{\cal W}_{L-1}),\\
\hat{\cal S}_{LT'1}&=-\frac{i}{2\sqrt{Q^2}}\Big[ q^{*0}
\,(\widetilde W^{13}-\widetilde W^{31})-
|\vec{q}\,^*|\,(\widetilde W^{10}-\widetilde W^{01})
\Big]=-\frac{i}{2\sqrt2}(\widetilde{\cal W}_{-1L}-
\widetilde{\cal W}_{L-1}-\widetilde{\cal W}_{+1L}+\widetilde{\cal W}_{L+1}),\\
\hat{\cal S}_{LT'2}&=-\frac{1}{2\sqrt{Q^2}}\Big[ q^{*0}
\,(\widetilde W^{23}+\widetilde W^{32})-
|\vec{q}\,^*|\,(\widetilde W^{20}+\widetilde W^{02})
\Big]=-\frac{i}{2\sqrt2}(\widetilde{\cal W}_{-1L}-
\widetilde{\cal W}_{L-1}+\widetilde{\cal W}_{+1L}-\widetilde{\cal W}_{L+1}).
\label{eq:eses}
\end{aligned}
\end{gather}
and in analogy to Eq.~(\ref{eq:eps_tensor}), $\widetilde{\cal W}_{rs}=\epsilon_{r\mu}\,\widetilde W^{\mu\nu}\,\epsilon^*_{s\nu}$.
Finally, the differential cross section can be written as\footnote{$CC\pm$ corresponds to CC
 neutrino/antineutrino induced reactions.}
\bea
&&\hspace*{-1cm}\frac{d\sigma_{CC\pm}}{d\Omega'dE'd\Omega^*_\pi}=\Gamma\Big\{
\frac{d\sigma_{T1}}{d\Omega^*_\pi}\Big|_{\phi^*_\pi=0}\pm
\sqrt{1-\epsilon^2}\,\frac{d\sigma_{T2}}{d\Omega^*_\pi}\Big|_{\phi^*_\pi=0}+\varepsilon\,
\frac{d\sigma_{L}}{d\Omega^*_\pi}\Big|_{\phi^*_\pi=0}\nonumber\\
&&\hspace{2cm}+
\Big(\sqrt{2\varepsilon(1+\varepsilon)}\ 
\frac{d\sigma_{LT1}}{d\Omega^*_\pi}\Big|_{\phi^*_\pi=0}
\pm\sqrt{2\varepsilon(1-\varepsilon)}\ 
\frac{d\sigma_{LT2}}{d\Omega^*_\pi}\Big|_{\phi^*_\pi=0}\Big)\,\cos\phi^*_\pi
\nonumber\\&&\hspace{2cm}+\Big(\pm\sqrt{2\varepsilon(1-\varepsilon)}\ 
\frac{d\sigma_{LT'1}}{d\Omega^*_\pi}\Big|_{\phi^*_\pi=0}
+\sqrt{2\varepsilon(1+\varepsilon)}\ 
\frac{d\sigma_{LT'2}}{d\Omega^*_\pi}\Big|_{\phi^*_\pi=0}
\Big)\,\sin\phi^*_\pi\nonumber\\
&&\hspace{2cm}
+\varepsilon\,
\frac{d\sigma_{TT}}{d\Omega^*_\pi}\Big|_{\phi^*_\pi=0}
\,\cos2\phi^*_\pi+\varepsilon\,
\frac{d\sigma_{TT'}}{d\Omega^*_\pi}\Big|_{\phi^*_\pi=0}
\,\sin2\phi^*_\pi\Big\},
\label{eq:fin1}
\eea
where 
\bea
\Gamma=\frac{G_F}{2\sqrt2\, \pi^3 M^2_W}\frac{|\vec k\,'|}{|\vec k\,|}
\frac{Q^2}{1-\varepsilon}k_\gamma
\eea
with $M_W$ the $W$ boson mass and
\bea
k_\gamma=\frac{W^2_{\pi N}-M^2}{2M},
\eea
and  where
\bea
\frac{d\sigma_{b}}{d\Omega^*_\pi}\Big|_{\phi^*_\pi=0}=\frac{\pi G_F M_W^2}{\sqrt2}
\frac1{k_\gamma}
\int\frac{|\vec k^*_\pi|^2d|\vec k^*_\pi|}{E^*_\pi}\hat S_b,\hspace{1cm}b=
T1,T2,L,TT,TT',LT1,LT2,LT'1,LT'2. \label{eq:ccmultipoles}
\eea
correspond to $W^* N\to\pi N'$  differential cross sections for a virtual 
$W$ boson for given polarization states  evaluated at $\phi^*_\pi=0$. 
We have used the factor $k_\gamma$, that
has been chosen to be the same as the one that is  used
in the case of pion electroproduction (see below), and that represents the 
laboratory energy of a real photon that would give rise to the same $W_{\pi N}$ 
final pion-nucleon invariant mass. The changes appropriate for the case of NC processes are straightforward to
make.

Note that Eq.~(\ref{eq:fin1}) can also be obtained from the expressions given in 
Eqs.~(\ref{eq:dcsabcde}) and (\ref{eq:abcde}), taking advantage that in the zero lepton 
mass limit, the non-zero components of the lepton tensor $L^{\mu\nu} (k^*, k^{\prime *})$ 
read,
\begin{align}
&
\begin{aligned}
L^{00}=& |\vec{q}\,^*|^2 \frac{\varepsilon}{1-\varepsilon}, & L^{11}&= 
\frac{Q^2}{2}\, \frac{1+\varepsilon}{1-\varepsilon},  & L^{22}&= 
\frac{Q^2}{2} ,  & L^{33} &= (q^{*0})^2 \frac{\varepsilon}{1-\varepsilon},\nonumber \\
\end{aligned}\\
&
\begin{aligned}
L^{03}&= q^{*0}|\vec{q}\,^*| \frac{\varepsilon}{1-\varepsilon},& L^{31} &= \frac{q^{*0} 
\sqrt{Q^2}}{2}\, \frac{\sqrt{2\varepsilon(1+\varepsilon)}}{1-\varepsilon}  ,
&L^{01} &= \frac{ |\vec{q}\,^*|\sqrt{Q^2}}{2}\, \frac{\sqrt{2\varepsilon(1+\varepsilon)}}
{1-\varepsilon}  \\
L^{12}&= i\frac{Q^2}{2}\, \sqrt{\frac{1+\varepsilon}{1-\varepsilon}}, & L^{32} &= i 
\frac{q^{*0} \sqrt{Q^2}}{2}\, \sqrt\frac{{2\varepsilon}}{1-\varepsilon}  ,& L^{02} &= 
i \frac{ |\vec{q}\,^*|\sqrt{Q^2}}{2}\,
\sqrt\frac{{2\varepsilon}}{1-\varepsilon}\ .
\end{aligned}
 \label{eq:lepton-easy}
\end{align}
%

%
\section{Pion electroproduction}
\label{app:helicity2}
For the pure electromagnetic case,  current conservation  implies
 ${\cal S}^{em}_{tt}={\cal S}^{em}_{tT}={\cal S}^{em}_{tT'}={\cal S}^{em}_{tL}=0$. Besides, 
 since for
 that case  one  has that
$\widetilde W^{a2}_{em}=\widetilde W^{2a}_{em}=0$ for $a=0,1,3$, then 
also ${\cal S}^{em}_{TT'}=0$ and the only possible
$\phi^*_\pi$ dependencies
 are $1,\cos\phi^*_\pi,\sin\phi^*_\pi$ and 
$\cos2\phi^*_\pi$. One would then get
\bea
g_{rr}g_{ss}{\cal L}^{em}_{rs}{\cal W}^{em}_{rs}&=&({\cal S}^{em}_{T}+
{\cal S}^{em}_{L})+{\cal
S}^{em}_{LT}\,\cos\phi^*_\pi+{\cal
 S}^{em}_{LT'}\,\sin\phi^*_\pi+{\cal S}^{em}_{TT}\,\cos2\phi^*_\pi.
\label{eq:csem}
\eea
 ${\cal S}^{em}_{LT'}$ appears only in the presence of
lepton polarization.  This is the reason 
 why in this case the corresponding term is not PV despite the
 presence of $\sin\phi^*_\pi$: spin 1/2 polarization vectors are in fact
 pseudovectors\footnote{For instance, in the case of a ultrarelativistic
 lepton with well defined helicity,  the  polarization vector 
 is given by
 $h\frac{k^\mu}{m_l}$ with $h$ the helicity that changes sign under parity.} 
 and their transformation under parity involves an extra minus
 sign that  compensates  the change  of  sign of 
$\sin\phi^*_\pi$ under parity. 

If we take the case in which both  electrons are ultrarelativistic and the
initial one has
well defined  helicity $h$, the lepton tensor is\footnote{\ Note that for 
$h=\mp1$, $L^{em}_{\mu\nu}$ coincides, up to the factor 1/4, with the leptonic tensor
given  in Eq.~(\ref{eq:lt}) for the
CC neutrino/antineutrino case}
\bea
L^{em}_{\mu\nu}=\frac14 \left(k_\mu k'_\nu+k_\nu k'_\mu-g_{\mu\nu}k\cdot
k'-ih\epsilon_{\mu\nu\alpha\beta}k^{\prime\alpha}k^\beta\right)
\eea
The factor 1/4 appears because the helicity projector is 
$\left(1+h\gamma_5\right)/2$, and therefore in addition to the 
differences between coupling constants that will be discussed below, 
there is an extra factor 1/2 between the  $\nu_{\ell} \ell^- W^+$ and the 
 $ee'\gamma$ vertexes, when the initial electron is polarized. Hence, we find
\bea
&&\hspace{-1cm}g_{rr}g_{ss}{\cal L}^{em}_{rs}{\cal W}^{em}_{rs}\nonumber\\
&&=\frac14 \frac{Q^2}{1-\varepsilon}\Big\{(\hat{\cal S}^{em}_{T}+
\varepsilon\,\hat{\cal S}^{em}_{L})+\sqrt{2\varepsilon(1+\varepsilon)}\,\hat{\cal
S}^{em}_{LT}\,\cos\phi^*_\pi+h\sqrt{2\varepsilon(1-\varepsilon)}\,\hat{\cal
 S}^{em}_{LT'}\,\sin\phi^*_\pi\nonumber+\varepsilon\,\hat{\cal S}^{em}_{TT}\,
 \cos2\phi^*_\pi\Big\}.\nonumber\\ \label{eq:auxcc1}
\eea
and the differential cross section for polarized initial electron reads
\begin{equation}
\frac{d\sigma_{em}}{d\Omega'dE'd\Omega^*_\pi}=\Gamma_{em}\Big\{
\sigma_T+\varepsilon\,\sigma_L+
\sqrt{2\varepsilon(1+\varepsilon)}\,\sigma_{LT}
\,\cos\phi^*_\pi +h\sqrt{2\varepsilon(1-\varepsilon)}\,\sigma_{LT'}
\,\sin\phi^*_\pi+\varepsilon\,\sigma_{TT}
\,\cos2\phi^*_\pi
\Big\},
\label{eq:finem1}
\end{equation}
where
\bea
\Gamma_{em}=\frac{\alpha}{2\pi^2}\frac{|\vec k\,'|}{|\vec k\,|}
\frac{1}{Q^2}\frac1{1-\varepsilon}k_\gamma \label{eq:defGamma}
\eea
and
\bea
\sigma_b\equiv\frac{d\sigma^{em}_{b}}{d\Omega^*_\pi}\Big|_{\phi^*_\pi=0}=
\frac{4\pi^2\alpha}{k_\gamma}
\int\frac{|\vec k^*_\pi|^2d|\vec k^*_\pi|}{E^*_\pi}\,\hat S^{em}_b,\hspace{1cm}b=
T,L,TT,LT,LT'.
\eea
is the $\gamma^* N\to\pi N'$ differential cross section for a  virtual
photon evaluated at $\phi^*_\pi=0$.  The above expressions are easily obtained  from Eqs.~(\ref{eq:fin1})-(\ref{eq:ccmultipoles}) 
replacing\footnote{Note that for the electromagnetic case 
$\hat{\cal S}^{em}_{T2}=\hat{\cal S}^{em}_{LT2}=\hat {\cal S}^{em}_{LT'2}=0$,
since $\widetilde W^{a2}_{em}=\widetilde W^{2a}_{em}=0$, for $a=0,1,3$. We
 have also defined $\hat{\cal S}^{em}_{LT} = \hat{\cal S}_{LT1}$ and
 $\hat{\cal S}^{em}_{LT'} = -\hat{\cal S}_{LT'1}$, adding 
 in the latter an extra sign with respect to the CC case, to take into 
 account  the fact that
 the neutrino lepton tensor is recovered when the helicity $h$ is set to $-1$.} 
\begin{eqnarray}
\left(\frac{G_F M_W^2}{\sqrt{2}}\right)^\frac12 = \frac{g}{2\sqrt{2}} &\to & e = \sqrt{4\pi\alpha} \nonumber\\
M_W^2&\to & Q^2,
\end{eqnarray}
and including the factor 1/4 of Eq.~(\ref{eq:auxcc1}) in the definition of  
$\Gamma_{em}$ in Eq.~(\ref{eq:defGamma}), while the $\hat S^{em}_b$ terms
 are given in Eqs.~(\ref{eq:eses}), but 
using the electromagnetic hadron tensor associated to the gauge invariant 
electromagnetic vector current. Note that the above replacements account 
for the change in the couplings and propagators between  CC and 
electromagnetic processes.
The different contributions  in Eq.~(\ref{eq:finem1}) 
read,
\begin{eqnarray}
 \sigma_T &=& \sigma_0
\,\frac{\widetilde{\cal H}_{em}^{11}+\widetilde{\cal H}_{em}^{22}}{2},\quad
\sigma_L=\sigma_0 \frac{Q^2}{(q^{*0})^2}\,\widetilde{\cal H}_{em}^{33},\quad
\sigma_{TT}=\sigma_0
\,\frac{\widetilde{\cal H}_{em}^{11}-\widetilde{\cal H}_{em}^{22}}{2} \nonumber \\
\sigma_{LT}&=&-\sigma_0
\sqrt{\frac{Q^2}{(q^{*0})^2}}\,{\rm Re}\,\widetilde{\cal H}_{em}^{13},\quad
\sigma_{LT'}=\sigma_0
\sqrt{\frac{Q^2}{(q^{*0})^2}}\,{\rm Im}\,\widetilde{\cal H}_{em}^{13},\quad
\label{eq:stletc}
\end{eqnarray}
where 
\begin{equation}
 \sigma_0=\frac{\alpha}{16\pi Mk_\gamma}\frac{|\vec k^*_\pi|_0}{W_{\pi N}}
\end{equation}
with $|\vec k^*_\pi|_0$ defined after Eq.~(\ref{eq:phase-space}). Further,   the $\phi_\pi^*= 0$ electromagnetic
 nucleon tensor is given by
\bea
\widetilde{\cal H}_{em}^{\mu\nu}&=&{\cal H}_{em}^{\mu\nu}(p^*,
p^{\prime *}= q^*+ p^*-\hat R^{-1} k^*_\pi,\hat R^{-1} k^*_\pi)\nonumber\\
&=&
\frac12\sum_{s,s'}\langle N'( p^{\prime *},s')\,\pi(\hat R^{-1} k^*_\pi)
|j^\mu_{em}(0)|N( p^*,s)\rangle\langle N'( p^{\prime *},s')\,\pi( \hat R^{-1} k^*_\pi)
|j^\nu_{em}(0)|N( p^*,s)\rangle^*,
\eea
with $j^\mu_{em}(0)$, the electromagnetic current operator (note that 
we have already factorized out the electron charge $e$ in $\sigma_0$) 
and we have made use of  current conservation, which implies that
\bea
q_0^{*}\widetilde{\cal H}_{em}^{0\nu}=|\vec q\,^*|\widetilde{\cal H}_{em}^{3\nu},\ \ 
q_0^*\widetilde{\cal H}_{em}^{\mu0}=|\vec q\,^*|\widetilde{\cal H}_{em}^{\mu3},
\eea

Extracting the $Q^2/(q^{*0})^2$ dependence in the above expressions
and defining $\varepsilon_L= \varepsilon Q^2/(q^{*0})^2$, it is also common to write~\cite{Drechsel:1992pn}
\bea
&&\hspace*{-1cm}\frac{d\sigma_{em}}{d\Omega'dE'd\Omega^*_\pi}=\Gamma_{em}\Big\{
\sigma_{T}+\varepsilon_L\,
\hat\sigma_{L}+
\sqrt{2\varepsilon_L(1+\varepsilon)}\
\hat\sigma_{LT}
\,\cos\phi^*_\pi\nonumber
+h\sqrt{2\varepsilon_L(1-\varepsilon)}\
\hat\sigma_{LT'}
\,\sin\phi^*_\pi+\varepsilon\,
\sigma_{TT}
\,\cos2\phi^*_\pi
\Big\}.
\eea
with $\hat\sigma_L,\hat\sigma_{LT}$ and $\hat\sigma_{LT'}$ modified accordingly as
\begin{equation}
\hat \sigma_L=\sigma_0\, \widetilde{\cal H}_{em}^{33},\quad
\hat\sigma_{LT}=-\sigma_0 \,{\rm Re}\,\widetilde{\cal H}_{em}^{13},\quad
\hat\sigma_{LT'}=\sigma_0\,{\rm Im}\,\widetilde{\cal H}_{em}^{13}.
\label{eq:stletc2}
\end{equation}
 In this
work we have used, however, the expression in Eq.~(\ref{eq:finem1}) with the 
definitions given in 
Eq.~(\ref{eq:stletc}).

\bibliography{biblio_pions}

\end{document}